\documentclass{jfm}
\usepackage{graphicx}
\usepackage{epstopdf, epsfig}
\usepackage{color}
\usepackage{amsmath}

\shorttitle{Spectral mode decomposition}
\shortauthor{V. J. Shinde}

\title{Spacetime-spectral analysis of flowfields}

\author{Vilas J. Shinde
\corresp{\email{vshinde@ae.msstate.edu}}}

\affiliation{Department of Aerospace Engineering, \\ Mississippi State University, Mississippi State, MS 39762, USA}

\begin{document}

\maketitle

\begin{abstract}
The classical Fourier analysis of a time signal, in the discrete sense, provides the frequency content of signal under the assumption of periodicity.
Although the original signal can be exactly recovered using an inverse transform, the time dependence of the spectrum remains inaccessible.
There exist various time-frequency analysis techniques, such as the short time fast Fourier and wavelet transforms, but those are fundamentally limited in achieving high resolution in both the time and frequency domains concurrently.
For spatiotemporal flowfields, the frequency based modal decompositions generally provide spatial modes with a temporal counterpart that evolves at a constant frequency.
However, an accurate time-local spectral contribution and its variation over time are highly desired to better understand the intermittent/extreme events, for instance, in turbulent flowfields.
To this end, this paper presents a spectral mode decomposition that yields spectral-space and spectral-time modes, where the latter along with the associated spectral energies provide an instantaneous spectrogram.
The spatiotemporal spectral modes get both the frequency and energy ranks.
The numerical examples demonstrate the use of the proposed technique not only in spacetime-frequency analysis but also in reduced-order modeling and denoising applications.
\end{abstract}


\section{Introduction} \label{sec:intro}

Spatiotemporal fields are omnipresent in all branches of sciences, and for that matter humanities.
The study of physical phenomena typically involves a set of variables ({\it e.g.}, velocity, pressure, concentration, etc. in fluid sciences) with a spatial distribution that evolves in time; above all, it includes understanding, analysis, modeling, and prediction of these spatiotemporal variable fields.
In fluid mechanics, turbulence is a prime example that exhibits complex spatiotemporal flowfields.
Needless to say, turbulence plays a key role in transport phenomena, environmental flows, plasma physics, to mention a few.
A turbulent flow is characterized by the presence of a wide range of length/time scales, comprising complex interactions across the scales~\citep{batchelor1953theory,tennekes1972first}.
Further, it manifests spatiotemporal coherence, energy bursts, and intermittent/extreme events whose identification and understanding are of critical importance from practical standpoint.
To this end, the spectral mode decomposition (SMD) presented in this paper facilitates the identification of spectral spatial coherence and its non-sinusoidal temporal dynamics, providing a highly resolved spectrogram.

Time-frequency analysis methods have been extensively used in signal processing, primarily to identify time-local spectral events.
A more popular Fourier transform (FT) when applied to the entire time domain provides overall frequency contributions; however, the time-local presence of frequencies is not retrieved.
The Fourier transform consists of a constant amplitude sine/cosine waves, thus the time varying spectral information, {\it i.e} spectrogram, remains inaccessible.
The spectrogram is commonly obtained by employing a short time Fourier transform (STFT) repeatedly.
Generally, STFT spectrogram provides only the spectral magnitudes, losing track of the phase information; the reconstruction of the original signal based on spectrogram is an intricate process.
To account for the frequency variation over time, the sine/cosine waves are modulated to change over time, leading to wavelets.
It employs a specific multi-frequency kernel, known as a mother wavelet, to identify the time-local spectral presence.
The more common Morlet is a combination of the sine and Gaussian functions.
Wavelet transforms are effective for non-stationary signals, and are particularly effective with non-harmonic functions.
Among several limitations of time-frequency analyses~\citep{cohen1995time,morales2022time}, the limitation of trade-off between the time and frequency can be significant, as resolving the lower frequencies require longer time duration.
In addition, the time-frequency analyses are commonly employed to local/sparse spatial information and seldom to extract a highly-resolved spatial coherence.

With the availability of scale-resolved experimental and numerical simulations of turbulent flows, numerous modal decomposition techniques have emerged in order to extract the spatial/temporal coherence in seemingly chaotic fluid motions~\citep{rowley2017model,taira2017modal}.
These techniques differ in their definition of the coherence and ranking of modes, as well as specific physical characteristics of their dynamics.
The more fundamental modal decomposition techniques include: Fourier (or spectral) decomposition, proper orthogonal decomposition (POD)~\citep{lumley1970stochastic,sirovich1987}, and dynamics mode decomposition (DMD)~\citep{schmid2010dynamic,rowley2009spectral}.
A common factor among all modal decomposition techniques is that they decompose spatiotemporal flowfields into spatial coherent patterns and their associated temporal dynamics.
Although essentially these techniques build on Fourier transform/eigen-/singular-value decompositions, these bring out distinct spatiotemporal dynamics that pertains to their construct.

Fourier spatial modes represent the spectral weights at specific frequency and the corresponding time coefficients are monotones of the associated frequencies.
By definition, the temporal coefficients form an orthogonal set of temporal modes, whereas the spatial Fourier modes need not be orthogonal.
POD is another popular technique that distills most energetic spatial/temporal modes for a given rank which is based on the energy.
It employs the eigenvalue decomposition, leading to bi-orthonormal sets of space and time modes.
Importantly, the POD modes may (and generally) comprise multiple frequencies in their spatiotemporal dynamics.
The variance associated with the modes, {\it i.e.} the modal energy, defines the mode rank.
Lastly, the DMD procedure seeks for spatial modes at a specific frequency and associated growth/decay rate.
The temporal modes of DMD exhibit a single frequency that is growing or decaying in time at a constant rate.
Similar to Fourier modes, the time modes are orthogonal by definition, whereas the spatial DMD modes need not be orthogonal.
For stationary flows, the DMD modes show sustained dynamics or zero growth rate, similar to the Fourier modes.
Nonetheless, the construct of Fourier and DMD modes are characteristically different; the Fourier modes comprise a constant spectral spatial coherence, while the DMD modes exhibit spectral spatial coherence that grows/decays with time.
The FT/POD/DMD and their variants have been extremely useful in decoding the spatiotemporal dynamics, size reduction, and constructing efficient predictive models of complex turbulent flows.

The two classes of techniques, namely the time-frequency analysis and the modal decomposition, separately seek to obtain, respectively, a well-resolved time-varying spectral contribution, and a converged spatial coherent patterns with associated temporal dynamics.
In this regard, a modal decomposition technique with time-varying spectral dynamics is highly desired in order to investigate the spatiotemporal flowfields in a unified manner.
The frequency based modal decomposition methods, {i.e.} FT, DMD, and spectral variants of POD~\citep{lumley67,towne2018spectral}, provide spatial modes corresponding to a unique frequency and associated temporal coefficients that evolve at a constant frequency.
In turbulent spatiotemporal flowfields, it is crucial to know the time-local spectral contribution and its variation over time, which need not be at a fixed frequency that is associated with the mode~\citep{sieber2016spectral,nekkanti2021frequency}.
In this quest, a spectral mode decomposition procedure is developed in this paper, where the time coefficients that are associated with the spectral spatial modes capture the exact time-local modal contribution and its variation over time.
The spectral spatial modes are essentially the Fourier modes at a unique frequency and with a specific spectral energy, which provides another measure for the modal rank.

The proposed SMD method for the spacetime-frequency analysis of flowfields is demonstrated considering the following three numerical examples with increasing level of complexity: 1. transitional lid-driven cavity (LDC) at Mach $0.5$ and Reynolds number $Re_L=12000$ based on the cavity length $L$, 2. a turbulent LDC at Mach $0.5$ and $Re_L=15000$, and 3. a fully turbulent shockwave boundary layer interaction (SBLI) at Mach $2.7$ and Reynolds number $Re_{\delta_{in}}=54600$ based on the inflow boundary layer thickness ${\delta_{in}}$.
In the following section (Sec.~\ref{sec:math_framework}), a mathematical framework of Fourier transform and the SMD procedure in a matrix form are presented.
Next, the numerical examples are presented in Sec.~\ref{sec:trans_flows}, and Sec.~\ref{sec:turb_flows}.
A potential use of SMD in full/partial flow reconstruction and denoising applications is explored in the next section (Sec.~\ref{sec:rom}).
Some concluding remarks are provided in Sec.~\ref{sec:concl}.

\section{Mathematical framework}\label{sec:math_framework}


\subsection{Spacetime flowfields and Fourier transform}\label{sec:math_flow}

A spectral description of fluid flows evokes complex vector spaces that provide suitable mathematical framework to handle basis functions with infinite degrees of freedom.
Here, we consider a vector space $\textsf{E}$ as a $d$-dimensional point space with a coordinate system and a reference frame, on which the Euclidean space with $\mathbb{C}^{d=3}$ can be realized by considering an orthonormal basis.
A complex spacetime flowfield, ${\pmb{u}}$, in a suitable closed domain $\mathsf{X} \subseteq \mathsf{E}$ in $\mathbb{R}^{d=3}$ may be represented through the mapping 
\[ {\pmb{u}}:\mathsf{X}\times [0,\mathsf{T}] \rightarrow \mathbb{C}^3:({\pmb{x}},{t})\mapsto {\pmb{u}}({\pmb{x}},{t})\text{ with } {\pmb{x}}\in \mathsf{X},\]
where ${t}$ is an instant from the total time $\mathsf{T}\subset \mathbb{R}$.
The definition of flowfields include all physical quantities (scalar, vector or tensor) that are expressed at each instant and at every spatial location.
Note that the real flowfields, {\it e.g.} $\pmb{u} \in \mathbb{R}^3$, over a complex space consist zero imaginary parts.

The complex inner product space can be given by the map:
\[ \langle \cdot, \cdot \rangle: \mathsf{X} \times \mathsf{X} \rightarrow \mathbb{C}^3, \]
satisfying the properties, namely, linearity, conjugate symmetry, and positive definitiveness.
Thus, a complex inner product $\langle {\pmb{u}}, {\pmb{u}} \rangle > 0$ for a non-zero ${\pmb{u}}$, and the complex norm $\| {\pmb{u}} \|=\sqrt{\langle {\pmb{u}},{\pmb{u}} \rangle}$ for any ${\pmb{u}}$, providing its magnitude.
The inner product over space utilizes spatial weights, accounting for the non-uniformity of the spatial domain.
The flowfields are assumed to be square integrable, i.e., $\| {\pmb{u}} \|^2<\infty$, providing a suitable space for the Fourier transform.

The Fourier transform of, $\pmb{u}(\pmb{x},t)$, on a time line is commonly defined as,
\begin{equation} \label{eq:fft}
{\pmb{s}}(\pmb{x},f) = \int_\mathsf{T} \pmb{u}(\pmb{x},t) e^{-i 2\pi f t} dt,
\end{equation}
where $i=\sqrt{-1}$.
${\pmb{s}}(\pmb{x},f) \in \mathbb{C}^{d=3}$ is a complex valued flowfield, and $f$ is a frequency from the entire spectrum $\mathsf{F}\subset \mathbb{R}$.
The inverse Fourier transform is then defined as,
\begin{equation} \label{eq:ifft}
{\pmb{u}}(\pmb{x},t) = \int_\mathsf{F} {\pmb{s}}(\pmb{x},f) e^{i 2\pi f t} df,
\end{equation}
returning the original flowfield, ${\pmb{u}}(\pmb{x},t)$.
The conservation of energy in Fourier (time-frequency) transformation is ensured by the Parseval-Plancherel theorems,
\begin{equation}\label{eq:parseval-plancherel}
    \int_\mathsf{T} |\pmb{u}(\pmb{x},t)|^2 dt = \int_\mathsf{F}|{\pmb{s}}(\pmb{x},f)|^2 df,
\end{equation}
which states that the total time-integrated energy is equal to the total frequency-integrated (spectral) energy of a function.
In the context spacetime flowfields, we can rewrite Eq.~\ref{eq:parseval-plancherel} as
\begin{align}
    \int_\mathsf{T} \left( \int_\mathsf{X} |\pmb{u}(\pmb{x},t)|^2 d\pmb{x} \right) dt &= \int_\mathsf{F} \left(\int_\mathsf{X}|{\pmb{s}}(\pmb{x},f)|^2 d\pmb{x}\right) df \nonumber\\
     &= \int_\mathsf{F} \lambda_{f} df
\end{align}
where $\lambda_f$ is a space-integrated total spectral energy at frequency $f$.

The factors such as the discrete version of Fourier transform, the fast Fourier transform (FFT) algorithm, and its ability to analyze/reconstruct complex flowfields have been instrumental in the popularity of Fourier transform.
To be consistent with the discrete formulation, we present the development of the spectral mode decomposition in the matrix form in the following section.

\subsection{Spectral mode decomposition}\label{sec:math_smd}

Let us consider a tensor ${\pmb{X}}$ that comprises the spacetime flowfields, \textit{e.g.} the velocity vector $\pmb{u}(\pmb{x},t)$ with the discrete time $t \in [0,\mathsf{T}]$, such that $t=\{t_{1}, t_{2},\cdots,t_n\}$.
Similarly, the space coordinate vector $\pmb{x}$ is considered to be discrete of size $m$, thus ${\pmb{X}} \in \mathbb{R}^{m\times n}$.
The singular value decomposition (SVD) of $\pmb{X}$ can be given as
\begin{equation}\label{eq:svd}
    \pmb{X}=\pmb{U}\pmb{\Sigma}\pmb{V}^{T},
\end{equation}
where $\pmb{U}$ and $\pmb{V}$ are the real unitary matrices comprising the left and right eigenvectors, respectively.
The corresponding non-zero real singular values form the diagonal matrix $\pmb{\Sigma}$.
Equation~\ref{eq:svd} can be written as:
\begin{equation}\label{eq:svd_ff}
    \pmb{X}=\pmb{U}\pmb{V}^{T}\pmb{\mathcal{F}}\pmb{\mathcal{F}}^\dagger\pmb{V}\pmb{\Sigma}\pmb{V}^{T},
\end{equation}
where $\pmb{\mathcal{F}}\in \mathbb{C}^{n\times n}$ is the discrete Fourier transform operator with an adjoint $\pmb{\mathcal{F}}^\dagger \in \mathbb{C}^{n\times n}$, such that
\begin{equation}
    \pmb{\mathcal{F}}^\dagger \pmb{\mathcal{F}} = \pmb{I},
\end{equation} 
forming an orthonormal basis.
Here, $\pmb{I}\in \mathbb{R}^{n\times n}$ is an identity matrix.
The forward and inverse Fourier transform matrices, with $\varpi=e^{(-i2\pi/n)}$, are respectively given by:
\begin{equation}
\pmb{\mathcal{F}}=
\begin{bmatrix} 
\cdot & \cdot & \cdots & \cdot & \cdot\\
\cdot & \cdot & \cdots & \cdot & \cdot\\
\vdots & \vdots & \frac{\varpi^{jk}}{\sqrt{n}} & \vdots & \vdots \\
\cdot & \cdot & \cdots & \cdot & \cdot\\
\cdot & \cdot & \cdots & \cdot & \cdot
\end{bmatrix}, \text{ and }
\pmb{\mathcal{F}^\dagger}=
\begin{bmatrix} 
\cdot & \cdot & \cdots & \cdot & \cdot\\
\cdot & \cdot & \cdots & \cdot & \cdot\\
\vdots & \vdots & \frac{(\varpi^\ast)^{jk}}{\sqrt{n}} & \vdots & \vdots \\
\cdot & \cdot & \cdots & \cdot & \cdot\\
\cdot & \cdot & \cdots & \cdot & \cdot
\end{bmatrix},
\end{equation}
where $(j=0,1,\cdots,n-1)$ and $(k=0,1,\cdots,n-1)$ are the running indices that provide the elements of the matrices.
For instance, the first and last elements of the matrices have exponents $j,k=0$ and $j,k=(n-1)$, respectively.
$\varpi^\ast$ is the complex conjugate of $\varpi$, i.e., $\varpi^\ast=e^{(i2\pi/n)}$.

The spectral space modes ($\pmb{\Phi} \in \mathbb{C}^{m\times n}$) and time modes ($\pmb{\Psi} \in \mathbb{C}^{n\times n}$) of the spacetime flowfields $\pmb{X}$ are, respectively,
\begin{equation}\label{eq:smd_phi_psi}
    \pmb{\Phi}=\pmb{U}\pmb{V}^{T}\pmb{\mathcal{F}} \text{ and } \pmb{\Psi}\pmb{\Lambda}^{1/2}=\pmb{V}\pmb{\Sigma}\pmb{V}^T\pmb{\mathcal{F}}.
\end{equation}
The real diagonal matrix $\pmb{\Lambda} \in \mathbb{R}^{n\times n}$ comprises the spectral energies associated with the modes.
The space modes are orthonormal, {\it i.e.}, $\pmb{\Phi}^\dagger\pmb{\Phi}=\pmb{I}$, while the time modes are not necessarily orthonormal, {\it i.e.}, $\pmb{\Psi}^\dagger\pmb{\Psi}\neq\pmb{I}$.
The spacetime flowfields in, $\pmb{X}$, can be expressed as:
\begin{equation}\label{eq:rec}
    \pmb{X}={\pmb{\Phi}}\pmb{\Lambda}^{\frac{1}{2}}{\pmb{\Psi}}^\dagger.
\end{equation}

The matrix $\pmb{\varphi}={\pmb{\Psi}}\pmb{\Lambda}^{\frac{1}{2}} \in \mathbb{C}^{n\times n}$ comprises non-normalized time modes, where each mode corresponds to a spatial mode at a unique frequency.
The squared norm of the individual modes of $\pmb{\varphi}$ also provides the spectral energies, {\it i.e.} $\|\varphi_f\|^2_\mathsf{T}=\lambda_f$, with $\pmb{\Lambda}=\text{diag}\{\lambda_1,\cdots, \lambda_f, \cdots, \lambda_n \}$.
Notably, this provides with an access to the temporal variation of the spectral energies - the spectrogram.

The SMD procedure transforms the spatio-temporal flowfields of $\pmb{X}$ into a set of complex space ($\pmb{\Phi}$) and time ($\pmb{\Psi}$) modes, and associated real energies ($\lambda_f$).
The orthonormal space modes have a unique frequency associated with each mode, while the corresponding time modes exhibit temporal variations of the space modes.
Unlike the Fourier time coefficients, the SMD time modes are not uniform over the entire time span.
This enables the non-normalized time modes, $\pmb{\varphi}$, to provide the temporal variation of the spectral energy associated with each mode.
As such, the spectrogram produced is at a highest time resolution of the original flowfields.

Alternatively, the non-normalized time modes $\pmb{\varphi}$ can be obtained by projecting the flowfields on the spatial modes, following SVD of Eq.~\ref{eq:svd}, as:
\begin{eqnarray}\label{eq:varphi}
        {\pmb{\varphi}}&=&\pmb{X}^T\pmb{\Phi} \label{eq:spectro}\\
        &=&\pmb{V}\pmb{\Sigma}\pmb{U}^T\pmb{U}\pmb{V}^T\pmb{\mathcal{F}}\\
        &=&\pmb{V}\pmb{\Sigma}\pmb{V}^T\pmb{\mathcal{F}}
\end{eqnarray}
As with the Fourier transform, the spectral modes in Eq.~\ref{eq:spectro} can an be replaced by alternatives formulations to obtain the spectrogram associated with, namely, the absolute spectral modes, the power spectral density, or only the real or imaginary parts of the spectral modes.
Nonetheless, the standard complex spectral modes ensure an exact reconstruction of the original flowfields via Eq.~\ref{eq:rec}.

\subsubsection{Relation with proper orthogonal decomposition and its spectral variants}

The POD and its relation with the compact SVD (Eq.~\ref{eq:svd}) are well known.
While SVD can be performed on any $m\times n$ matrix, POD requires a diagonalizable matrix that can be factorized as $\pmb{C}_{mm}=\pmb{U}\pmb{\Sigma}^2\pmb{U}^{-1} \in \mathbb{R}^{m\times m}$ or $\pmb{C}_{nn}=\pmb{V}\pmb{\Sigma}^2\pmb{V}^{-1} \in \mathbb{R}^{n \times n}$.
The SVD and POD are related by the data matrix multiplication as:
\begin{equation}
    \pmb{C}_{mm}=\pmb{X}\pmb{X}^T=\pmb{U}\pmb{\Sigma}\pmb{V}^T (\pmb{U}\pmb{\Sigma}\pmb{V}^T)^T=\pmb{U}\pmb{\Sigma}^2\pmb{U}^T
\end{equation}
or
\begin{equation}
    \pmb{C}_{nn}=\pmb{X}^T\pmb{X}=(\pmb{U}\pmb{\Sigma}\pmb{V}^T)^T\pmb{U}\pmb{\Sigma}\pmb{V}^T=\pmb{V}\pmb{\Sigma}^2\pmb{V}^T,
\end{equation}
The spectral modes of Eq.~\ref{eq:smd_phi_psi} can be expressed in terms of the above POD correlation matrices as:
\begin{equation}
    \pmb{\Phi} = \pmb{U}\pmb{\Sigma}^{-1}\pmb{U}^T \pmb{X}\pmb{\mathcal{F}} = \pmb{C}_{mm}^{-1/2}\pmb{X}\pmb{\mathcal{F}}
\end{equation}
and 
\begin{equation}
    \pmb{\Psi}\pmb{\Lambda}^{1/2} = \pmb{V}\pmb{\Sigma}\pmb{V}^T \pmb{\mathcal{F}} = \pmb{C}_{nn}^{1/2}\pmb{\mathcal{F}},
\end{equation}
respectively.
These equations manifest the relations among the POD, the Fourier operator, and the SMD modes.

The POD procedure leads to spatial and temporal sets of orthonormal modes and associated modal energies that are optimal for a given number of modes~\citep{kosambi1943statistics,lumley67,berkooz1993proper}.
The spatio-temporal POD modes are uncorrelated in both space and time and may comprise multiple wavenumbers/frequencies, exhibiting a non-zero coherence among the modes.
A major point of difference between the POD and SMD is that each SMD mode is associated with a single frequency, which also has a negative counterpart.
The SMD modes appear in complex pairs, where a mode at a negative frequency has identical real part but negative imaginary part when compared to the corresponding mode at the positive frequency.
Unlike POD, the SMD temporal modes are not necessarily uncorrelated.
In terms of the similarities, both the POD and SMD spatial modes are orthonormal, where the modal energies can be utilized to define their ranks.
Furthermore, the spatial and temporal modes for both techniques can exhibit a non-zero coherence among the modes.

The spectral variants of POD typically utilize the spatial Fourier modes at a single frequency and subject those to the POD procedure, leading to energy optimal spectral modes at that frequency~\citep{lumley67,glauser1987coherent,citriniti2000reconstruction,braud2004low,towne2017spectral}.
Analogous to the standard space-time POD, the spectral POD (SPOD) constructs a space or time cross-spectral density tensor to solve an eigenvalue problem.
The resultant SPOD modes are associated with that single frequency, meaning that at a single frequency there exist a multitude of SPOD modes.
As with the Fourier modes, the SPOD modes at different frequencies are not orthogonal to each other by construction.
On the contrary, the SMD spatial modes at different frequencies are orthogonal as well as they are bijective in terms of the modes and associated frequencies.
While an SMD mode is associated with a single frequency, its temporal evolution need not exhibit a uniform amplitude, particularly in turbulent flows.

\subsubsection{Relation with dynamic mode decomposition}

The method of DMD aims at extracting coherent spatial modes where each mode evolves at a single frequency and has a characteristic modal growth or decay over the selected time horizon.
DMD has been extensively used in many applications and there exist several variants of the method~\citep{chen2012variants, jovanovic2014sparsity}.
For the comparison with SMD, we consider the DMD in its original form presented by~\citet{schmid2010dynamic}.

Briefly, a set of flow snapshots are arranged in two matrices $\pmb{X}$ and $\pmb{Y}$, where $\pmb{Y}$ comprises the time-shifted columns of $\pmb{X}$, and $\pmb{X},\pmb{Y}\in \mathbb{R}^{m\times n}$.
The DMD procedure then aims at finding a matrix $\pmb{A}\in \mathbb{R}^{m\times m}$ such that 
\begin{equation}\label{eq:ax=y}
\pmb{A}\pmb{X}=\pmb{Y}.
\end{equation}
The compact singular value decomposition of $\pmb{X}$, \textit{i.e.} $\pmb{X}=\pmb{U}\pmb{\Sigma}\pmb{V}^T$, provides a lower dimensional representation of $\pmb{A}$ as,
\begin{equation}\label{eq:Atilde}
\tilde{\pmb{A}}=\pmb{U}^T\pmb{A}\pmb{U}=\pmb{U}^T\pmb{Y}\pmb{V}\pmb{\Sigma}^{-1} \in \mathbb{R}^{n\times n}.
\end{equation}
The DMD modes, $\pmb{\phi}\in \mathbb{C}^{m\times n}$ are given by 
\begin{equation}\label{eq:dmd_phi}
\pmb{\phi}=\pmb{U}\pmb{\psi},
\end{equation}
where the eigenvectors $\pmb{\psi}\in \mathbb{C}^{n\times n}$ are obtained by solving an eigenvalue problem:
\begin{equation}
\tilde{\pmb{A}}\pmb{\psi}=\pmb{\psi}\pmb{\kappa}.
\end{equation}
The diagonal matrix, $\pmb{\kappa}$, consists of complex eigenvalues that lead to the growth rates and frequencies of the DMD modes.

A formal relation between the SMD (Eq.~\ref{eq:smd_phi_psi}) and DMD (Eq.~\ref{eq:dmd_phi}) space modes can be given as:
\begin{equation}\label{eq:smd_dmd_phi}
    \pmb{\Phi}\pmb{\mathcal{F}}^\dagger\pmb{V}=\pmb{\phi}\pmb{\psi}^\dagger.
\end{equation}
The DMD spatial modes are not necessarily orthogonal, as the companion matrix $\tilde{\pmb{A}}$ can be non-symmetrical (non-normal).
However, the corresponding temporal coefficients exhibit orthogonality, as they are pure sinusoids which linearly grow or decay over the considered time horizon.
On the contrary, the SMD space modes are orthogonal by construction, while their temporal evolution is not necessarily uncorrelated.
In other words, while the DMD modes grow or decay linearly, whereas the SMD modes exhibit coherent temporal evolution pertinent to the instantaneous flow physics.
In the following sections, we demonstrate the use of SMD for spacetime-frequency analysis of spatiotemporal flowfields, in terms of high-fidelity direct numerical/large eddy simulations (DNS/LES) databases of flows with increasing complexity.

\section{Lid-driven cavity flow}\label{sec:trans_flows}

The LDC flow is a canonical case study in fluid dynamics.
It is simple in terms of its geometry yet it exhibits a fairly complex flow physics.
The flow remains steady until a critical value of Reynolds number, and becomes unsteady via Hopf bifurcation~\citep{ghia1982high,shen1991hopf,ramanan1994linear}.
In the post-bifurcation regime, the flow three-dimensionality alongside the end-wall effects become significant from the flow physics viewpoint~\citep{koseff1983three,sheu2002flow,albensoeder2005accurate,lopez2017transition}.
Nonetheless, DNS of two-dimensional LDC serves as a canonical benchmark for many efforts~\citep{bruneau20062d,shinde2021lagrangian,shinde2025lagrangian}, including the present.
In addition, the flow compressibility is known to have stabilizing effect on the LDC flow dynamics~\citep{bergamo2015compressible,ohmichi2017compressibility,ranjan2020robust,shinde2021lagrangian,shinde2025lagrangian}, increasing the value of the critical Reynolds number.
For Mach $0.5$, the critical Reynolds number based on the cavity length is $Re_L\approx 11000$~\citep{ohmichi2017compressibility,shinde2021lagrangian,shinde2025lagrangian}.

To demonstrate the use of SMD, we consider the LDC flow at post-critical Reynolds numbers of $Re_L=12000$ and $Re_L=15000$, anticipating a few frequency dominant unsteadiness at the lower Reynolds number case and an increased level of turbulence at the higher Reynolds number case.
The computational details on the geometry and grid convergence are provided in an appendix of~\cite{shinde2021lagrangian}; thus, those details are not reproduced here for brevity.
Here, we utilize a database that has an LDC flow domain discretized using the grid resolution of $501\times 501$, while the time domain comprises $1024$ snapshots that are separated by a time-step of $\delta t=0.05$.
This leads to the frequency resolution of $\delta f\approx 0.01955$ over $f\in[0,10]$, set by the Nyquist criterion.


\begin{figure}
\centering
\begin{minipage}{0.24\textwidth}
\centering {\includegraphics[width=1.0\textwidth]{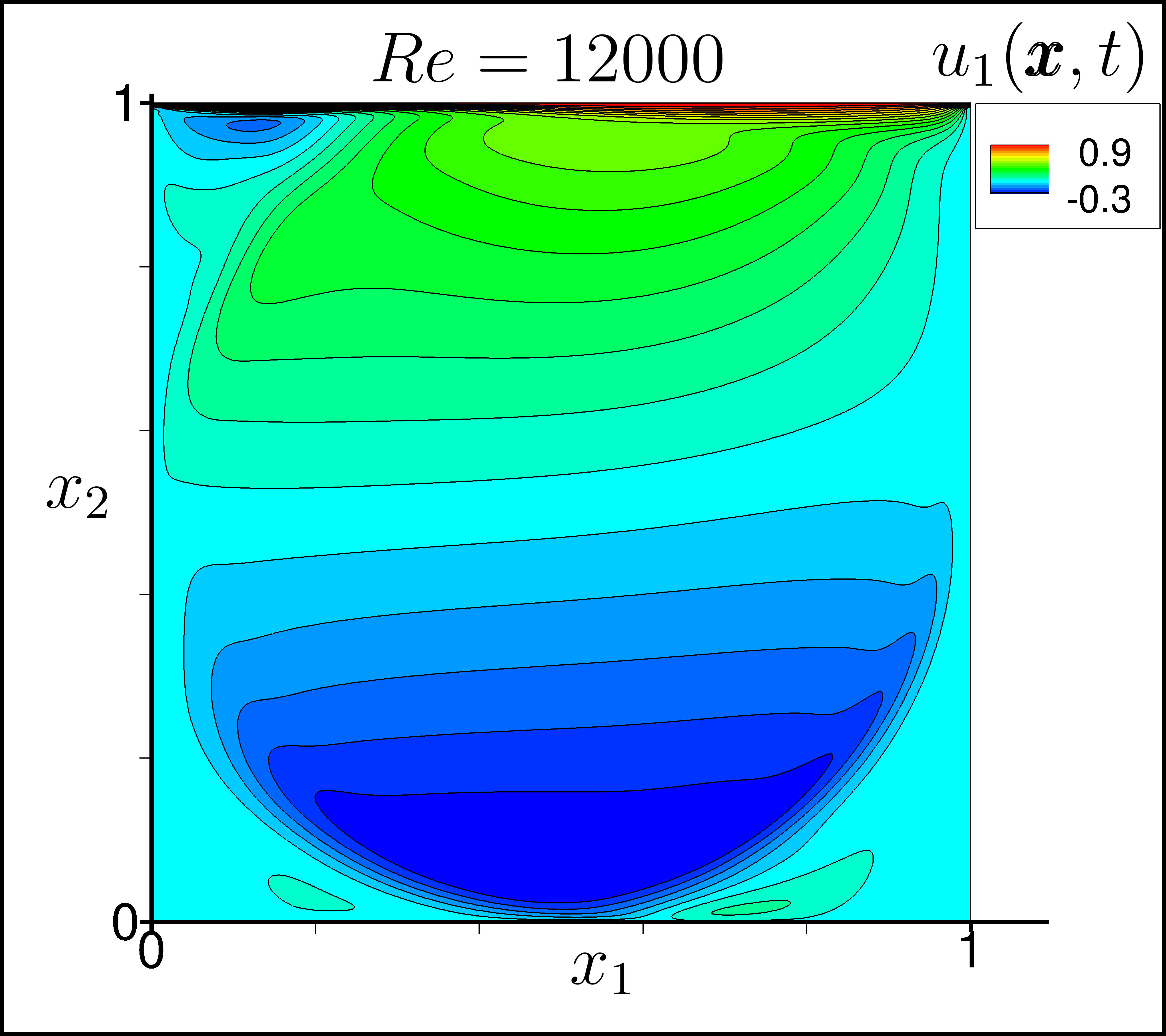}}\\(a) 
\end{minipage}
\begin{minipage}{0.24\textwidth}
\centering {\includegraphics[width=1.0\textwidth]{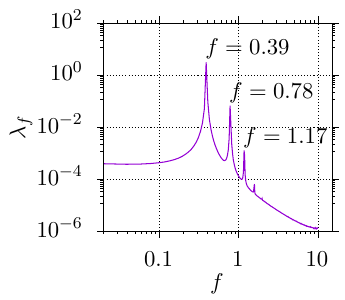}}\\(b) 
\end{minipage}
\begin{minipage}{0.24\textwidth}
\centering {\includegraphics[width=1.0\textwidth]{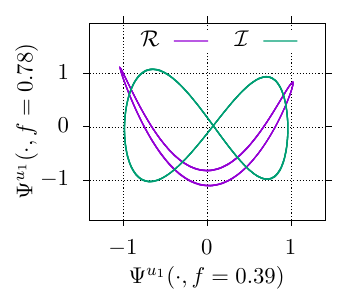}}\\(c) 
\end{minipage}
\begin{minipage}{0.24\textwidth}
\centering {\includegraphics[width=1.0\textwidth]{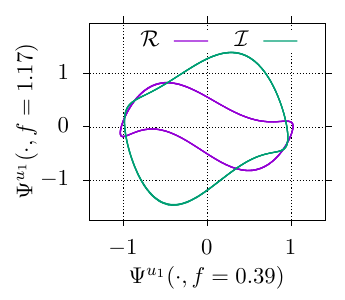}}\\(d) 
\end{minipage}\\
\begin{minipage}{0.24\textwidth}
\centering {\includegraphics[width=1.0\textwidth]{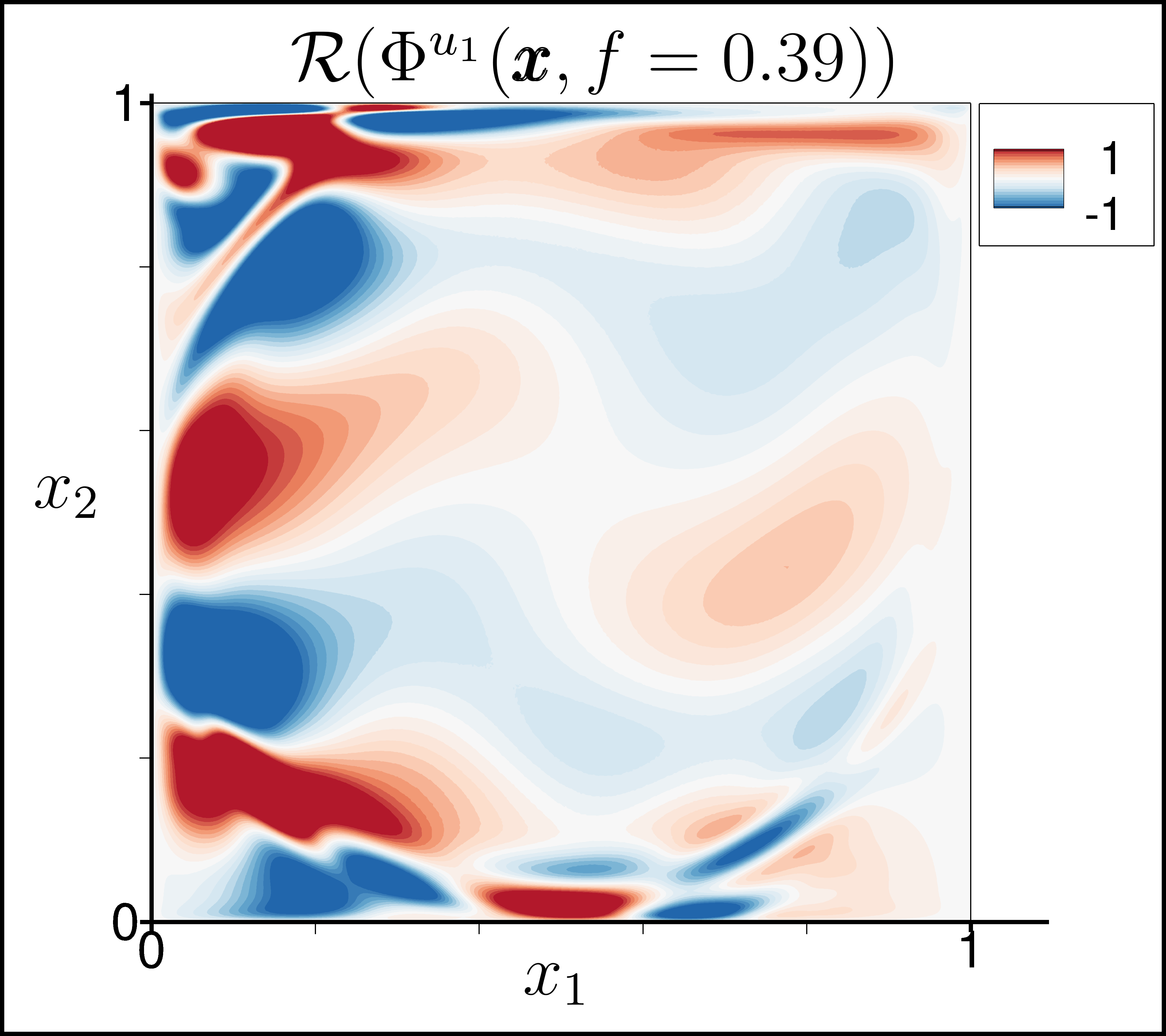}}\\(e)
\end{minipage}
\begin{minipage}{0.24\textwidth}
\centering {\includegraphics[width=1.0\textwidth]{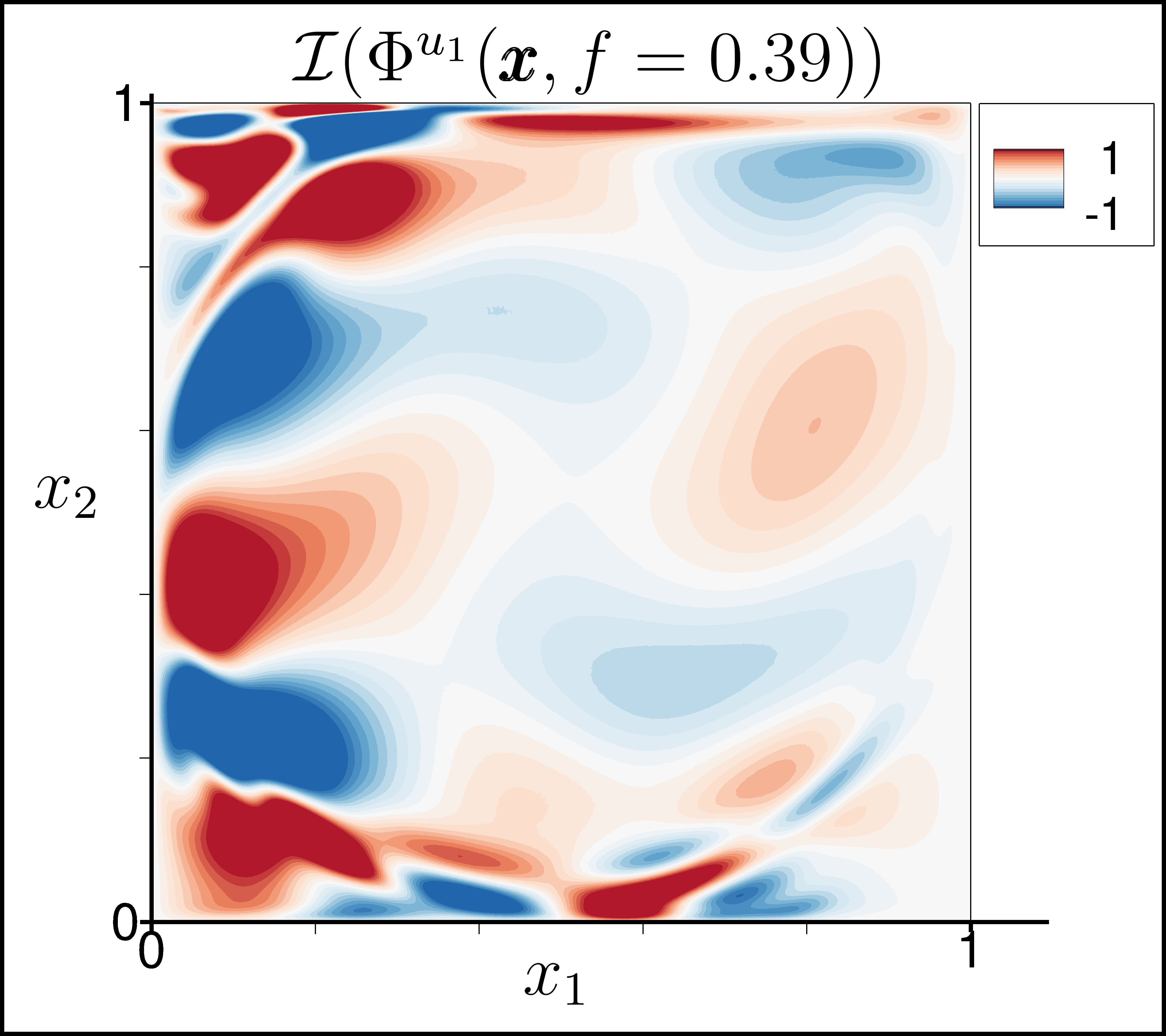}}\\(f) 
\end{minipage}
\begin{minipage}{0.24\textwidth}
\centering {\includegraphics[width=1.0\textwidth]{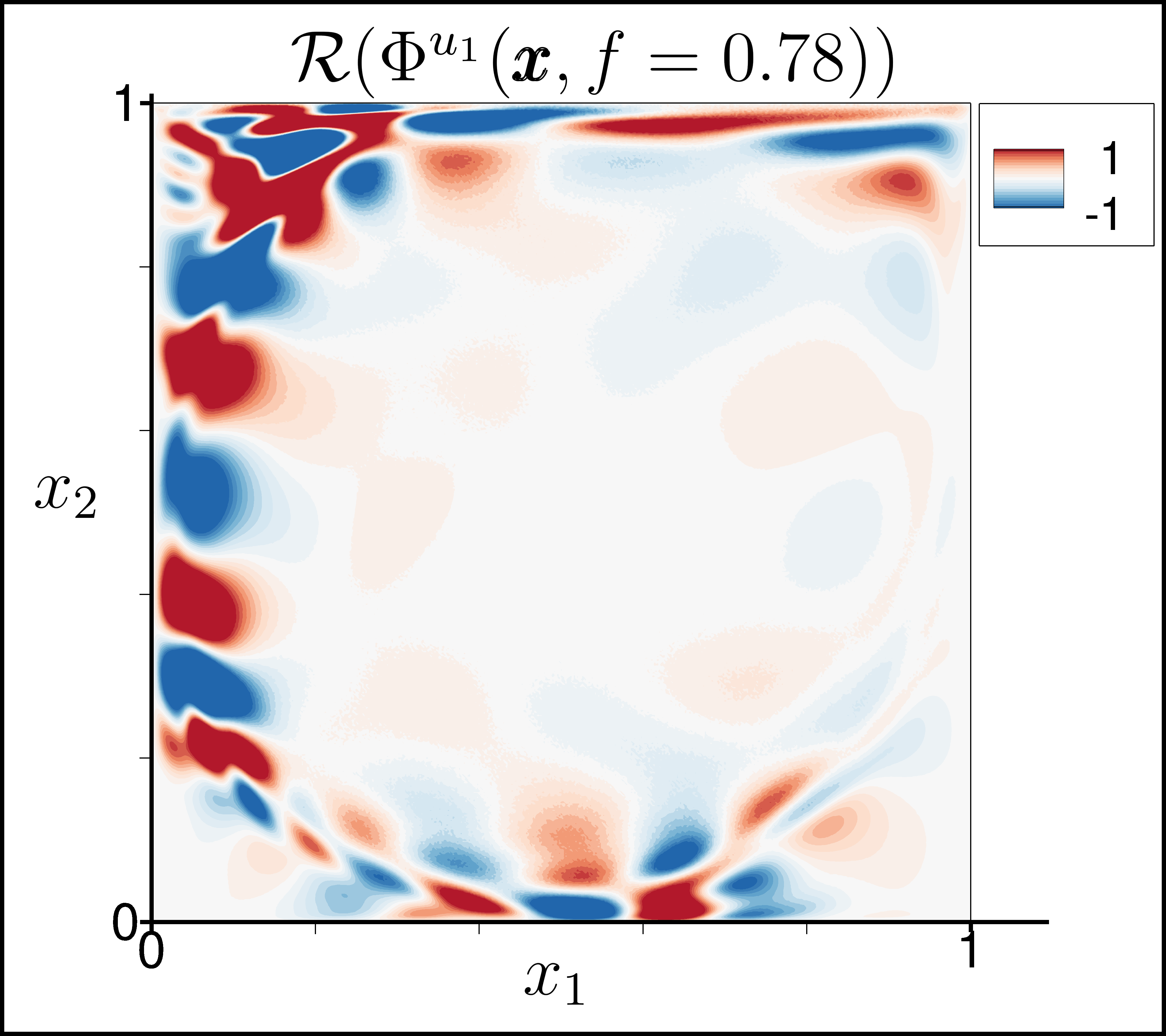}}\\(g) 
\end{minipage}
\begin{minipage}{0.24\textwidth}
\centering {\includegraphics[width=1.0\textwidth]{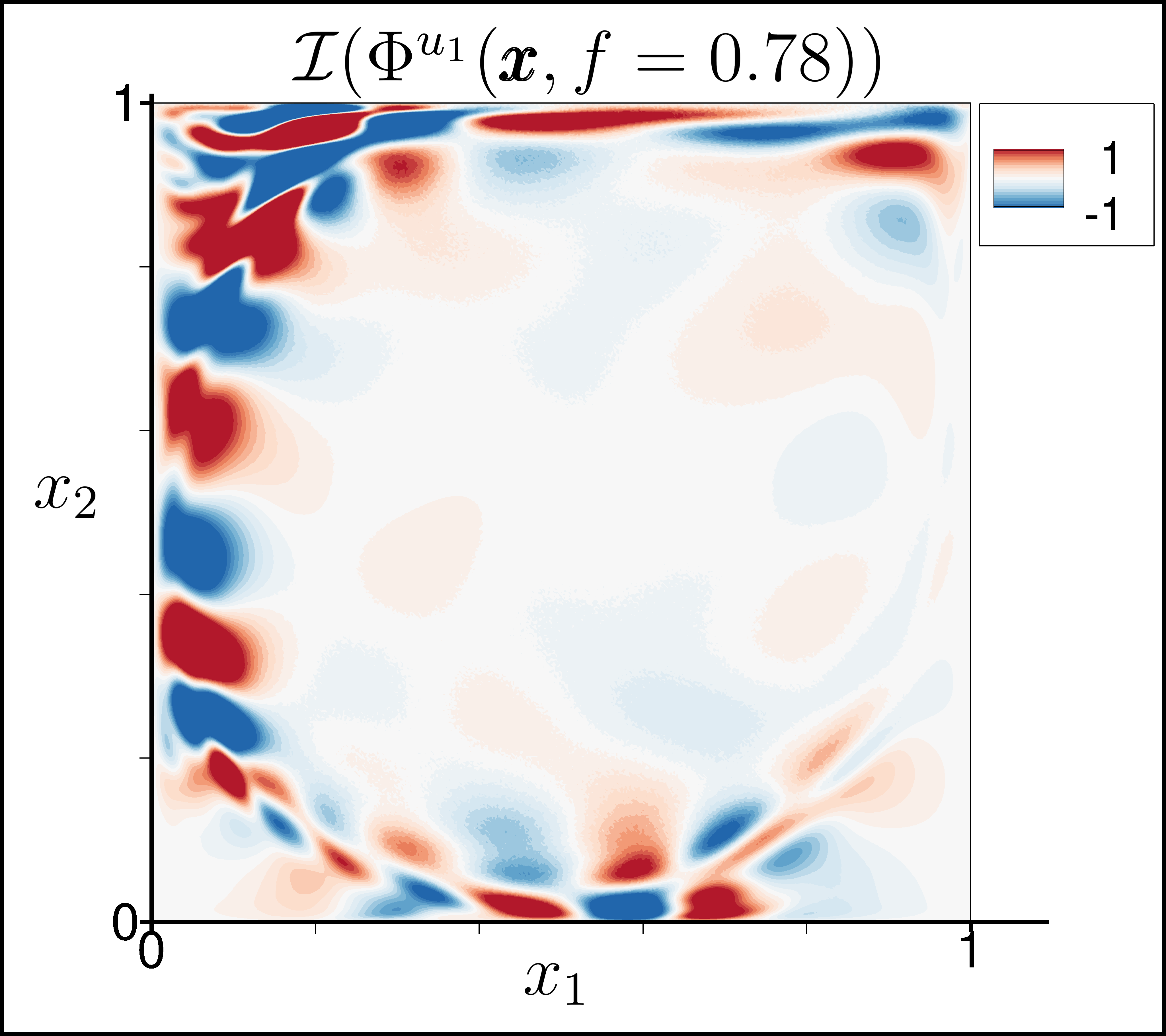}}\\(h)
\end{minipage}
\caption{Spectral mode decomposition of LDC at $Re=12000$. (a) An instantaneous DNS flowfield of the horizontal velocity component. (b) Spectral energies of the SMD modes. (c) Phase portraits of the SMD time-modes. (d) Real part of the SMD mode at $f=0.39$. (e) Imaginary part of the SMD mode at $f=0.39$. (d) Real part of the SMD mode at $f=0.78$. (e) Imaginary part of the SMD mode at $f=0.78$.}
\label{fig:smd_ldc_12k}
\end{figure}

The solution vector, of the full compressible Navier-Stokes equations, comprises the three velocity components, the pressure, and the density of the flow.
For the SMD analysis and its comparison with POD/DMD, we consider spacetime flowfields of the horizontal velocity ($u_1$).
An instance of the horizontal velocity for $Re=12000$ is displayed in Fig.~\ref{fig:smd_ldc_12k}(a).
At this post-critical Reynolds number, the LDC flow exhibits a prominent spectral peak at $f=0.39$ and its two super-harmonics at $f=0.78$, and $f=1.16$, which are evident in the SMD modal energy graph of Fig.~\ref{fig:smd_ldc_12k}(b).
Interestingly, the stability analysis of a pre-critical~\citep{shinde2021lagrangian,shen1991hopf,ohmichi2017compressibility} LDC baseflow at $Re_L=7000$ and a post-critical~\citep{shinde2025lagrangian} LDC flow at $Re_L=12000$ indicate that the first and second instability modes have frequencies of $f=0.13$, $f=0.26$, respectively.
Here, at $Re=12000$, the most energetic SMD mode appears at the third super-harmonic of the first instability mode frequency.

\begin{figure}
\centering
\begin{minipage}{0.24\textwidth}
\centering {\includegraphics[width=1.0\textwidth]{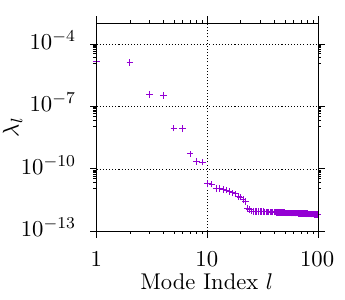}}\\(a)
\end{minipage}
\begin{minipage}{0.24\textwidth}
\centering {\includegraphics[width=1.0\textwidth]{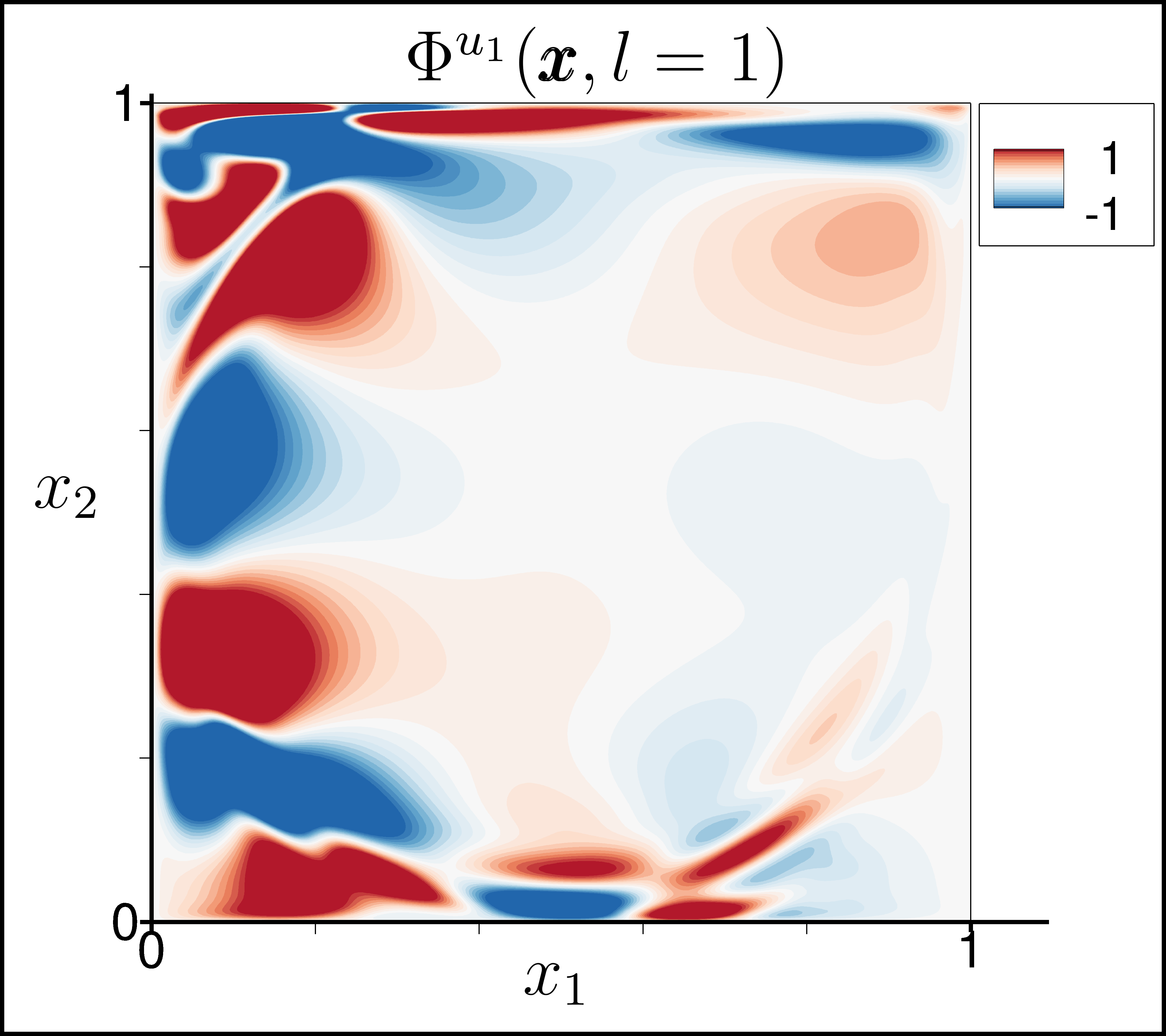}}\\(b) 
\end{minipage}
\begin{minipage}{0.24\textwidth}
\centering {\includegraphics[width=1.0\textwidth]{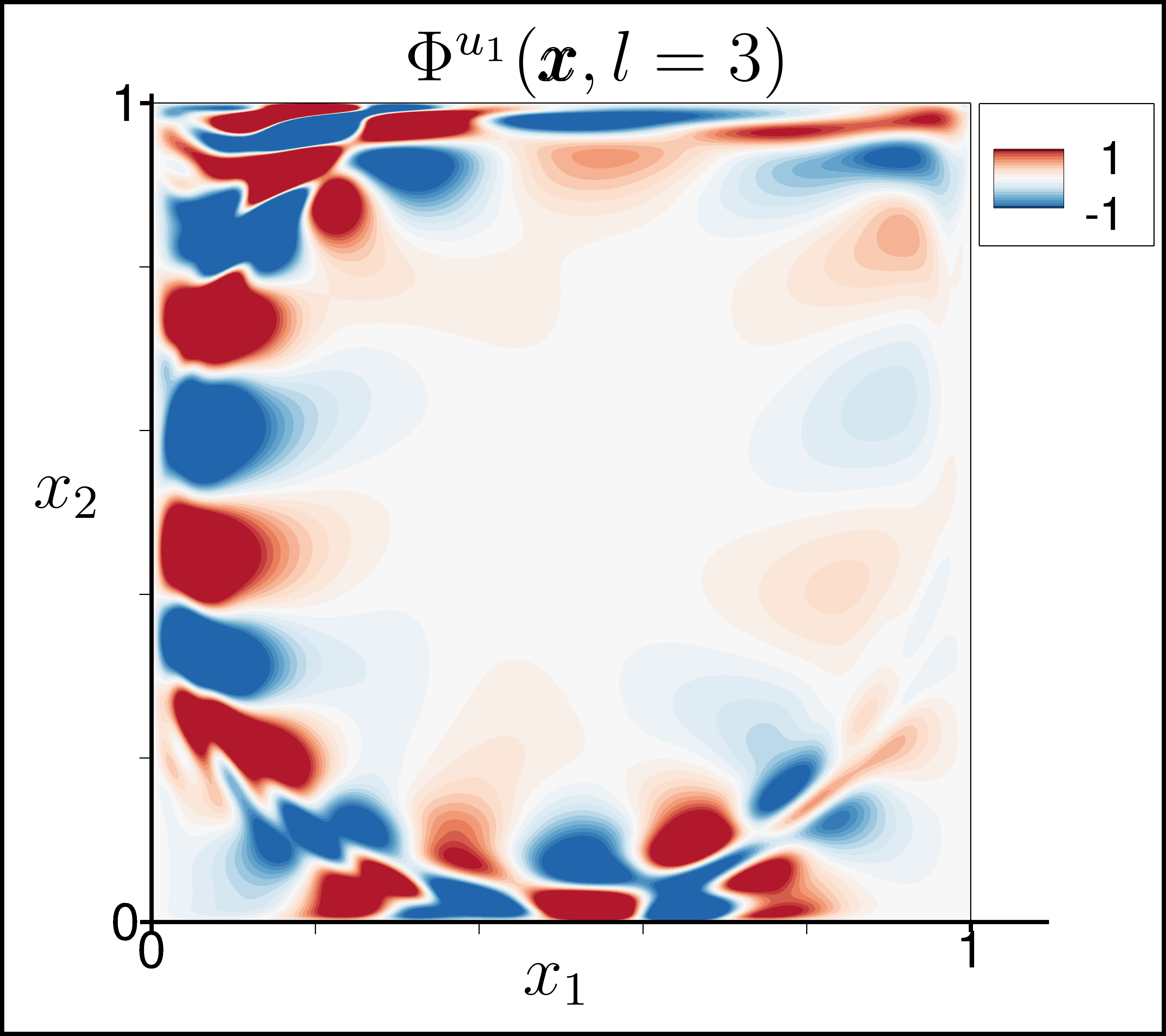}}\\(c) 
\end{minipage}
\begin{minipage}{0.24\textwidth}
\centering {\includegraphics[width=1.0\textwidth]{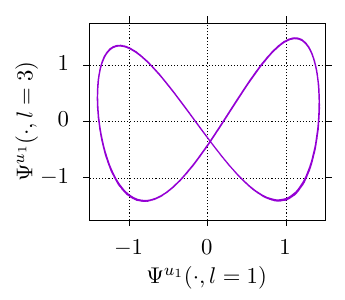}}\\(d)
\end{minipage}
\begin{minipage}{0.24\textwidth}
\centering {\includegraphics[width=1.0\textwidth]{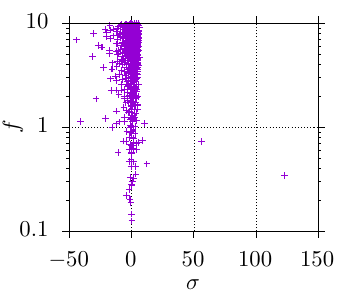}}\\(e)
\end{minipage}
\begin{minipage}{0.24\textwidth}
\centering {\includegraphics[width=1.0\textwidth]{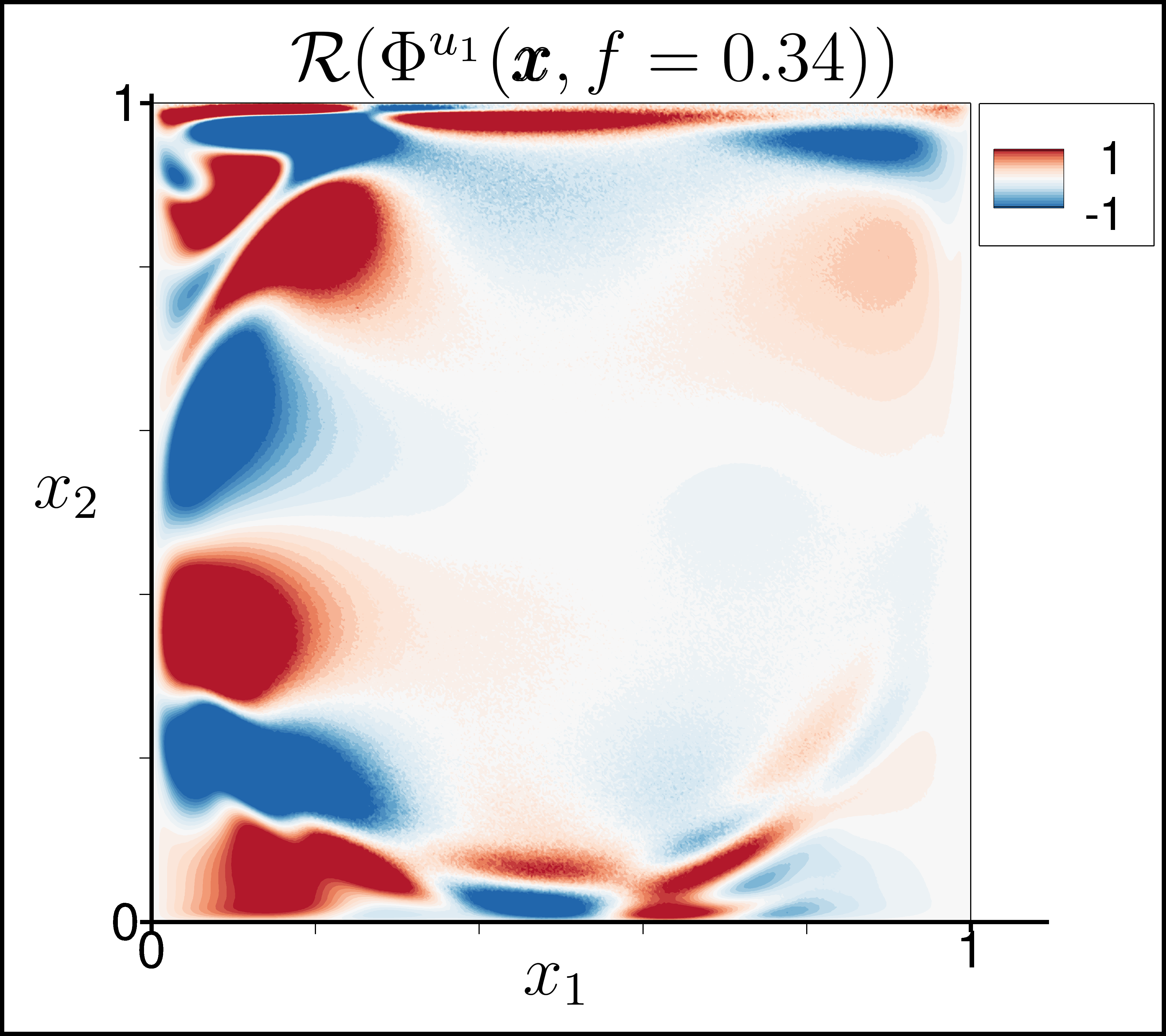}}\\(f) 
\end{minipage}
\begin{minipage}{0.24\textwidth}
\centering {\includegraphics[width=1.0\textwidth]{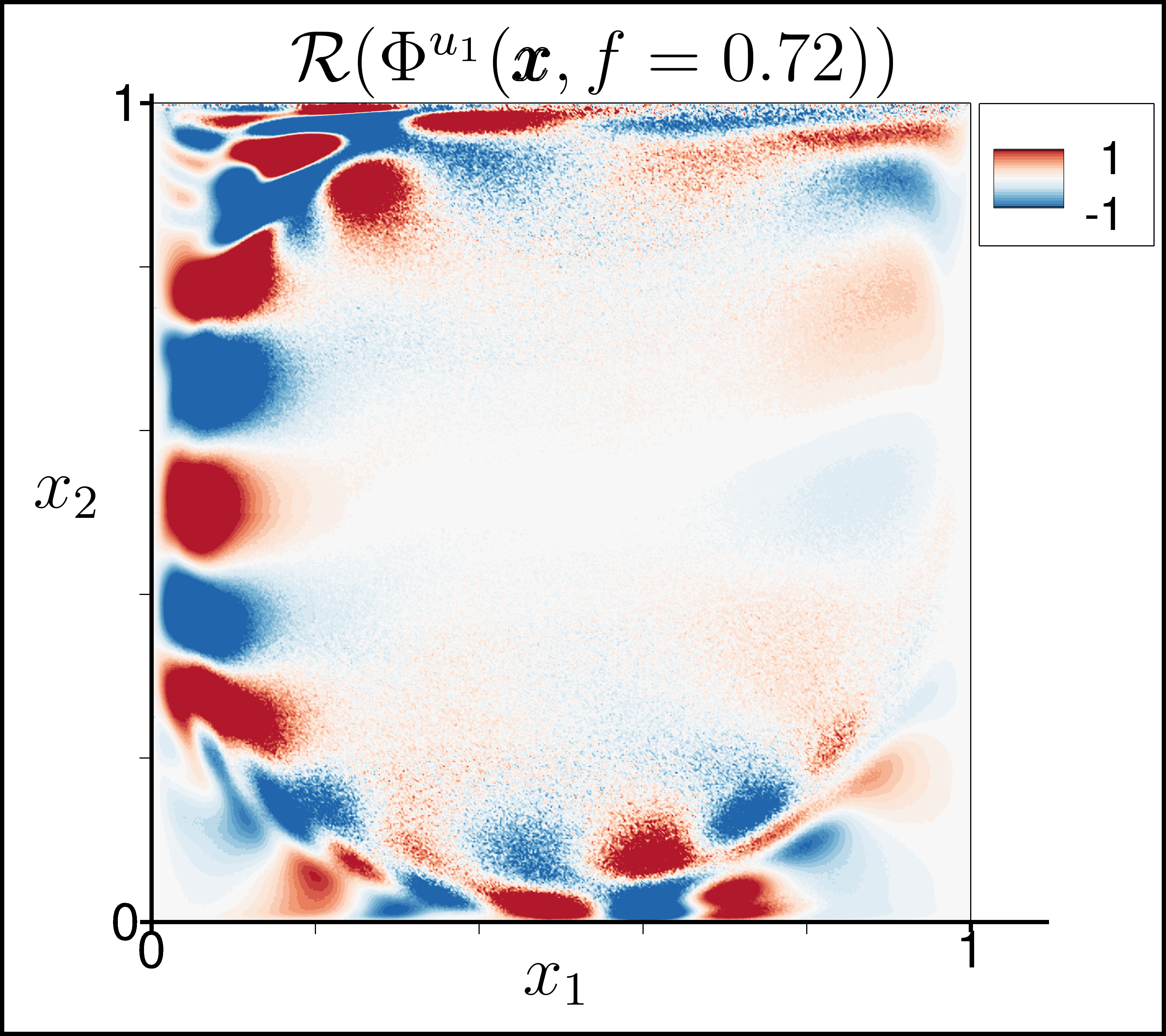}}\\(g) 
\end{minipage}
\begin{minipage}{0.24\textwidth}
\centering {\includegraphics[width=1.0\textwidth]{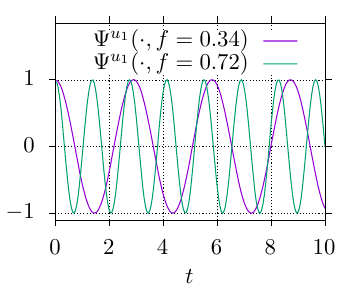}}\\(h)
\end{minipage}
\caption{POD (top row) and DMD (bottom row) of the horizontal velocity of LDC at $Re=12000$. (a) POD modal energies. (b) POD spatial mode at rank $l=1$. (c) POD spatial mode at rank $l=3$ that corresponds to the SMD modes at $f=0.78$. (d) Phase portrait of the POD first and third time modes. (e) Eigenspectrum of DMD, indicating growth/decay ($\sigma$) of modes at a frequency $f$. (d) Real part of the DMD mode at $f=0.34$. (e) Real part of the DMD mode at $f=0.72$. (f) Time coefficients of DMD modes at $f=0.34$ and $f=0.72$}
\label{fig:poddmd_ldc_12k}
\end{figure}

The spatial SMD modes (${\Phi}^{u_1}(\pmb{x},f=0.39)$ and ${\Phi}^{u_1}(\pmb{x},f=0.78)$) that correspond to the spectral energy peaks of Fig.~\ref{fig:smd_ldc_12k}(b) are displayed in Fig.~\ref{fig:smd_ldc_12k}(d) through (g) in terms of their real and imaginary parts.
The spatial shift in the modal peaks of the real and imaginary parts for both modes at $f=0.39$ (Fig.~\ref{fig:smd_ldc_12k}(d) and (e)) and $f=0.78$ (Fig.~\ref{fig:smd_ldc_12k}(f) and (g)) is evident.
As expected, the peaks/troughs of the spatial modes are distributed along the shear layer region of the LDC., consistent with the energy based POD and the frequency/growth-rate based DMD modes of Fig.~\ref{fig:poddmd_ldc_12k}.
The SMD temporal modes (${\Psi}^{u_1}(t,f=0.39)$ and ${\Psi}^{u_1}(t,f=0.78)$) are displayed in Fig.~\ref{fig:smd_ldc_12k}(c) in terms of phase portraits.
The figure shows both the real and imaginary parts of the time-modes, exhibiting limit cycle oscillations (LCO), indicating that the LDC flow at Reynolds number $12000$ consists of a primary frequency and its sub/super-harmonics in totality.

The POD and DMD of the LDC flow in terms of the horizontal velocity are presented in Fig.~\ref{fig:poddmd_ldc_12k} in the top and bottom rows, respectively.
These modal decompositions comprise eigenvalues (left figures), spatial modes (in the middle), and their temporal modes (right sub-figures).
The LDC flow at $Re=12000$ exhibits a low rank behavior, where only a few modes capture most of the system dynamics.
The first few ($l\lessapprox 6$) POD modes (Fig.~\ref{fig:poddmd_ldc_12k}a) comprise most of flow the flow energy, where modes $l=1$ and $l=3$ correspond to the SMD modes at $f=0.39$ and $0.78$, respectively.
The spatial POD modes (Fig.~\ref{fig:poddmd_ldc_12k}b, c) and their temporal dynamics (Fig.~\ref{fig:poddmd_ldc_12k}d) compare closely with the SMD modes of Fig.~\ref{fig:smd_ldc_12k}, which are also the leading modes in terms of their spectral energy ranks.

The DMD modes that correspond to the SMD modes are at $f=0.34$ and $f=0.72$, exhibiting similar modal shapes (Fig.~\ref{fig:poddmd_ldc_12k}f, g) and temporal dynamics (Fig.~\ref{fig:poddmd_ldc_12k}h).
For the considered number of snapshots ($n=1024$) and time duration ($51.2$), these DMD modes also stand out for their higher growth rates, {\it i.e.} $\sigma=125$ and $\sigma=55$, respectively (Fig.~\ref{fig:poddmd_ldc_12k}e).
In a statistically stationary flow, the growth/decay rates of the DMD modes should be equal to zero~\citep{towne2018spectral}; however, the numerical time-averaging and its subtraction from the flow may introduce some non-zero growth/decay rates to the DMD modes.
While the growth rate and frequency are the key identifier of DMD modes, uncertainties in those values can lead to spurious modes.
The energy based ranking of the SMD modes is as unambiguous as the POD mode ranking, in addition to the frequency based ranking.

\begin{figure}
\centering
\begin{minipage}{1.0\textwidth}
\centering {\includegraphics[width=1.0\textwidth]{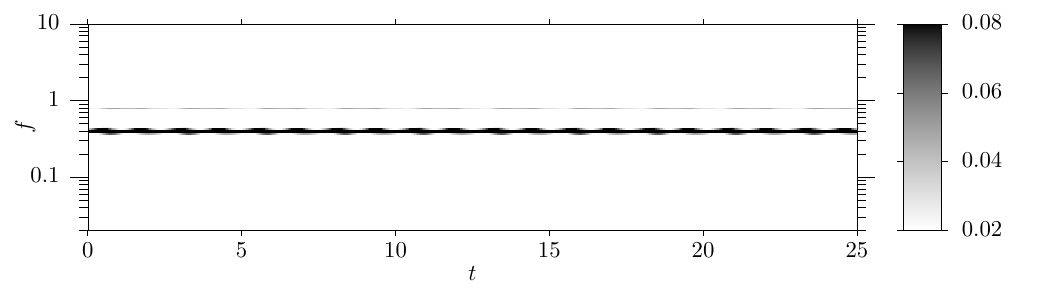}}
\end{minipage}
\caption{Spectrogram, $|\varphi_f|^2$, of LDC flow at $Re=12000$.}
\label{fig:sg_ldc_12k}
\end{figure}

A key distinguishing feature of the SMD is the non-dimensionalized time modes, $\pmb{\varphi}$, which provide an instantaneous spectral energy as a function of time - a spectrogram.
The spectrogram, $|\pmb{\varphi}|^2$, of the LDC at $Re=12000$ is displayed in Fig.~\ref{fig:sg_ldc_12k}, where the time variation of the two leading SMD modes $f=0.39$ and $f=0.78$ is evident.
The amplitudes of time variation are smaller compare to the difference of the spectral energies at these two peaks.
Given that the second SMD mode is a higher harmonic of the first mode, it is expected that the two modes are largely synchronized in terms of their peaks and troughs of the time variations, as depicted in the spectrogram (Fig.~\ref{fig:sg_ldc_12k}).


\begin{figure}
    \centering
    \begin{minipage}{0.24\textwidth}
    \centering {\includegraphics[width=1.0\textwidth]{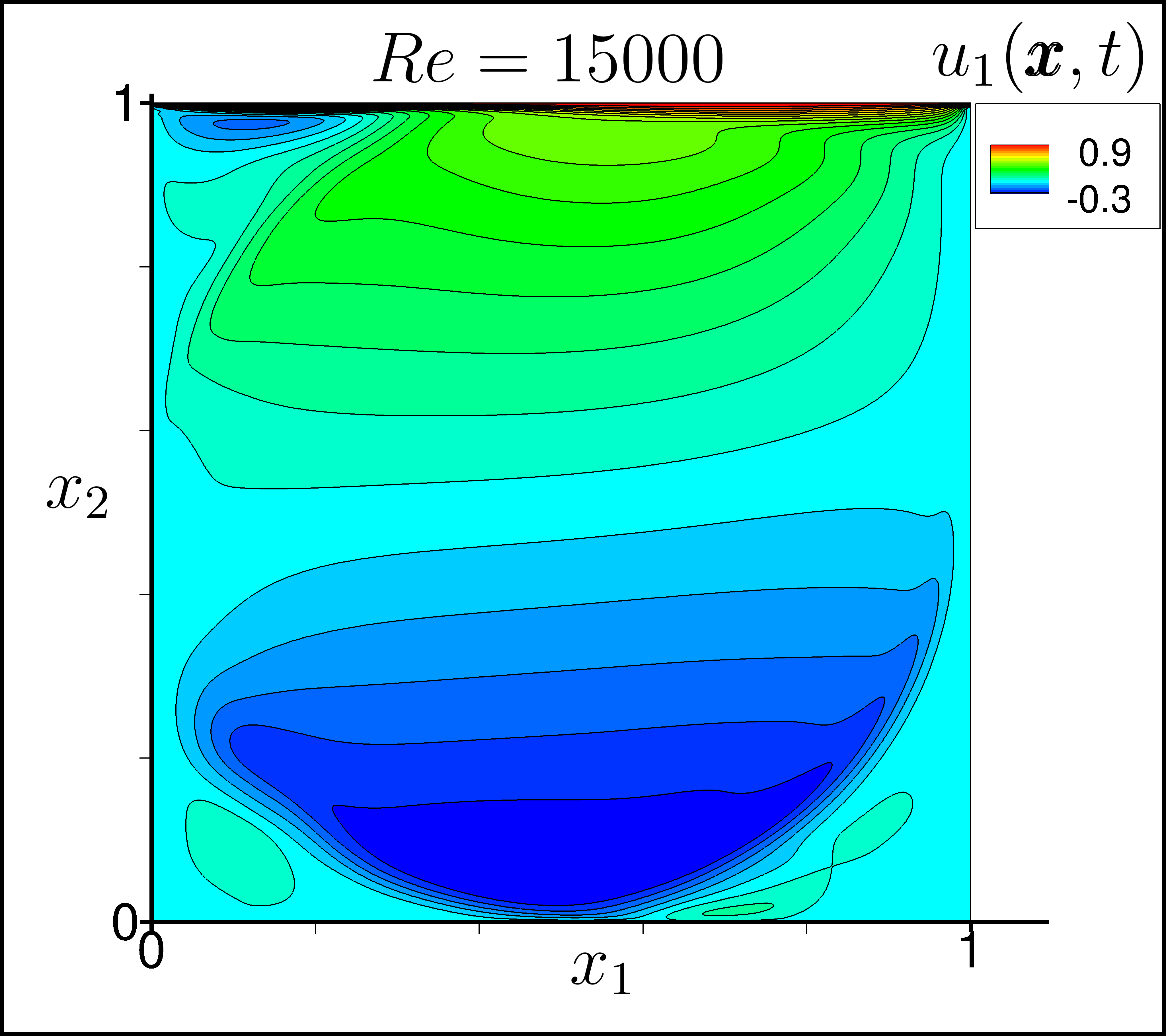}}\\(a)
    \end{minipage}
    \begin{minipage}{0.24\textwidth}
    \centering {\includegraphics[width=1.0\textwidth]{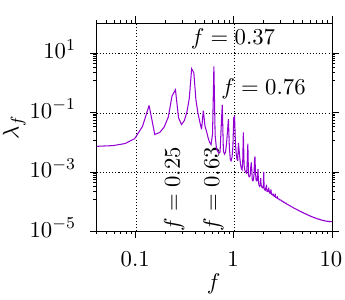}}\\(b)
    \end{minipage}
    \begin{minipage}{0.24\textwidth}
    \centering {\includegraphics[width=1.0\textwidth]{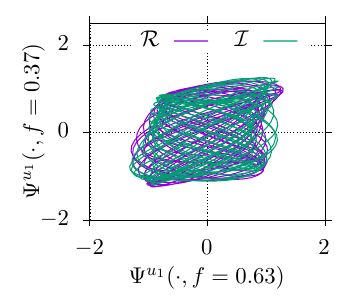}}\\(c)
    \end{minipage}
    \begin{minipage}{0.24\textwidth}
    \centering {\includegraphics[width=1.0\textwidth]{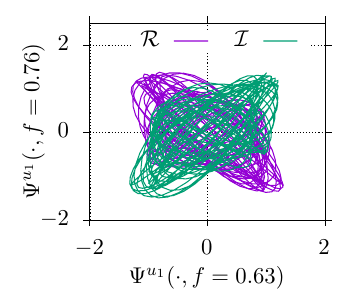}}\\(d)
    \end{minipage}\\
    \begin{minipage}{0.24\textwidth}
    \centering {\includegraphics[width=1.0\textwidth]{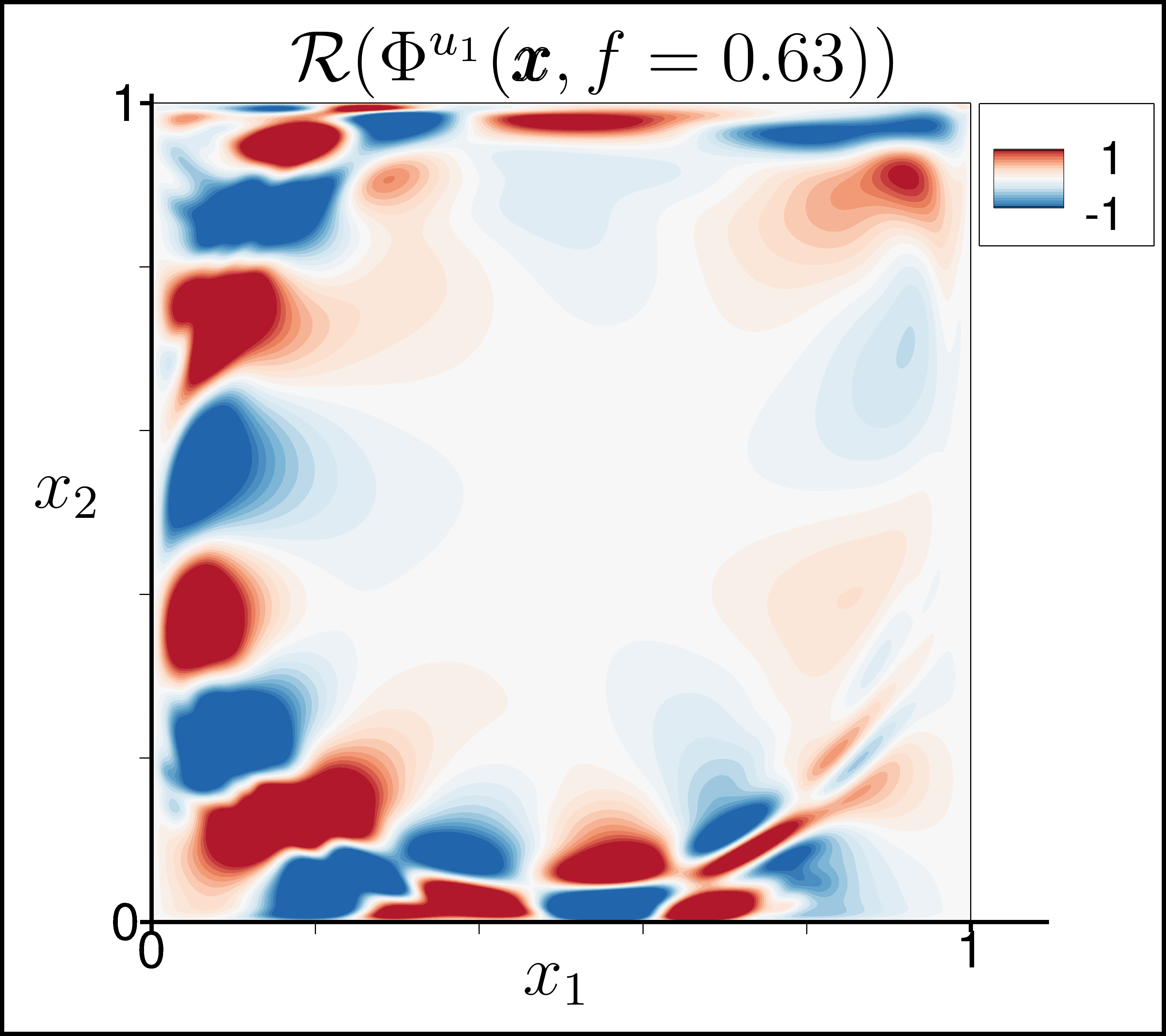}}\\(e) 
    \end{minipage}
    \begin{minipage}{0.24\textwidth}
    \centering {\includegraphics[width=1.0\textwidth]{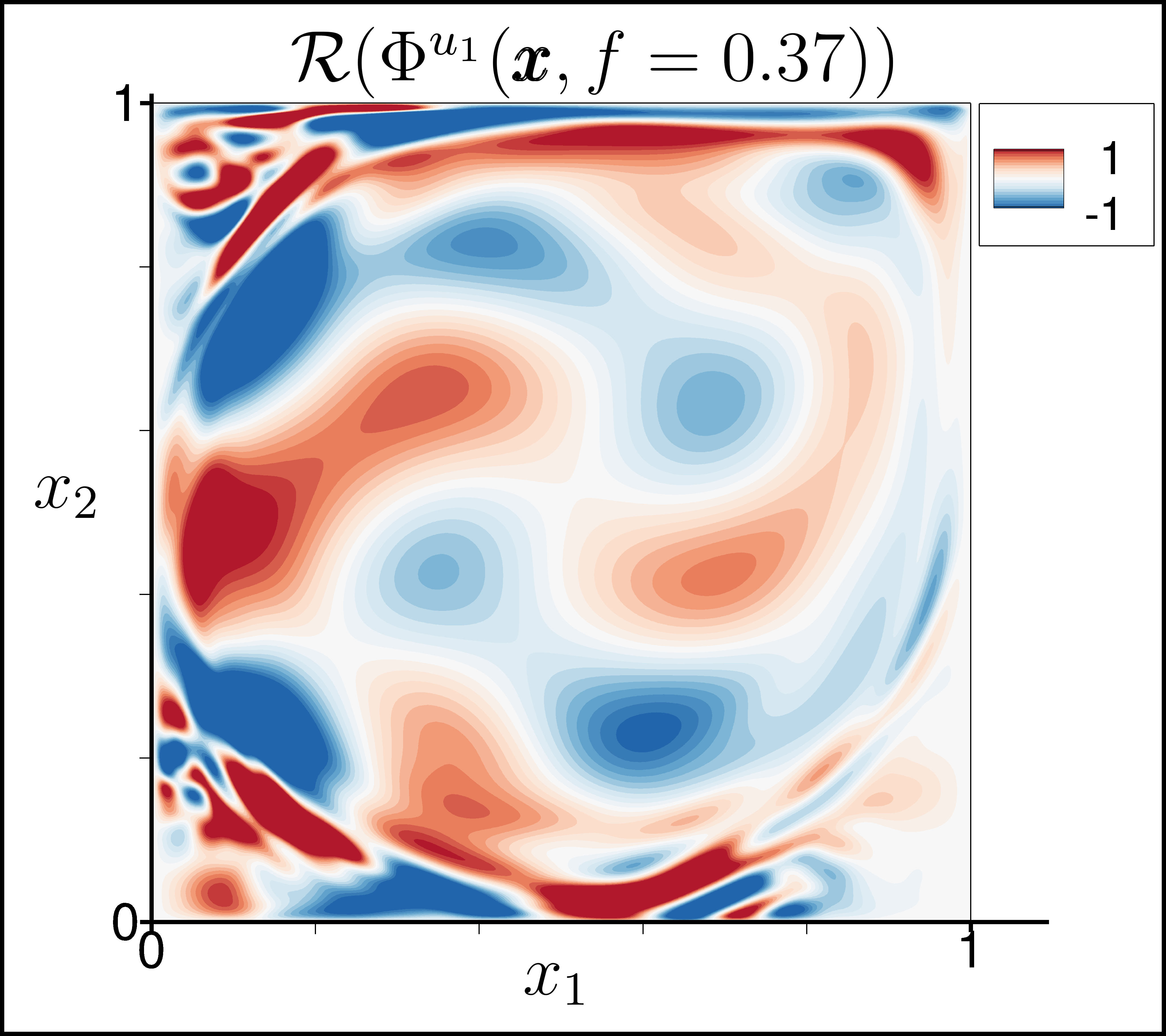}}\\(f) 
    \end{minipage}
    \begin{minipage}{0.24\textwidth}
    \centering {\includegraphics[width=1.0\textwidth]{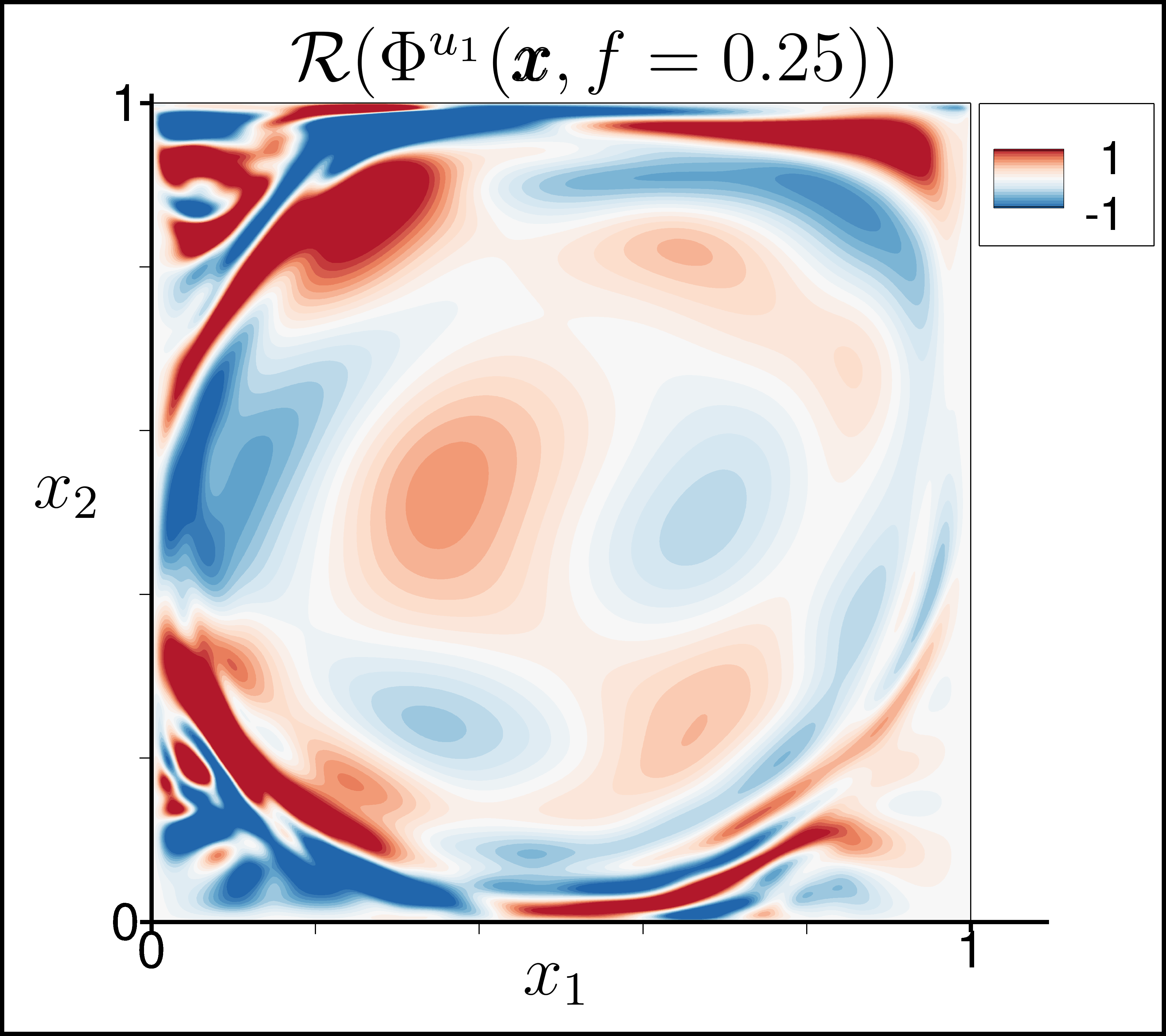}}\\(g) 
    \end{minipage}
    \begin{minipage}{0.24\textwidth}
    \centering {\includegraphics[width=1.0\textwidth]{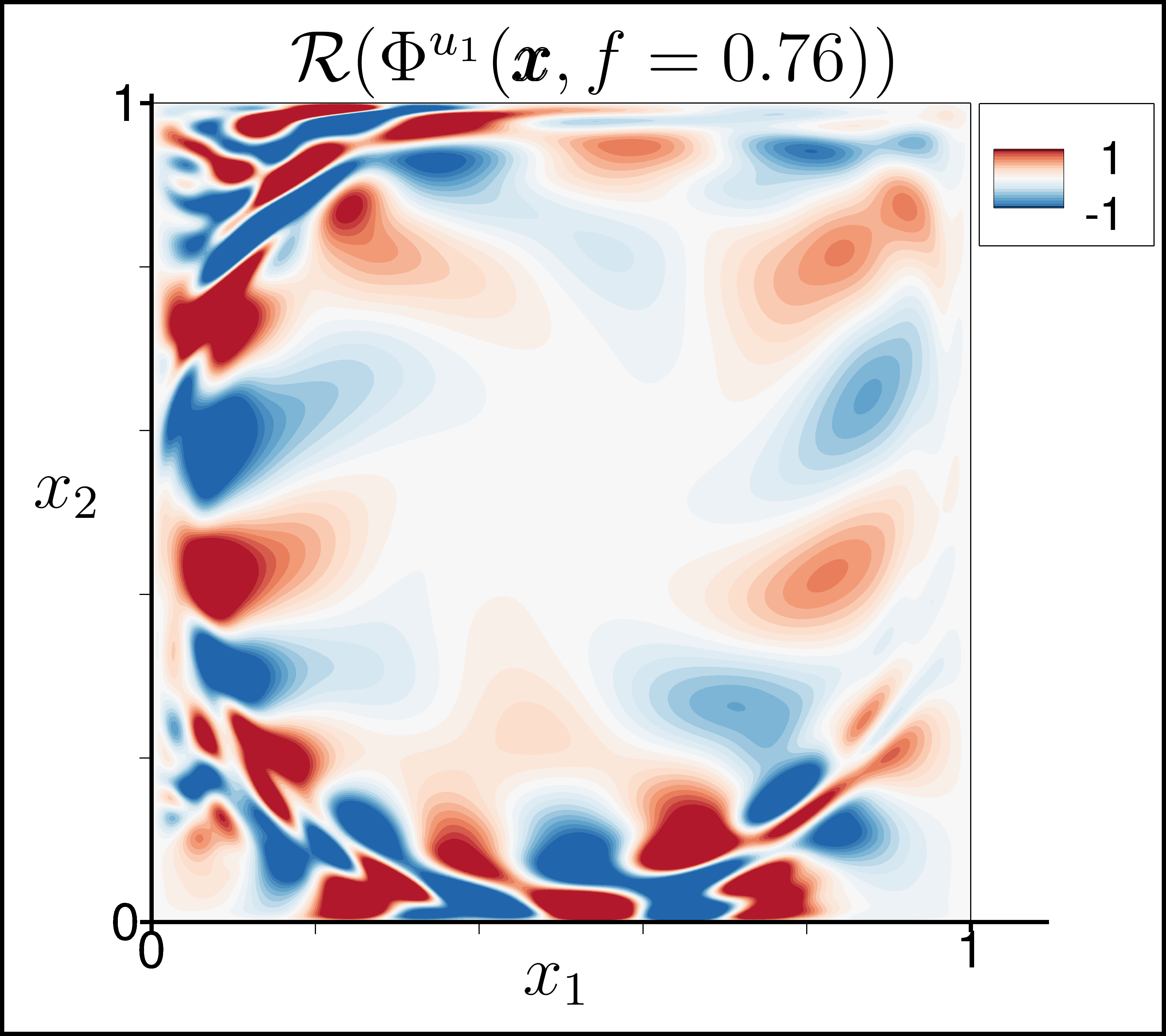}}\\(h) 
    \end{minipage}
    \caption{Spectral mode decomposition of LDC at $Re=15000$. (a) Spectral energies of the SMD modes. (b) Phase portraits of the SMD time-modes at $f=0.63$ versus $f=0.37$. (c) Phase portraits of the SMD time-modes at $f=0.63$ versus $f=0.76$. (d) Real part of the spatial spectral mode at $f=0.63$. (e) Real part of the spatial spectral mode at $f=0.37$. (d) Real part of the spatial spectral mode at $f=0.25$. (e) Real part of the spatial spectral mode at $f=0.76$.}
    \label{fig:smd_ldc_15k}
\end{figure}

\begin{figure}
\centering
\begin{minipage}{0.24\textwidth}
\centering {\includegraphics[width=1.0\textwidth]{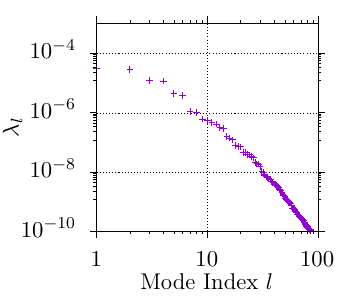}}\\(a)
\end{minipage}
\begin{minipage}{0.24\textwidth}
\centering {\includegraphics[width=1.0\textwidth]{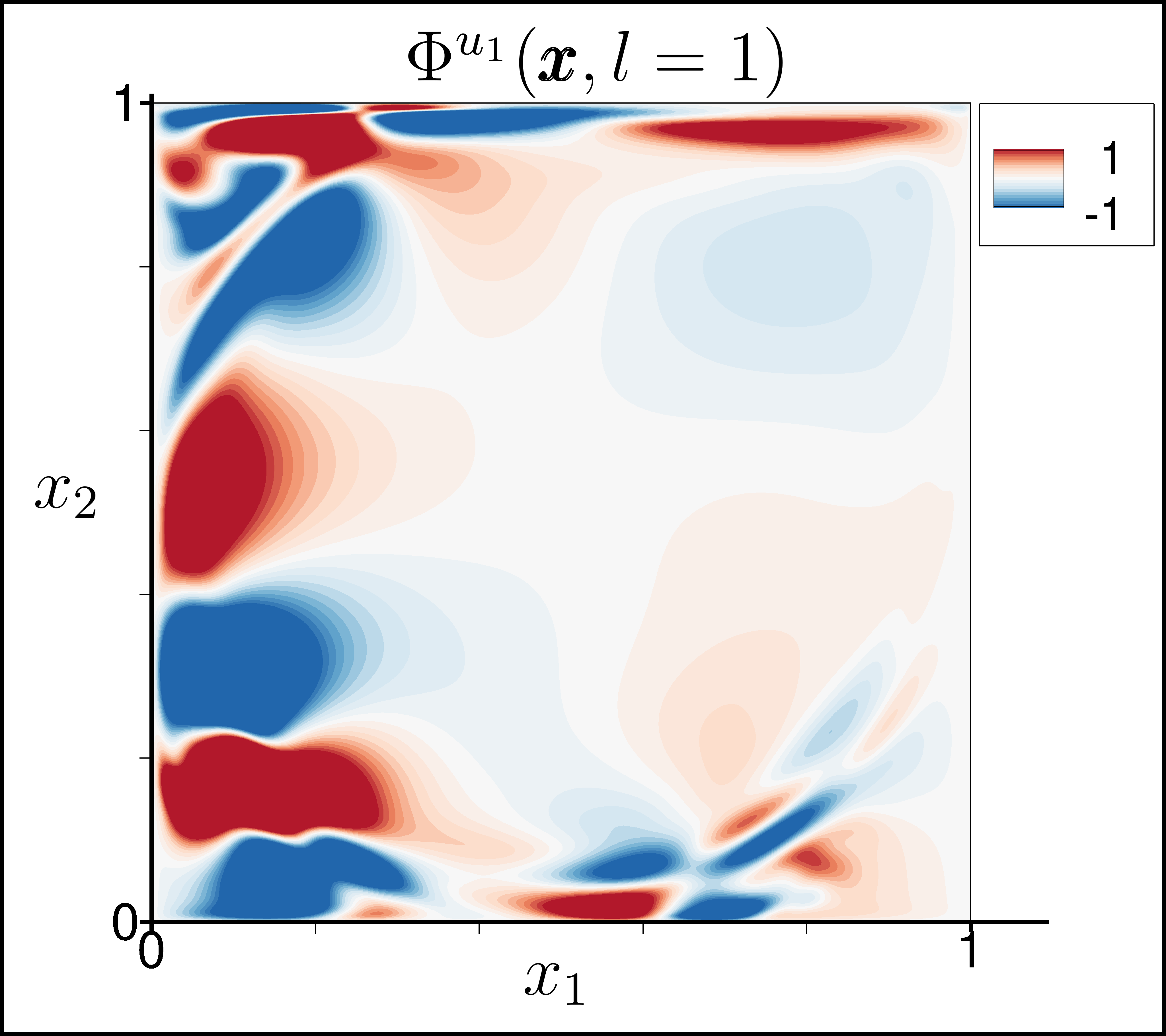}}\\(b) 
\end{minipage}
\begin{minipage}{0.24\textwidth}
\centering {\includegraphics[width=1.0\textwidth]{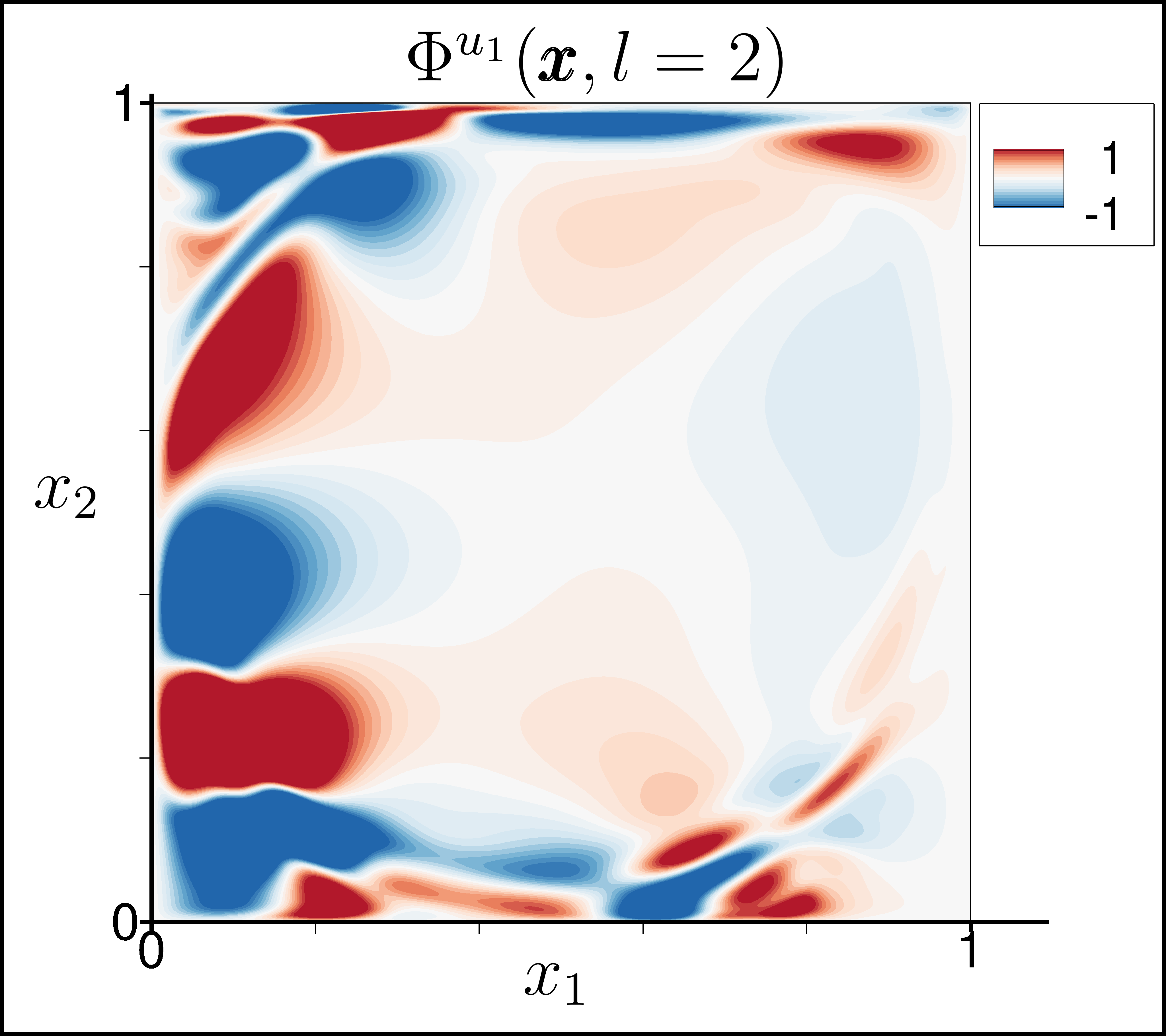}}\\(c) 
\end{minipage}
\begin{minipage}{0.24\textwidth}
\centering {\includegraphics[width=1.0\textwidth]{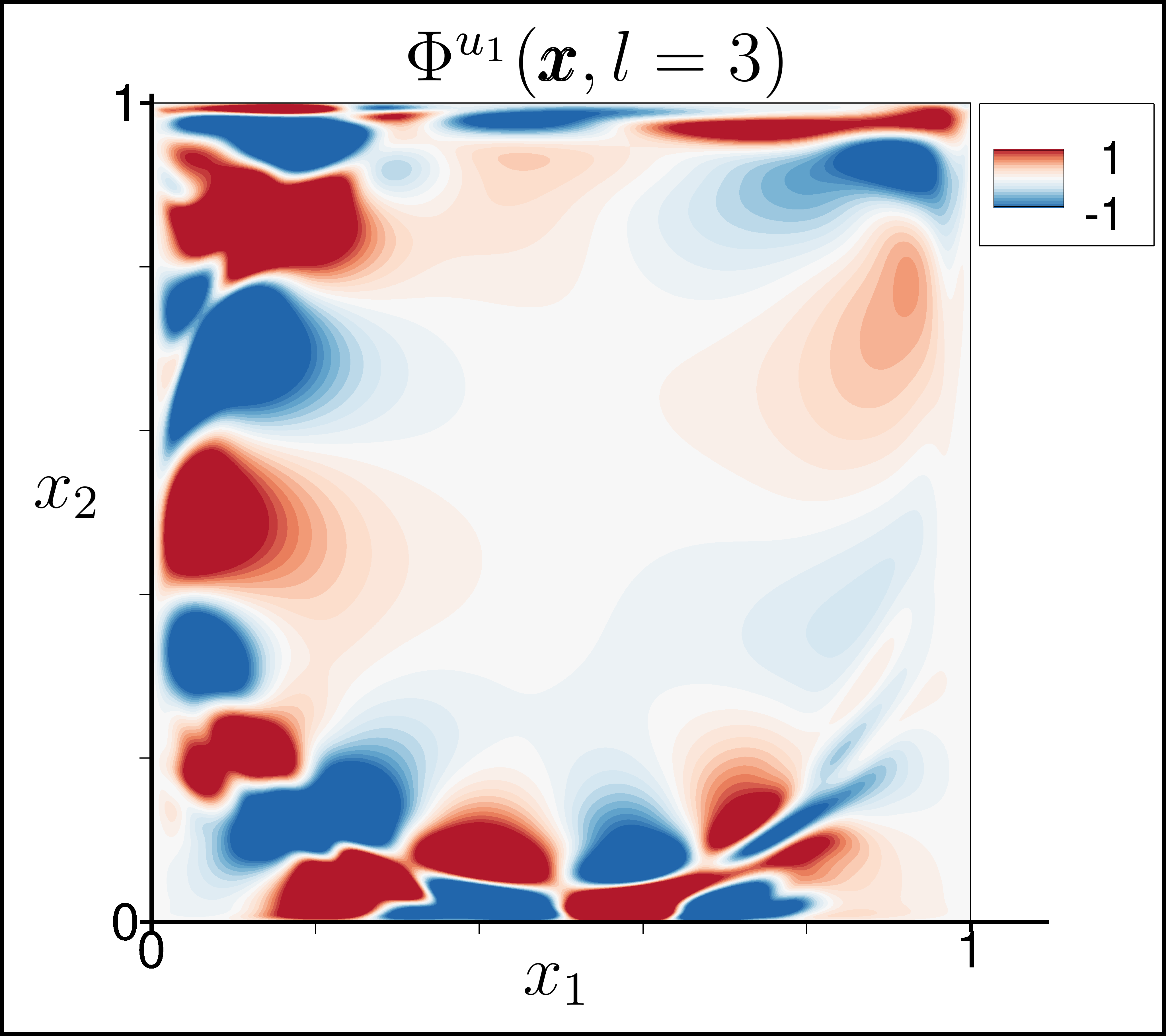}}\\(d)
\end{minipage}
\begin{minipage}{0.24\textwidth}
\centering {\includegraphics[width=1.0\textwidth]{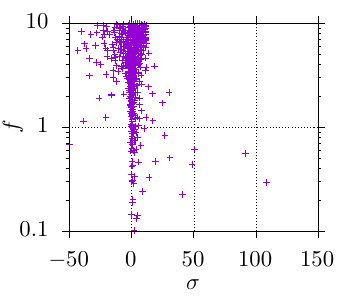}}\\(e)
\end{minipage}
\begin{minipage}{0.24\textwidth}
\centering {\includegraphics[width=1.0\textwidth]{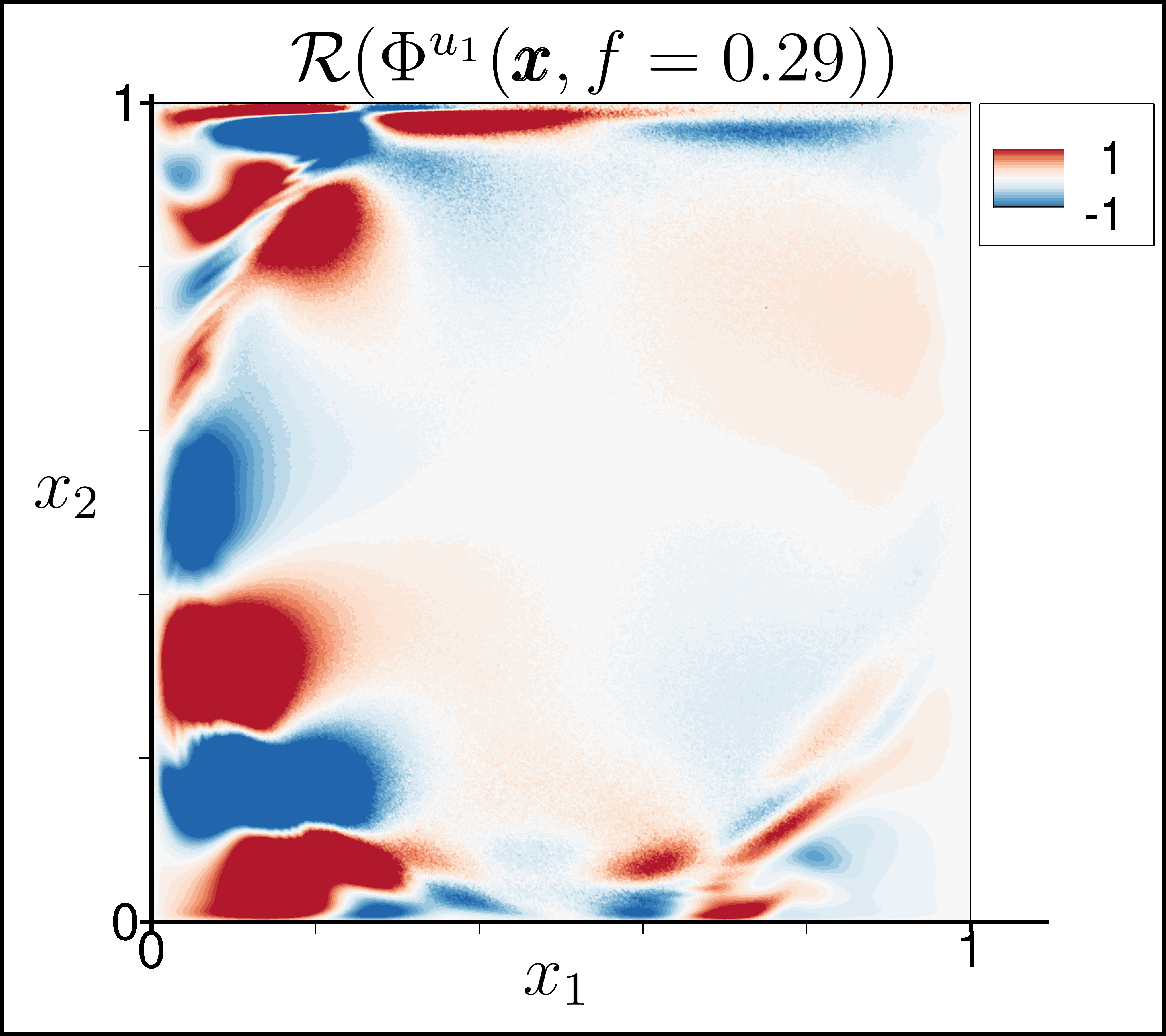}}\\(f) 
\end{minipage}
\begin{minipage}{0.24\textwidth}
\centering {\includegraphics[width=1.0\textwidth]{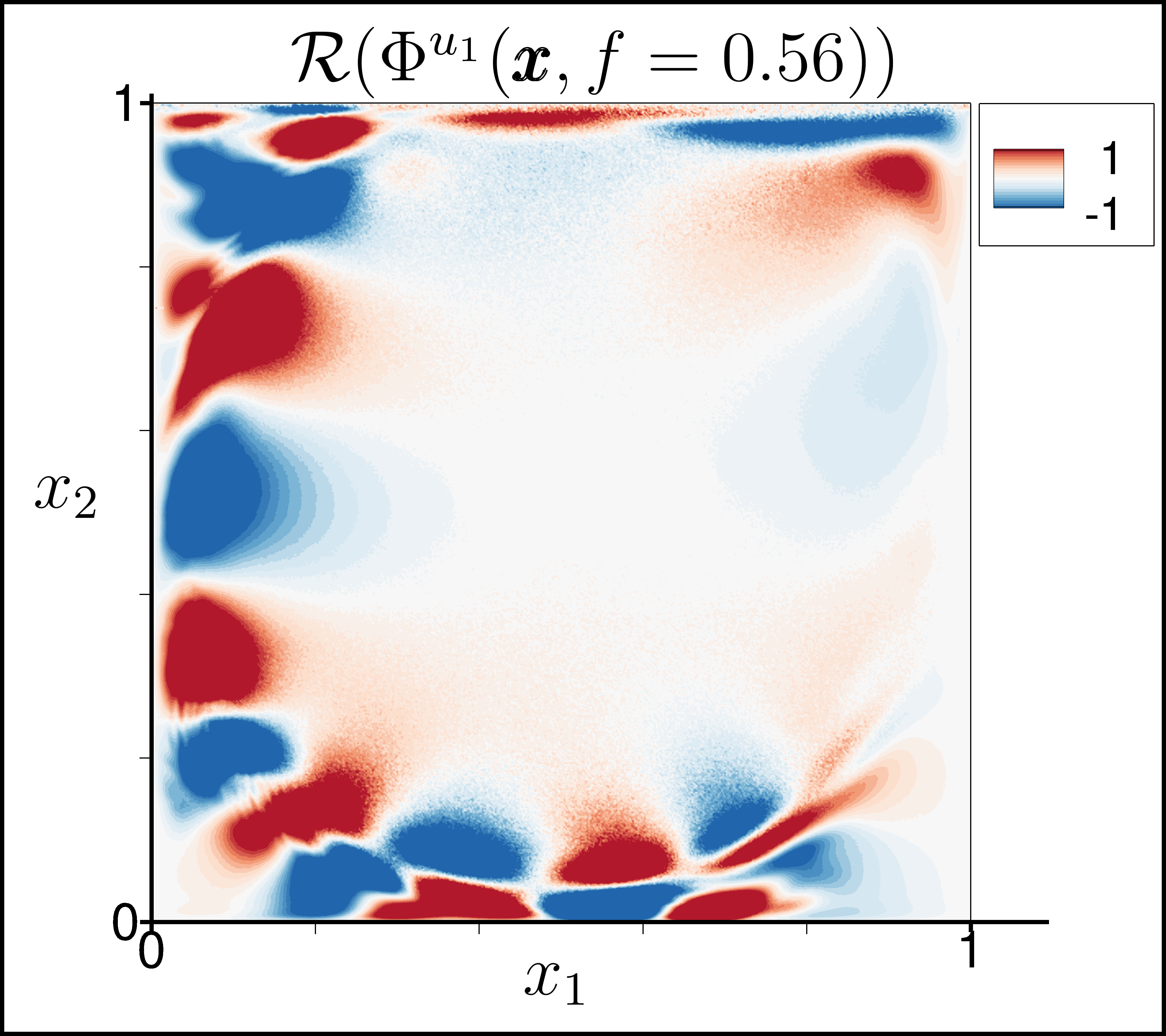}}\\(g) 
\end{minipage}
\begin{minipage}{0.24\textwidth}
\centering {\includegraphics[width=1.0\textwidth]{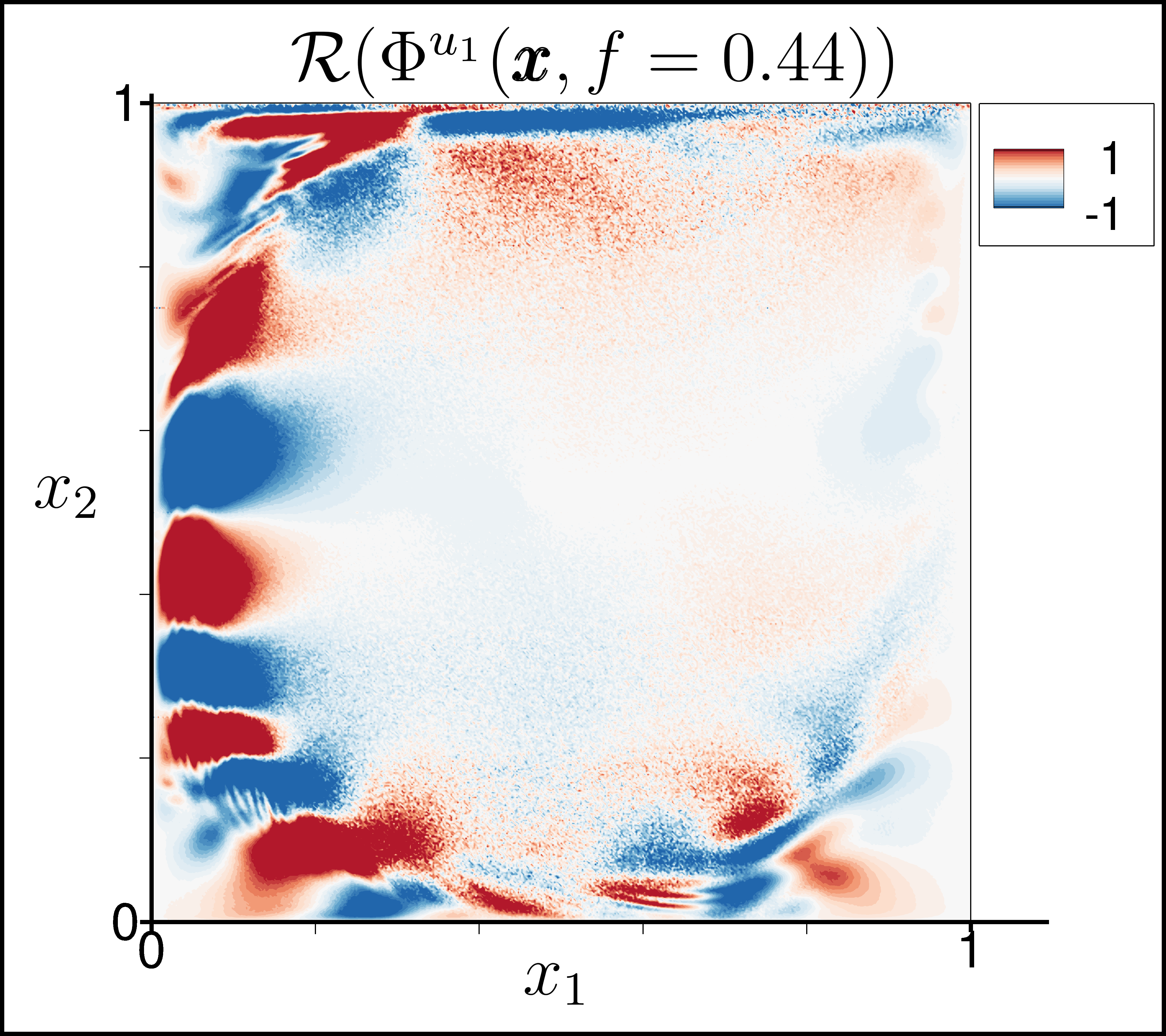}}\\(h)
\end{minipage}
\caption{POD (top row) and DMD (bottom) of LDC at $Re=15000$. (a) POD modal energies. (b) POD rank $l=1$ mode. (c) POD rank $l=2$ mode (d) POD rank $l=3$ mode.  (e) DMD eigenvalues. (f) DMD mode at $f=0.29$. (d) DMD mode at $f=0.56$. (e) DMD mode at $f=0.44$.}
\label{fig:poddmd_ldc_15k}
\end{figure}

Next, we consider the LDC flow at a higher Reynolds number of $Re_L=15000$, increasing the level of flow turbulence.
The SMD modal energies are displayed in Fig.~\ref{fig:smd_ldc_15k}(a), where the increased number of frequency peaks is evident.
The frequency peaks at $f=0.37$ and $f=0.76$ are reminiscent of the $f=0.39$ and $0.78$ of the LDC flow at $Re_L=12000$.
The increase of Reynolds number has introduced two energetically prominent modes at $f=0.25$ and $f=0.63$.
In terms of the associated spectral energies, these modes can be ranked in the decreasing order are as follows: $f=0.63$, $f=0.37$, $f=0.25$, and $f=0.76$.
The next lower peak in the energy rank is at $f=0.14$, which corresponds to the instability mode typically predicted by the stability analysis techniques~\citep{ohmichi2017compressibility,shinde2025lagrangian}, using baseflows at pre-critical Reynolds numbers.
In addition, there exist several higher frequency peaks, albeit at lower spectral energy.

The SMD modes corresponding to the spectral peaks of Fig.~\ref{fig:smd_ldc_15k}(a) are displayed in Fig.~\ref{fig:smd_ldc_15k}(d) through (g).
These figures show the real parts of the modes at $f=0.63$, $f=0.37$, $f=0.25$, and $f=0.76$, respectively.
The modal structure of these modes, to a great degree, resembles with that of the POD and DMD modes of the same LDC flow configuration ({\it i.e.}, LDC at Mach $0.5$ and $Re_L=15000$) in Fig.~\ref{fig:poddmd_ldc_15k}.
However, there are subtle differences in the modal energy ranks, their spatial and temporal significance pertinent to the decomposition technique.
The top three energy ranked modes of POD (Fig.~\ref{fig:poddmd_ldc_15k}b,c,d), the dynamically relevant prominent modes of DMD (Fig.~\ref{fig:poddmd_ldc_15k}f,g,h), and the top four spectral energy based SMD modes (Fig.~\ref{fig:poddmd_ldc_15k}d,e,f,g) manifest the same physics of LDC flow in three distinct spatio-temporal dynamics.

In contrast to the POD and DMD modes, each SMD mode is at a single frequency and an associated unique energy rank.
The modal structure of SMD modes closely relates to that of POD and DMD modes that are in the same energy or frequency proximity, respectively. 
However, by construction the POD modes do not represent a single frequency; while DMD modes have an associated growth/decay rate, but not a unique energy ranking.
For instance, the third POD mode (Fig.~\ref{fig:poddmd_ldc_15k}d) has a modal structure that is similar to the leading SMD mode at $f=0.63$ (Fig.~\ref{fig:smd_ldc_15k}d), which compares well with the third DMD mode at $f=0.56$ (Fig.~\ref{fig:poddmd_ldc_15k}g).
In SMD, the frequency criterion on top of the energy ranking helps extracting the most (energetically and dynamically) relevant modes in a straightforward manner.
In general, the modal structures at higher frequency modes consist of smaller lobes of peaks and troughs (Fig.~\ref{fig:smd_ldc_15k}d through~\ref{fig:smd_ldc_15k}g), which are representative of the smaller scales of flow turbulence that engender at higher Reynolds numbers.

The phase portraits of the SMD time modes at $f=0.63$, $f=0.37$, and $f=0.76$ are shown in Fig.~\ref{fig:smd_ldc_15k}(b) and (c) for both the real and imaginary parts.
The chaotic yet organized nature of these spectral modes is evident in these figures.
The LDC flow at $Re_L=15000$ evolves from the LCO at $Re_L=12000$ to more nonlinear oscillations of the low-rank high-energy SMD time-modes.
The temporal variation of the SMD time mode ($\Psi^{u_1}_f$) follows the contribution of the corresponding spatial SMD mode ($\Phi^{u_1}_f$) at that particular time.
Thus, the amplitude modulation of the these time modes need not exhibit a regular period that corresponds to the modal frequency, and may comprise time variation pertaining to the instantaneous flow physics.
To some degree, this is analogous to POD, where the temporal modes may comprise multi-frequency oscillations in order to seek spatial modes that are optimal in energy, albeit the SMD time modes are not uncorrelated by construction.

\begin{figure}
\centering
\begin{minipage}{1.0\textwidth}
\centering {\includegraphics[width=1.0\textwidth]{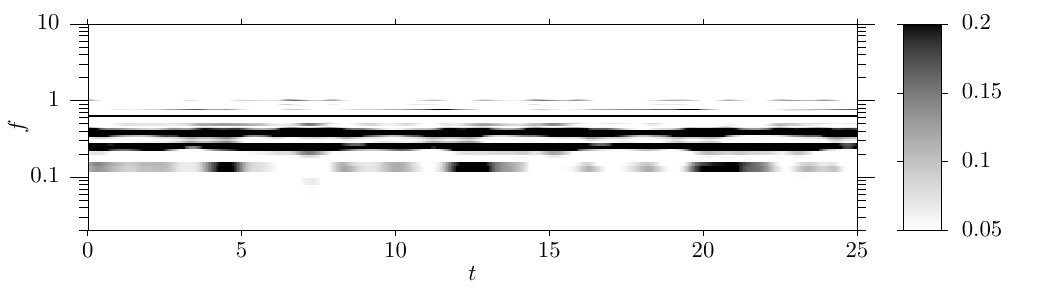}}
\end{minipage}
\caption{Spectrogram, $|\varphi_f|^2$, of LDC flow at $Re=15000$.}
\label{fig:sg_ldc_15k}
\end{figure}

The spectrogram of LDC flow at $Re_L=15000$ is shown in Fig.~\ref{fig:sg_ldc_15k}.
The instantaneous time variation of the prominent frequencies at, namely, $f=0.63$, $f=0.37$, and $f=0.25$ is evident in the spectrogram.
It also shows the peaks of spectral energy at $f=0.13$ and $f=0.76$.
The energy ranking of the spectral modes is also evident in the spectrogram.
The SMD modes $\Phi^{u_1}_{f=0.63}$ and $\Phi^{u_1}_{f=0.37}$ are close in their energy content, where the presence of the former mode is more consistent in the flow.
Instantaneously, the peaks at $f=0.25$ and $f=0.13$ also appear stronger but less regular as compared to the peaks of $f=0.62$ and $f=0.37$.
Thus, the total spectral energy of Fig.~\ref{fig:smd_ldc_15k}(a) displays the modes at $f=0.25$, $f=0.76$, and $f=0.13$ as the third, fourth and fifth in the spectral energy ranking, respectively.

The spectrogram of Fig.~\ref{fig:sg_ldc_15k} also highlights the relative temporal organization of the energy contribution by these modes.
On the time axis, there are zones of overall lower spectral energy, {\it e.g.} over time $7.5 \lessapprox t \lessapprox 9.5$.
For increasing time ($t\ge 9.5$), the spectral energy peaks at the previously noted frequencies appear with varying magnitudes in a sequence as follows: a weak peak of $f=0.63$, a strong peak of $f=0.37$, a weak peak at $0.25$, a much weaker peak at $0.13$, a strong peak at $0.63$, a strong peak at $0.37$, a strong peak at $0.25$, a weak peak of $f=0.13$, and so on.
The temporal organization of these spectral peaks appears as a repeating pattern, indicating potentially a signature pattern of the spectrogram for a stationary LDC flow at $Re_L=15000$.
Notably, such spectrogram (e.g., Fig.~\ref{fig:sg_ldc_15k}), elucidating the relative time-presence of frequencies, is not produced in the more common model decomposition techniques, such as the POD and DMD or their variants.

\section{Turbulent shockwave boundary layer interaction flow}\label{sec:turb_flows}

Next, we consider a more complex flow configuration of shockwave turbulent boundary layer interaction to perform spacetime spectral analysis using SMD.
The turbulent boundary layer flow is at a supersonic Mach number $M=2.7$ and Reynolds number of $Re_\delta=54600$ based on the inflow boundary layer thickness, $\delta_{in}$.
An oblique incident shock at an angle of $30.29\deg$ impinges on the boundary layer, engendering SBLI.
The incident shock is strong enough to produce flow separation, in the time-mean sense, giving rise to the characteristic low-frequency unsteadiness of SBLI as well as an increased level of turbulence. 

\begin{figure}
\centering
\begin{minipage}{0.48\textwidth}
\centering {\includegraphics[width=1.0\textwidth]{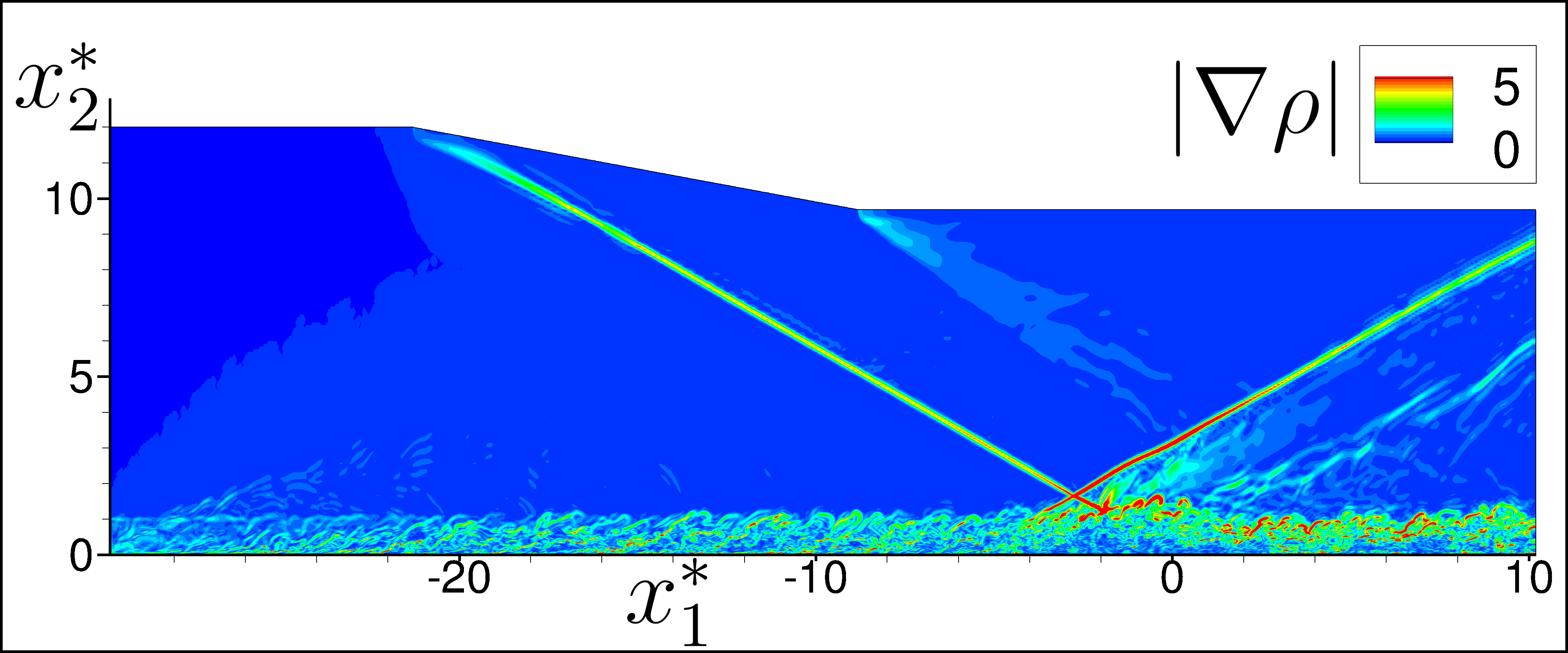}}\\(a)
\end{minipage}
\begin{minipage}{0.48\textwidth}
\centering {\includegraphics[width=1.0\textwidth]{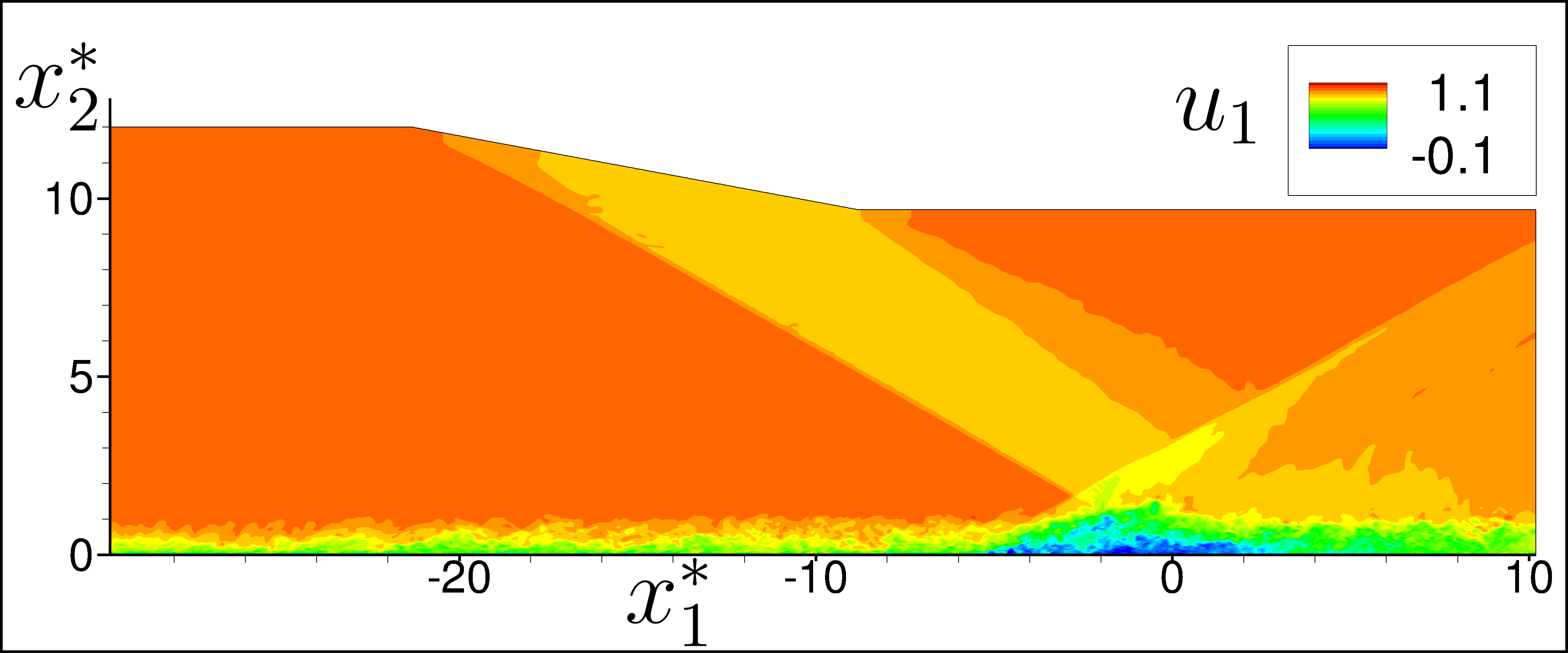}}\\(b)
\end{minipage}
\caption{Turbulent shockwave boundary layer interaction at Mach $M=2.7$ and Reynolds number of $Re_\delta=54600$, shown in terms of instantaneous flowfields of (a) the density gradient magnitude $|\nabla\rho|$, and (b) the streamwise velocity $u_1$.}
\label{fig:flow_sbli}
\end{figure}

The details of the flow configuration, computational domain, numerical grid, methodology along side the insights into the flow physics are provided in our prior works~\citep{shinde2021supersonic,shinde2022features,shinde2025distributed}.
Briefly, the flow geometry is shown in Fig.~\ref{fig:flow_sbli}(a), exhibiting flow structures in terms of the instantaneous density gradient magnitude ($|\nabla \rho|$).
The computational mesh consists of $1201\times 324\times 481$ points in the streamwise, wall normal, and spanwise directions, respectively, with a total of $\approx 187\times 10^6$ grid points, resulting in a highly resolved LES database.
The inflow is at $x_1^\ast=-29.8$, where an artificially generated turbulent boundary layer is introduced to the computation domain.
A fully developed turbulent boundary layer is achieved by $x_1^\ast=-21.3$.
The turbulent boundary layer enters into the SBLI region a few boundary layer thickness upstream the inviscid shock impingement location of $x_1^\ast=0$.
The incident shock causes a boundary layer separation of an extent $\approx 5\delta_{in}$ all upstream of $x_1^\ast=0$.


The key features of SBLI include: low-frequency unsteadiness or breathing of the separation bubble, a separation shock that oscillates with the separation bubble, formation of a shear layer and its interaction with the incident shock, reattachment of the shear layer, generation of compression shocks post shock impingement location, and importantly the feedback from the reattachment to the separation region through the separation bubble~\citep{dussauge2006unsteadiness,touber2009comparison,gaitonde2015progress,pasquariello2017unsteady,shinde2022features}.
In addition, SBLI typically leads to an increase of turbulence via distinct mechanisms in the upstream and the downstream regions of the shock impingement~\citep{andreopoulos2000shock,helm2014characterization,dupont2019compressible,shinde2022features}.
As a result, SBLI manifest a wide range of length and time scales, involving hydrodynamics mechanisms that exhibit unsteadiness orders of magnitude lower compared to the scales of incoming turbulent boundary layer.
In addition, there exist a mid-range frequency scales pertaining to the shear-layer vortical structures, as well as a higher-range frequency scales potentially involving acoustic mechanisms~\citep{pirozzoli2006direct}.

The spacetime-spectral analysis via SMD is expected to segregate the key features of SBLI, providing insights into the interplay among different scales.
As noted before, SBLI comprises a wide range of scales including an order of magnitude low-frequency oscillations compared to the incoming turbulence.
For the freestream velocity of $u_\infty=1$ and the inflow boundary layer thickness of $\delta_{in}=1$, a turbulent boundary layer manifests non-dimensional frequencies about $f\approx 1$; however, an SBLI with a flow separation length of $L\approx 5\delta_{in}$ (such as the one considered here) engenders unsteadiness typically around $f\approx 0.006$~\citep{clemens2014low,dussauge2006unsteadiness,gaitonde2015progress}.
Following the Nyquist criterion, a set of $n=1280$ snapshots with time step of $\delta t=0.8$ are gathered to perform SMD analysis.
Also, the SMD is performed over the 2D mid-plane (as shown in Fig.~\ref{fig:flow_sbli}) of the full 3D LES domain, treating the three components of velocity and the pressure.
The large flowfields and their SMD require parallel computing, where we adopt the recent methodologies of parallelization presented in~\cite{shinde2025distributed}.
In particular, SMD employs the parallel SVD version described in~\cite{shinde2025distributed} in the context of parallel DMD.
An instance of streamwise velocity is displayed in Fig.~\ref{fig:flow_sbli}(b), indicating the SBLI region with a flow recirculation bubble.

\begin{figure}
\centering
\begin{minipage}{0.34\textwidth}
\centering {\includegraphics[width=1.0\textwidth]{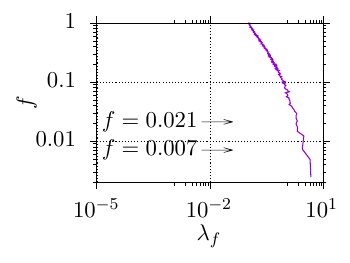}}\\(a)
\end{minipage}
\begin{minipage}{0.64\textwidth}
\centering {\includegraphics[width=1.0\textwidth]{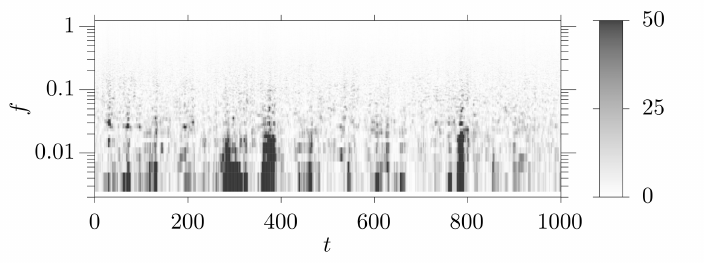}}\\(b)
\end{minipage}\\
\begin{minipage}{0.34\textwidth}
\centering {\includegraphics[width=1.0\textwidth]{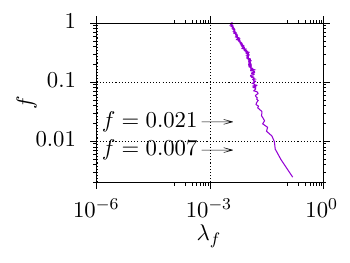}}\\(c)
\end{minipage}
\begin{minipage}{0.64\textwidth}
\centering {\includegraphics[width=1.0\textwidth]{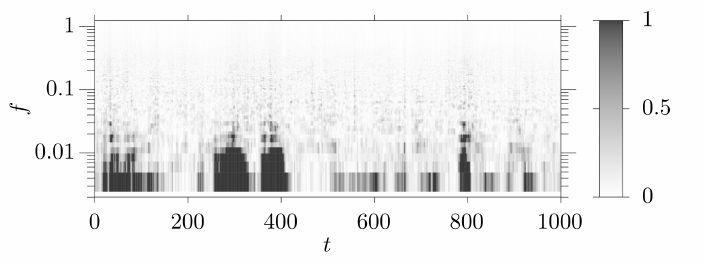}}\\(d)
\end{minipage}
\caption{Global spectral energies ($\lambda_f$) and spectrograms, $|\varphi_f|^2$, of the shockwave boundary layer interaction flow. (a) Spacetime integrated spectral energy of the streamwise velocity,  (b) spectrogram of the streamwise velocity, (c) total spectral energy of the pressure, and (d) spectrogram of the pressure.}
\label{fig:sg_sbli}
\end{figure}

The natural first step in SMD analysis of fluid flows is to examine the total spectral energies along side the spectrogram.
These provide insights into the prominent spectral peaks of the flow as well as their temporal presence.
The most energetic peaks and thus associated spatial modes can be then examined for the spatial appearance/structure of mode.
Figure~\ref{fig:sg_sbli} displays the total spectral energies associated with SMD modes (on the left) and their spectrograms (on the right) for the streamwise velocity (in the top row) and the pressure (in the bottom row).
As expected, the both spectra and spectrograms reveal a wide range of frequencies from less than $f=0.0025$ to $f\approx 1$.
The classical low-frequency peak at $f\approx 0.007$ (which corresponds to $f_L=0.035$ in terms of the  streamwise extent of flow separation, $L$) is observable in both the spectrograms and the total spectra.
A mid-range spectral peak at $f\approx 0.021$ (which corresponds to $f_L=0.11$) is indicated in the figures for both the streamwise velocity (Fig.~\ref{fig:sg_sbli}a,b) and the pressure (Fig.~\ref{fig:sg_sbli}c,d).
In addition, there are peripheral peaks in both in the higher/lower ranges of frequencies; however these are not prominent at least with the given spectral resolution.

The spectrograms of the streamwise velocity and the pressure bring out novel insights into SBLI in terms of the time synchronization of the different spectral peaks, and thus the spatial events in the SBLI.
The low-frequency peaks generally correspond to the streamwise breathing motion of the separation bubble, whereas the mid-range frequency unsteadiness is generally attributed to the shear layer vortex shedding~\citep{adler2018dynamic,shinde2022features}.
The intensity of the spectral peaks of the streamwise velocity spectrogram (Fig.~\ref{fig:sg_sbli}b) increases for all ranges of frequencies at the same time but in an intermittent manner.
The consistent presence of the mid-range spectral peak, which corresponds to the shear-layer vortex shedding, is evident in the both spectrograms.
Furthermore, the peaks of low-frequency unsteadiness for the streamwise velocity and the pressure do not appear at the same time instance, suggesting a lag/phase dynamics between the velocity and the pressure fields.

\begin{figure}
\centering
\begin{minipage}{0.48\textwidth}
\centering {\includegraphics[width=1.0\textwidth]{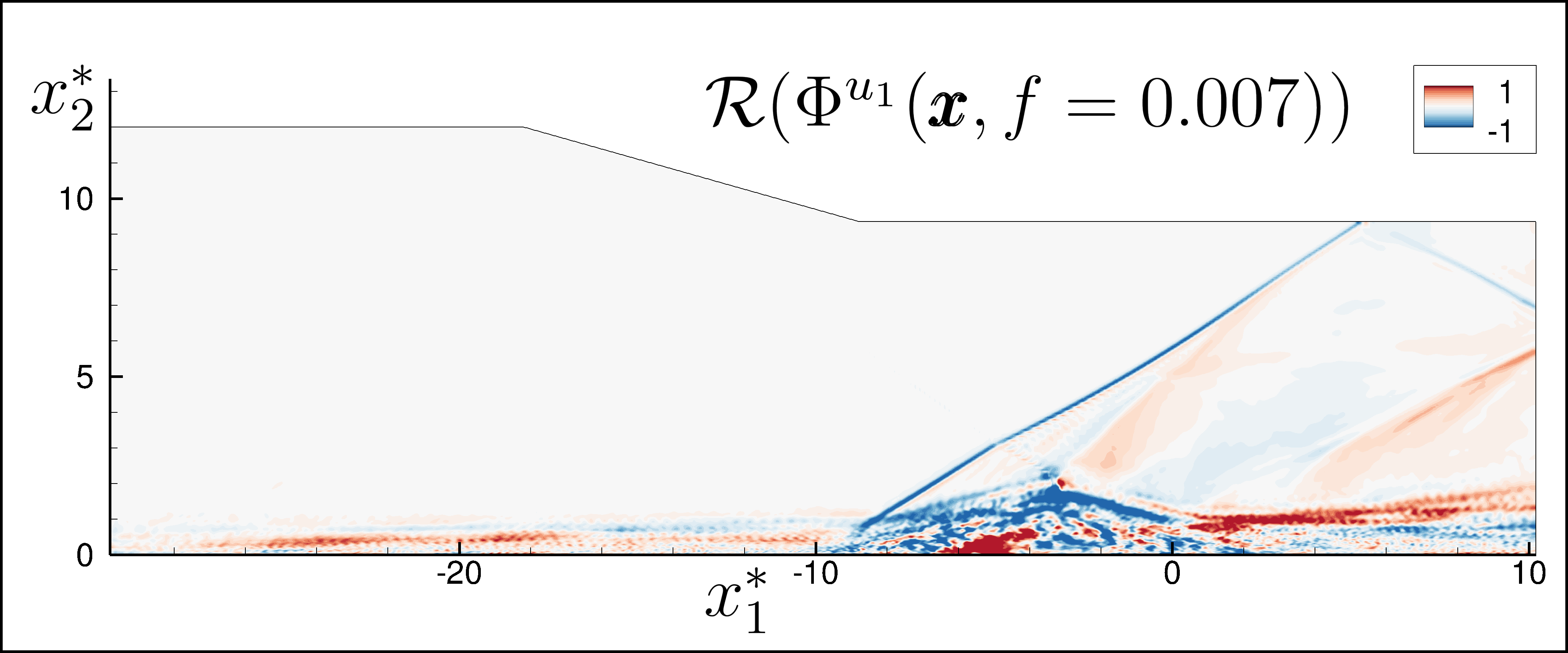}}\\(a)
\end{minipage}
\begin{minipage}{0.48\textwidth}
\centering {\includegraphics[width=1.0\textwidth]{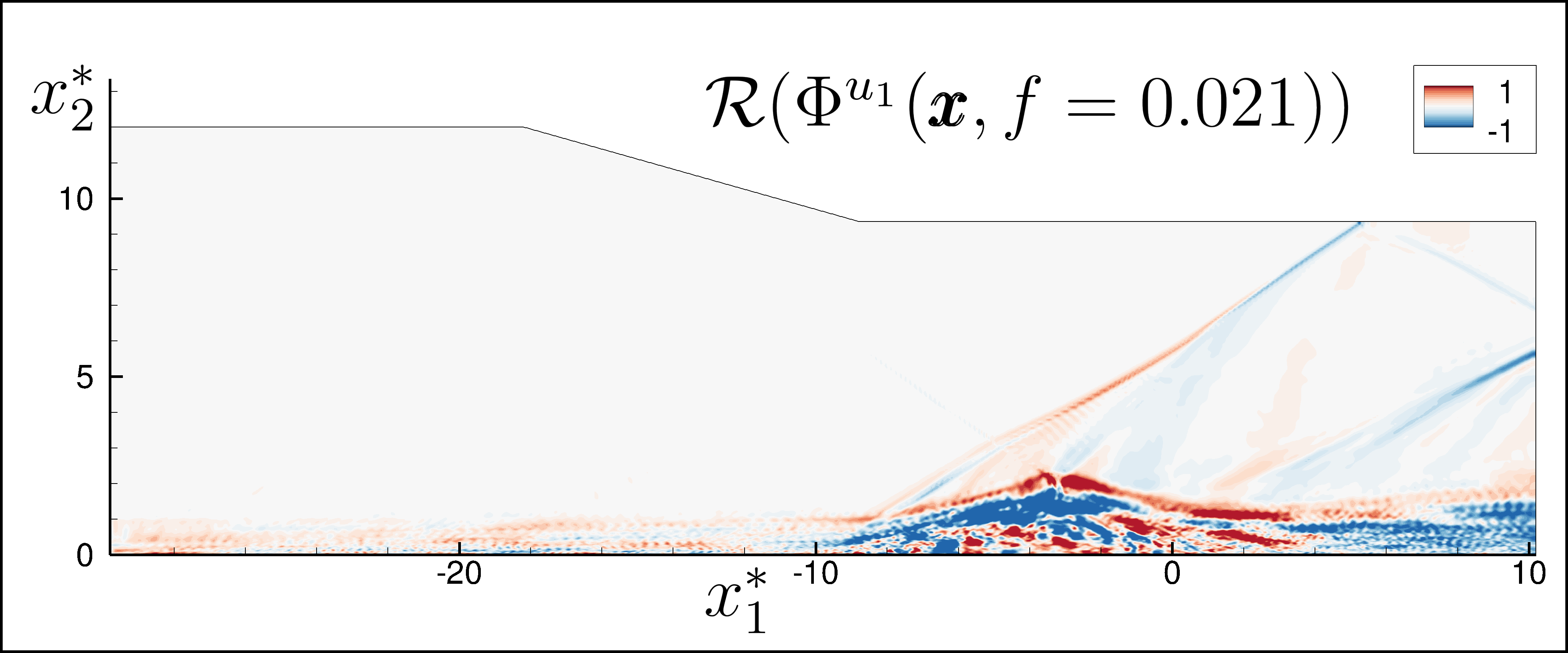}}\\(b) 
\end{minipage}\\
\begin{minipage}{0.48\textwidth}
\centering {\includegraphics[width=1.0\textwidth]{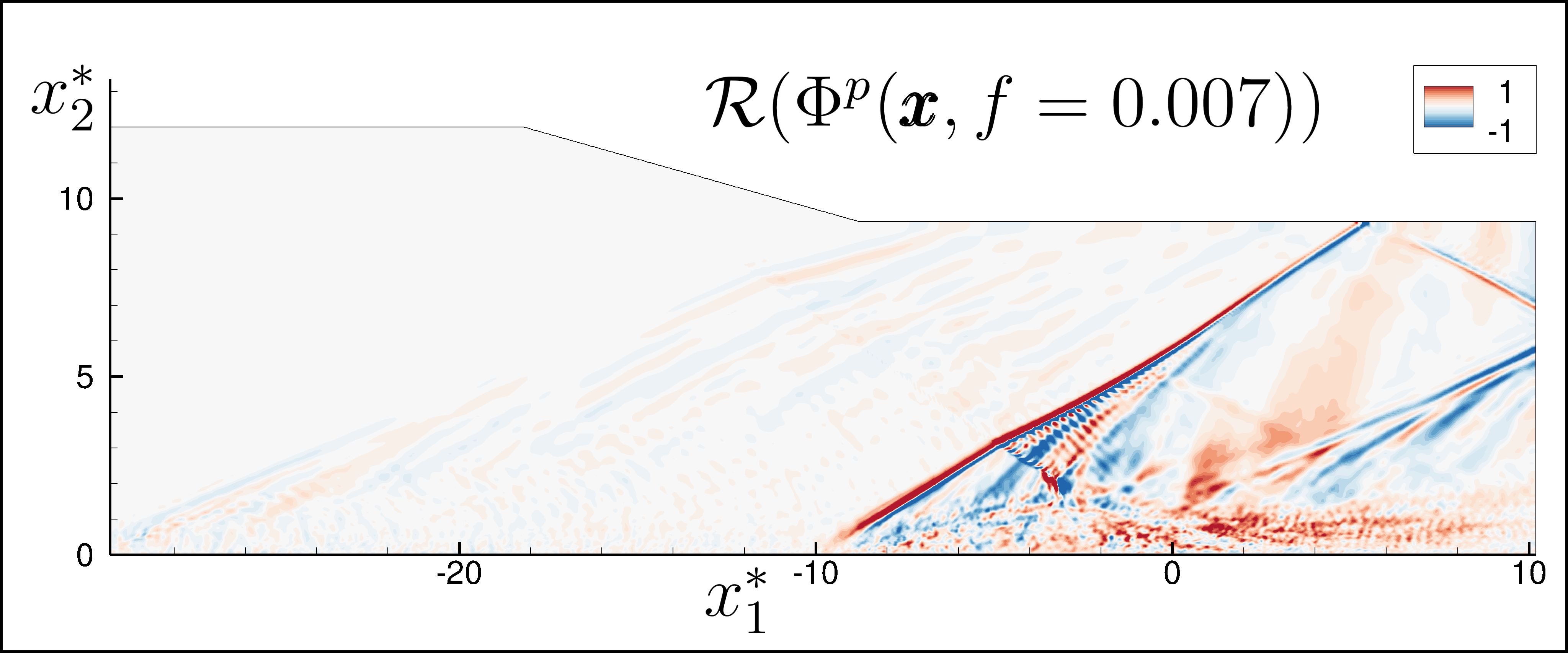}}\\(c) 
\end{minipage}
\begin{minipage}{0.48\textwidth}
\centering {\includegraphics[width=1.0\textwidth]{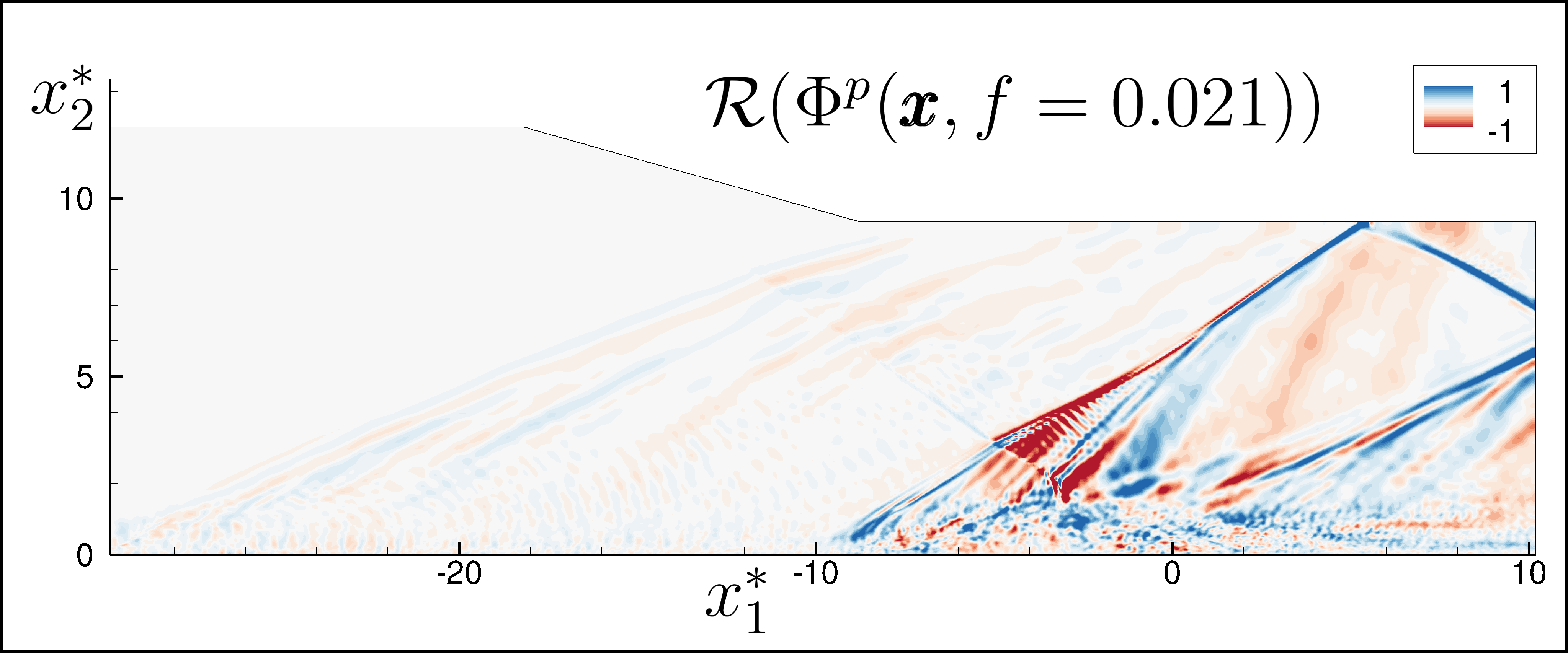}}\\(d)
\end{minipage}
\caption{Spectral spatial modes of SBLI (real parts) for streamwise velocity (top row) and pressure (bottom row) at $f=0.007$ (of the left) and $f=0.44$ (on the right).}
\label{fig:sbli_phi}
\end{figure}

The SMD modes of SBLI that correspond to the spectral peaks at $f=0.007$ and $f=0.021$ are presented in Fig.~\ref{fig:sbli_phi}, for the streamwise velocity (top row) and the pressure (bottom row).
Since the SBLI is the main exhibitor of the low-frequency unsteadiness in this flow, the energy dominant SMD modes peak in the vicinity of $x_1^\ast=0$, {\it i.e.} the shock impingement location, and the downstream region.
This is consistent with the modes of POD and DMD of SBLI, see for instance~\citep{pasquariello2017unsteady,nichols2017stability,shinde2022features,shinde2025distributed}.
The low frequency SMD mode at $f=0.007$ for the streamwise velocity delineates the flow separation region and the separation shock.
In addition, the modal structure is present in the downstream as well as much of the upstream region in the turbulent boundary layer.
The second mode of streamwise velocity at $f=0.021$ also displays its presence in the upstream region of the SBLI, highlighting the mid-range ($f_L=0.11$) frequency scales of the SBLI.
For the streamwise velocity fluctuations, it is long argued that the fluctuations in the upstream region of SBLI are the cause of SBLI low-frequency unsteadiness~\citep{beresh2002relationship,adams2000direct,ganapathisubramani2009low}, albeit the causality is not evident here.
The peaks of the pressure modes on the other hand are mainly situated in the separation and downstream regions (Fig.~\ref{fig:sbli_phi}c,d).
Notably, the emanating shock off the SBLI are exhibited by most pressure modes, with relatively less presence in the upstream boundary layer region.
The pressure mode dynamics is more aligned with the second mechanism of SBLI unsteadiness, which remains intrinsic to the separated flow and feedback from the reattachment to separation points~\citep{dussauge2006unsteadiness,piponniau2009simple,touber2009large}.
The spacetime-spectral dynamics of other modes for the streamwise velocity and the pressure as well as the spanwise and wall normal velocity components is expected to reveal more about the physics of SBLI.

\section{Full/reduced rank reconstruction of flowfields}\label{sec:rom}

Reconstruction of the original flowfields is an important facet of modal decomposition techniques.
This generally involves a linear combination of select spatial modes and corresponding time coefficients.
The modal reconstruction of flowfields enables various applications, including data reduction, denoising, reduced-order predictive modeling, among others.
In this section, we present the reconstruction of flowfields using the SMD procedure, specifically utilizing Eq.~\ref{eq:rec}.
The real part of the tensor $\pmb{X}$ of Eq.~\ref{eq:rec} consists of the original real flowfields with no imaginary part.
In reconstruction of $\pmb{X}$ however, the imaginary part may comprise machine-precision/discretization-related errors that can be neglected, unless the original flowfields had a non-zero imaginary part.
The equation Eq.~\ref{eq:rec} can be expressed as,
\begin{equation} \label{eq:smd_rec}
    \pmb{X}=\sum_{n=1}^{N}\pmb{\phi}_n\lambda_n^{\frac{1}{2}}\pmb{\psi}_n^\dagger,
\end{equation}
where $N$ is the maximum number of spectral modes.
The element-wise contribution of a particular spectral mode ($n$) at a unique frequency ($f$) in the reconstruction of $\pmb{U}$ can be given as
\begin{align}
    \mathcal{R}(U_{ij}) &=\lambda^{\frac{1}{2}} \left[ \mathcal{R}(\phi_i)\mathcal{R}(\psi_j)+\mathcal{I}(\phi_i)\mathcal{I}(\psi_j) \right], \text{ and} \label{eq:proj1} \\
    \mathcal{I}(U_{ij}) &=\lambda^{\frac{1}{2}} \left[ \mathcal{I}(\phi_i)\mathcal{R}(\psi_j)-\mathcal{R}(\phi_i)\mathcal{I}(\psi_j) \right],\label{eq:proj2}
\end{align}
where $\mathcal{R}$ and $\mathcal{I}$ stand for real and imaginary parts, respectively.

\begin{figure}
\centering
\begin{minipage}{0.32\textwidth}
\centering {\includegraphics[width=0.8\textwidth]{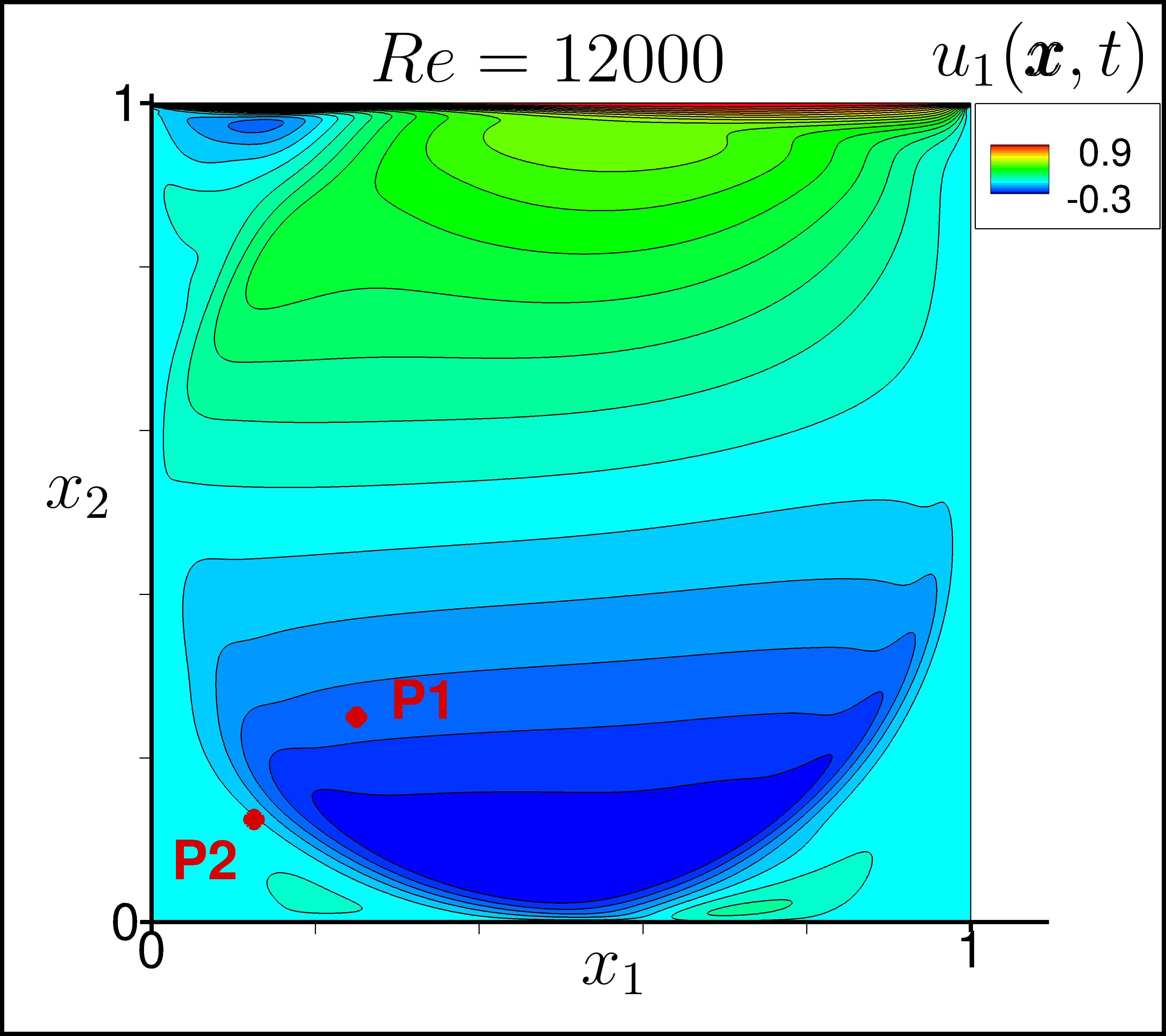}}\\(a)
\end{minipage}
\begin{minipage}{0.66\textwidth}
\centering {\includegraphics[width=1.0\textwidth]{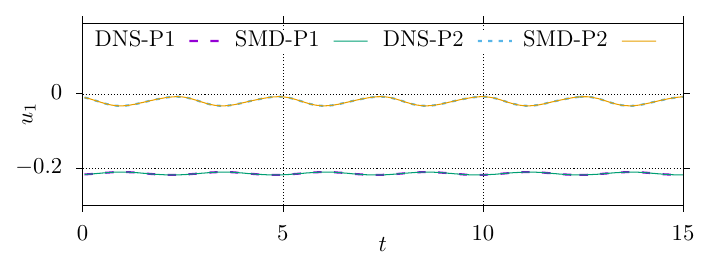}}\\(b)
\end{minipage}\\
\begin{minipage}{0.32\textwidth}
\centering {\includegraphics[width=0.8\textwidth]{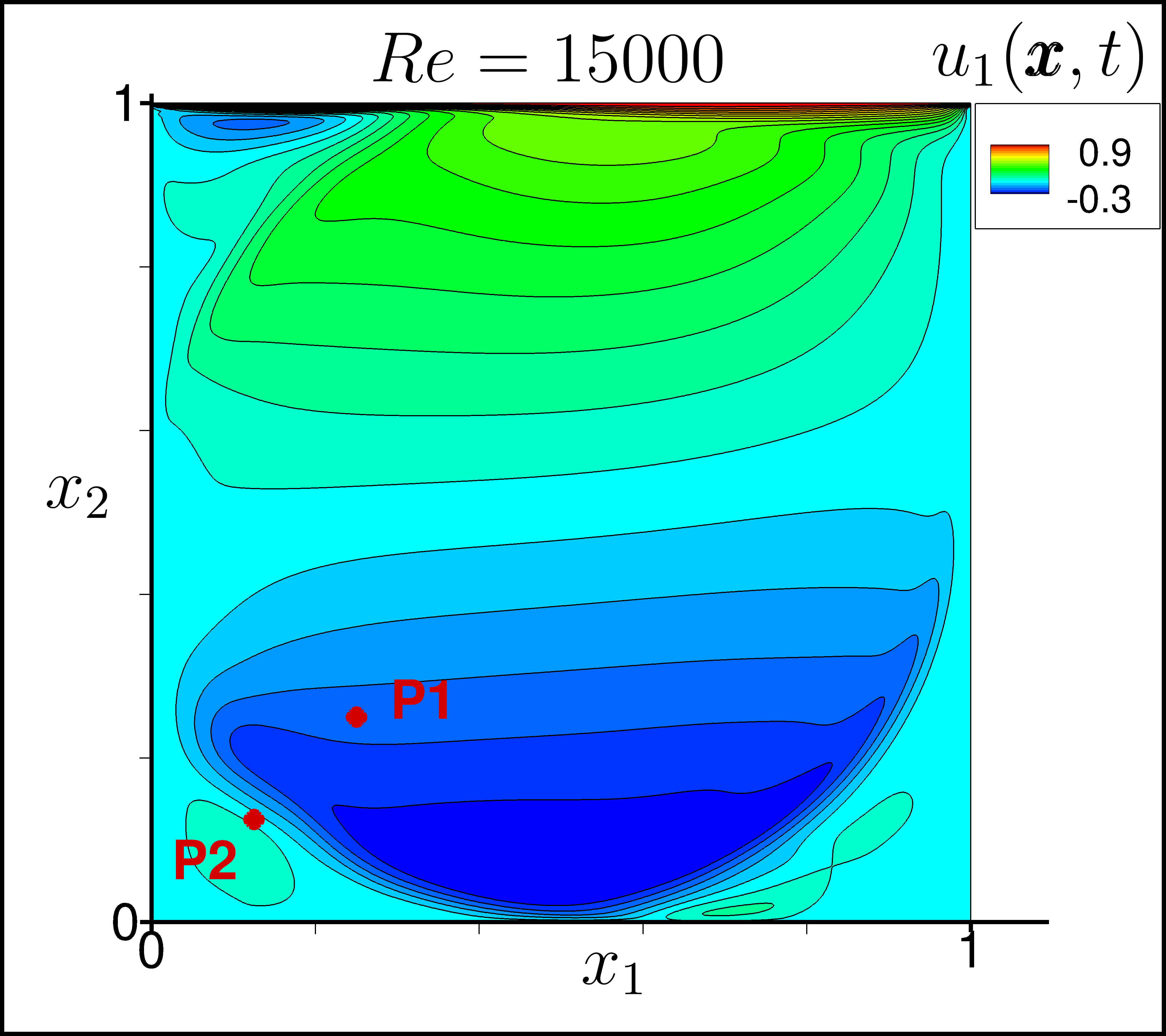}}\\(c)
\end{minipage}
\begin{minipage}{0.66\textwidth}
\centering {\includegraphics[width=1.0\textwidth]{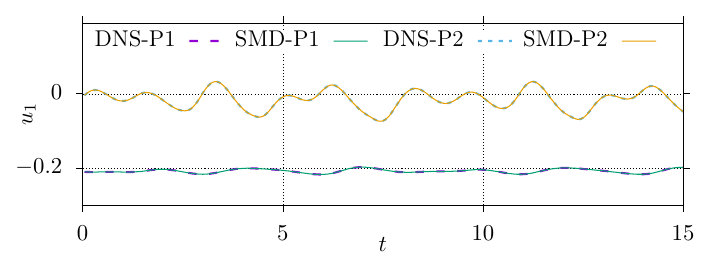}}\\(d)
\end{minipage}
\caption{Reconstruction of the horizontal velocity flowfields from the SMD modes of LDC at $Re=12000$ and $Re=15000$. (a) Reconstructed horizontal velocity field at the same time instance as in Fig.~\ref{fig:smd_ldc_12k}(a), (b) Comparison of the space-point time-variation of the horizontal velocity between DNS and SMD-reconstructed flowfield at the probe locations: P1=($0.25$, $0.25$) and P2=($0.125$, $0.125$)}
\label{fig:rec_ldc}
\end{figure}

The SMD reconstruction of the LDC flow at $Re=12000$ of Fig.~\ref{fig:smd_ldc_12k}(a) and the LDC flow at $Re=15000$ of Fig.~\ref{fig:smd_ldc_15k}(a) by means of Eq.~\ref{eq:smd_rec} is presented in Fig.~\ref{fig:rec_ldc}.
It is a full reconstruction where all spectral modes are utilized.
Fig.~\ref{fig:rec_ldc}(a) and Fig.~\ref{fig:rec_ldc}(c) display the reconstructed horizontal component of velocity in the LDC at $Re=12000$ and $Re=15000$, respectively, where the reconstruction is at the same time instance of Fig.~\ref{fig:smd_ldc_12k}(a) and Fig.~\ref{fig:smd_ldc_15k}(a), respectively.
The original and reconstructed LDC flowfields for both cases match to the machine precision.
Furthermore, the flow velocity is compared at two probes that are located in the core vortex ($P1$) and shear layer ($P2$) regions of the LDC flow at $Re=12000$ and $Re=15000$ in Fig.~\ref{fig:rec_ldc}(b) and Fig.~\ref{fig:rec_ldc}(d), respectively.
The instantaneous probe values of the velocity exactly matches between the original DNS computations and the reconstructed solution via SMD procedure.
The same of level of accuracy is present in the reconstruction of other solution state variables, including the pressure.

\begin{figure}
\centering
\begin{minipage}{0.48\textwidth}
\centering {\includegraphics[width=1.0\textwidth]{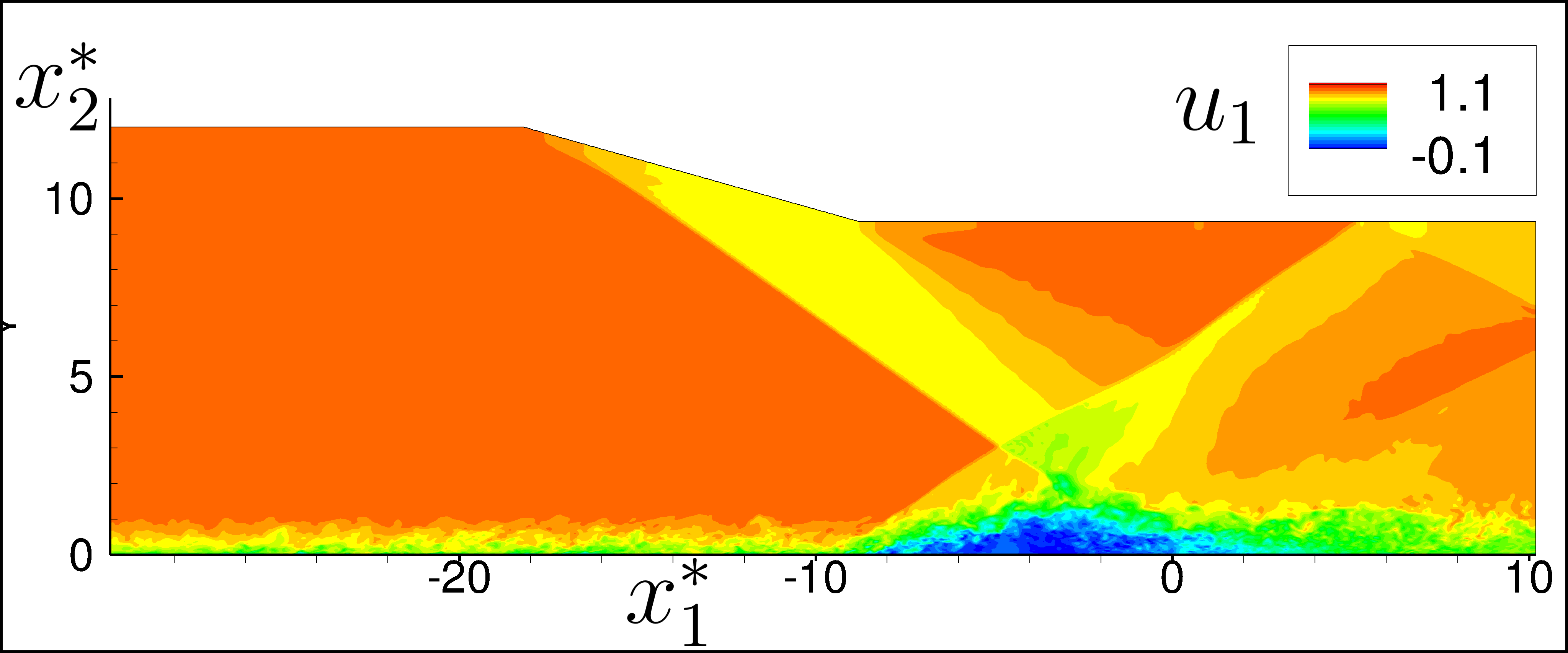}}\\(a)
\end{minipage}
\begin{minipage}{0.48\textwidth}
\centering {\includegraphics[width=1.0\textwidth]{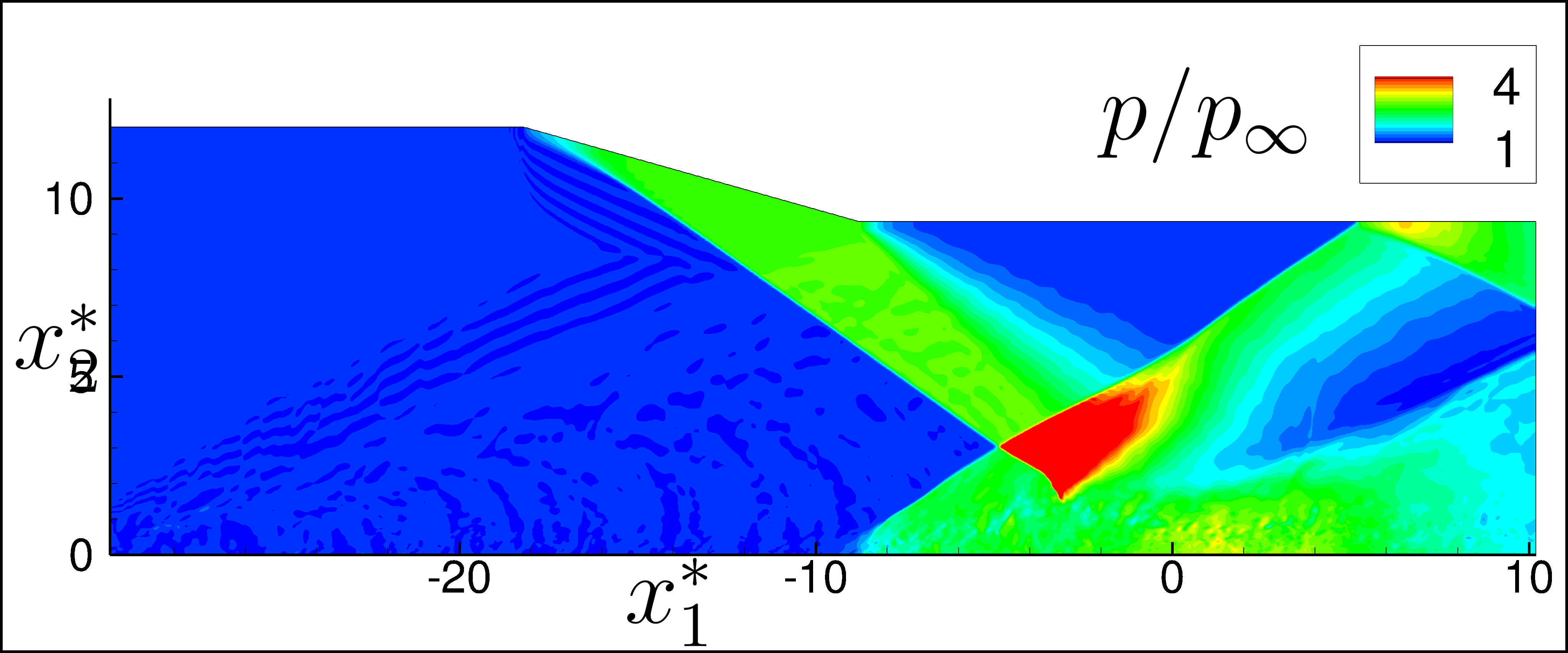}}\\(b)
\end{minipage}
\begin{minipage}{1.0\textwidth}
\centering {\includegraphics[width=1.0\textwidth]{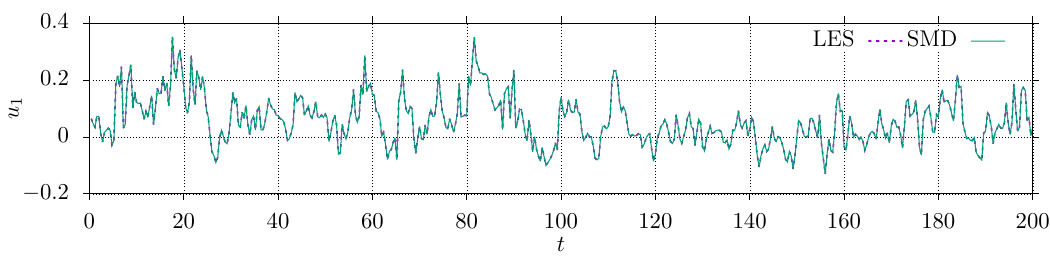}}\\(c)
\end{minipage}\\
\begin{minipage}{1.0\textwidth}
\centering {\includegraphics[width=1.0\textwidth]{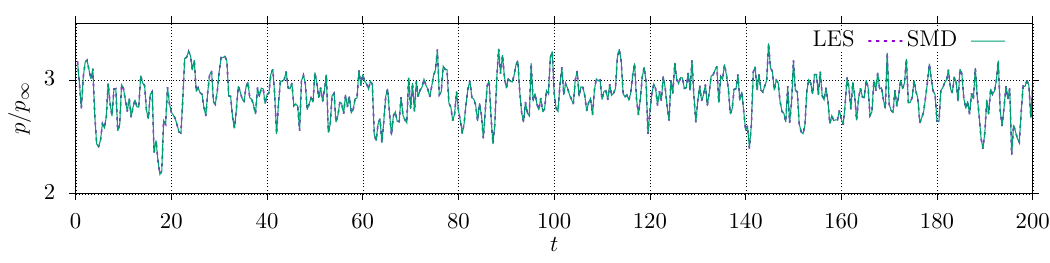}}\\(d)
\end{minipage}
\caption{Reconstruction of SBLI instantaneous flowfields using SMD procedure for (a) the streamwise velocity and (b) the pressure. Comparison of the instantaneous (c) streamwise velocity and (d) pressure near the shock impingement location (probe at $x^\ast_1=0.17$, $x^\ast_2=0.1$) between the LES and SMD.}
\label{fig:flow_rec_sbli}
\end{figure}

The full reconstruction (Eq.~\ref{eq:smd_rec}) of the SBLI flowfields is presented in Fig.~\ref{fig:flow_rec_sbli} in terms of the instantaneous streamwise velocity and the pressure fields.
The reconstructed velocity field (Fig.~\ref{fig:flow_rec_sbli}a) is at the same time instance as the original velocity field that is shown in Fig.~\ref{fig:flow_sbli}(b), exhibiting an agreement to the machine precision.
The corresponding pressure field of Fig.~\ref{fig:flow_rec_sbli}(b) is also reproduced with the same degree of accuracy.
To compare the reconstructed flowfields at a probe location, we consider a probe that is situated close to the shock impingement location, at ($x_1^\ast=0.17$, $x_2^\ast=0.1)$, where the flow is highly turbulent.
The original and reconstructed flowfields at this probe location are displayed in Fig.~\ref{fig:flow_rec_sbli}(c) for the streamwise velocity and Fig.~\ref{fig:flow_rec_sbli}(d) for the pressure.
In this full reconstruction, both the velocity and pressure probe signals are completely reproduced with no loss of information.

\begin{figure}
\centering
\begin{minipage}{0.32\textwidth}
\centering {\includegraphics[width=1.0\textwidth]{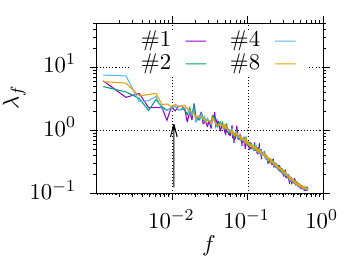}}\\(a)
\end{minipage}
\begin{minipage}{0.28\textwidth}
\centering {\includegraphics[width=1.0\textwidth]{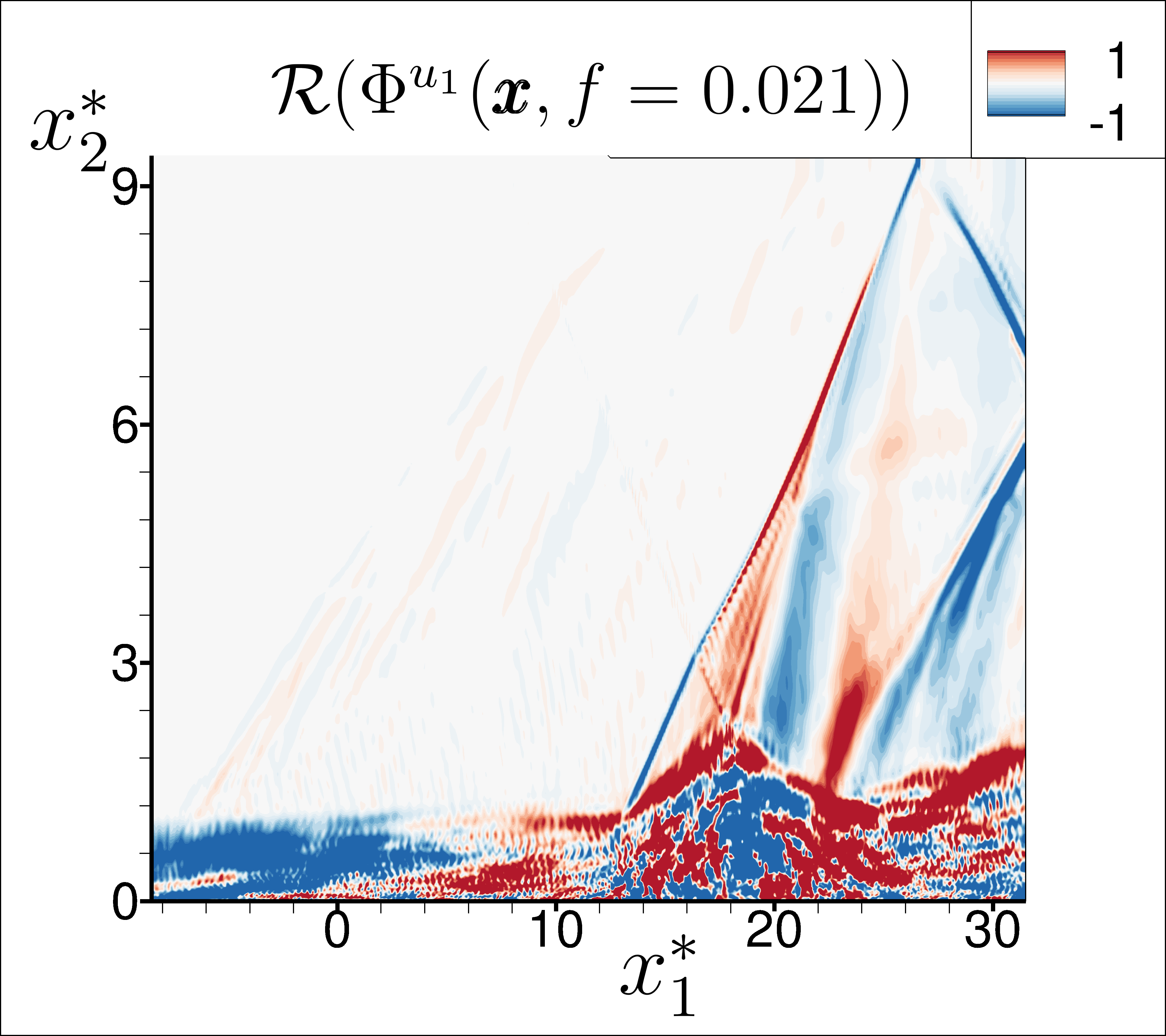}}\\(b)
\end{minipage}
\begin{minipage}{0.28\textwidth}
\centering {\includegraphics[width=1.0\textwidth]{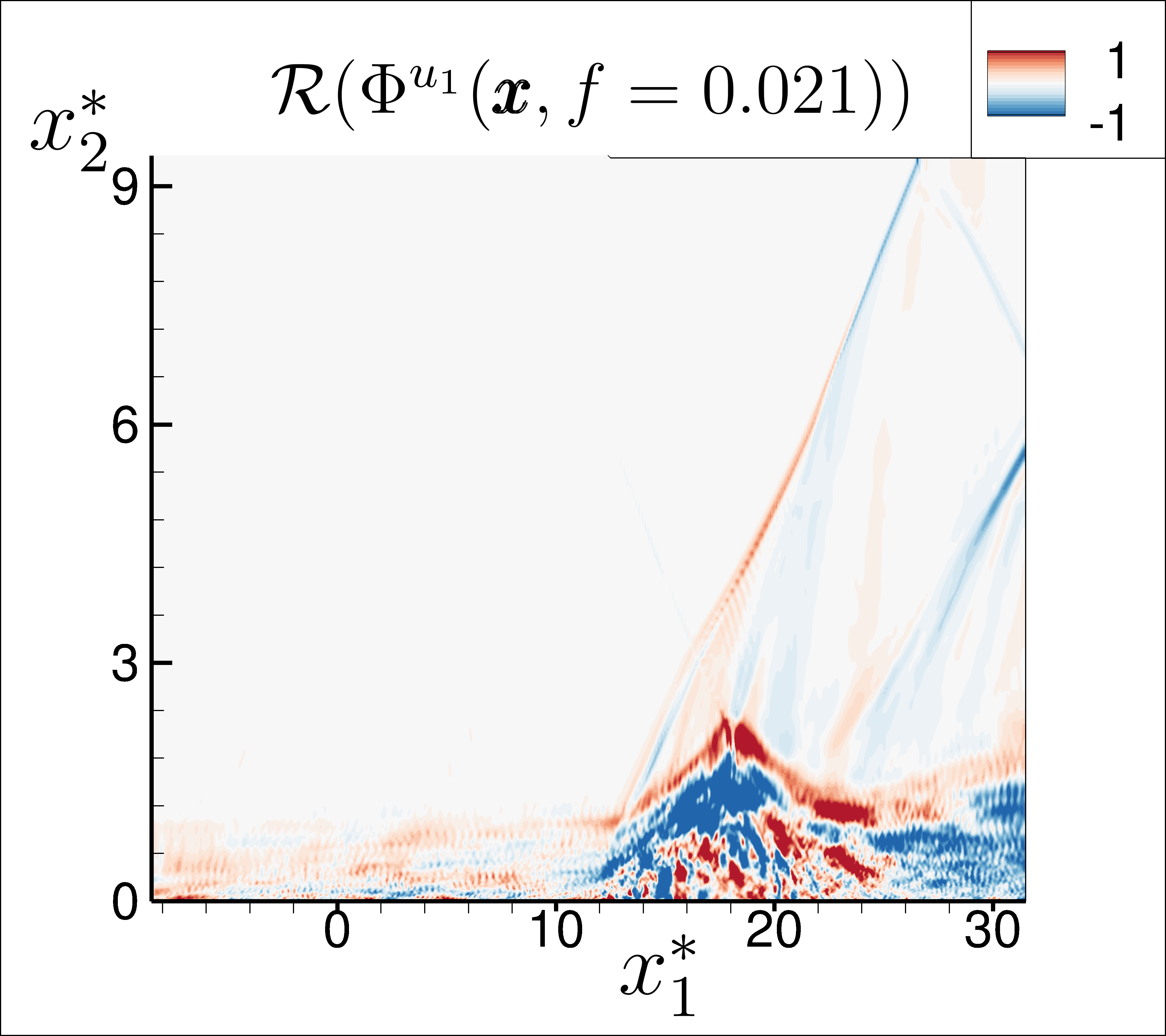}}\\(c)
\end{minipage}
\begin{minipage}{0.32\textwidth}
\centering {\includegraphics[width=1.0\textwidth]{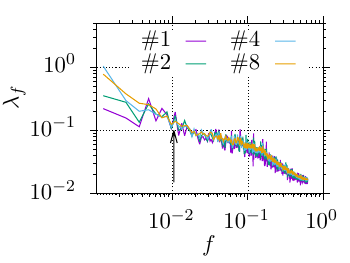}}\\(d)
\end{minipage}
\begin{minipage}{0.28\textwidth}
\centering {\includegraphics[width=1.0\textwidth]{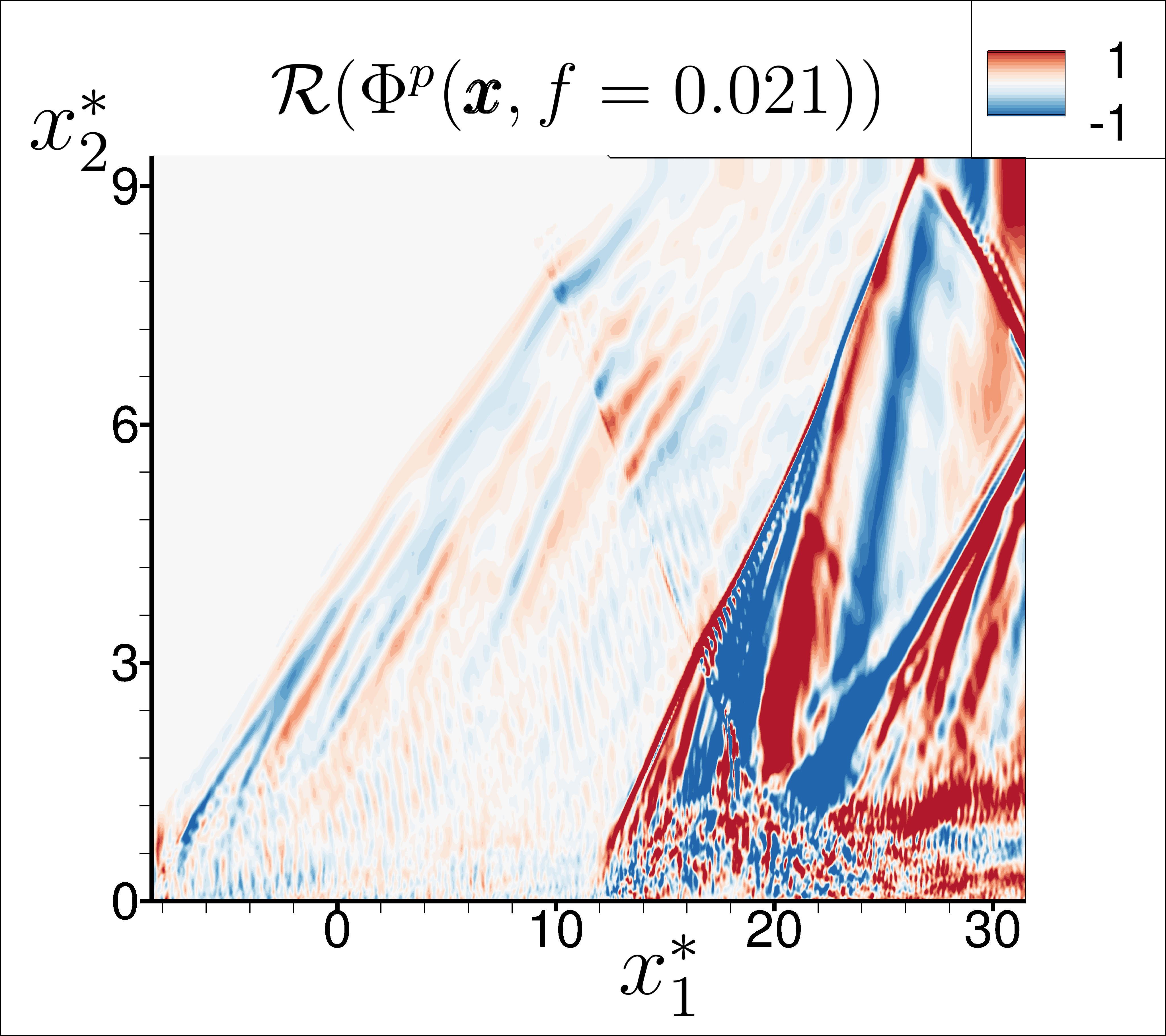}}\\(e)
\end{minipage}
\begin{minipage}{0.28\textwidth}
\centering {\includegraphics[width=1.0\textwidth]{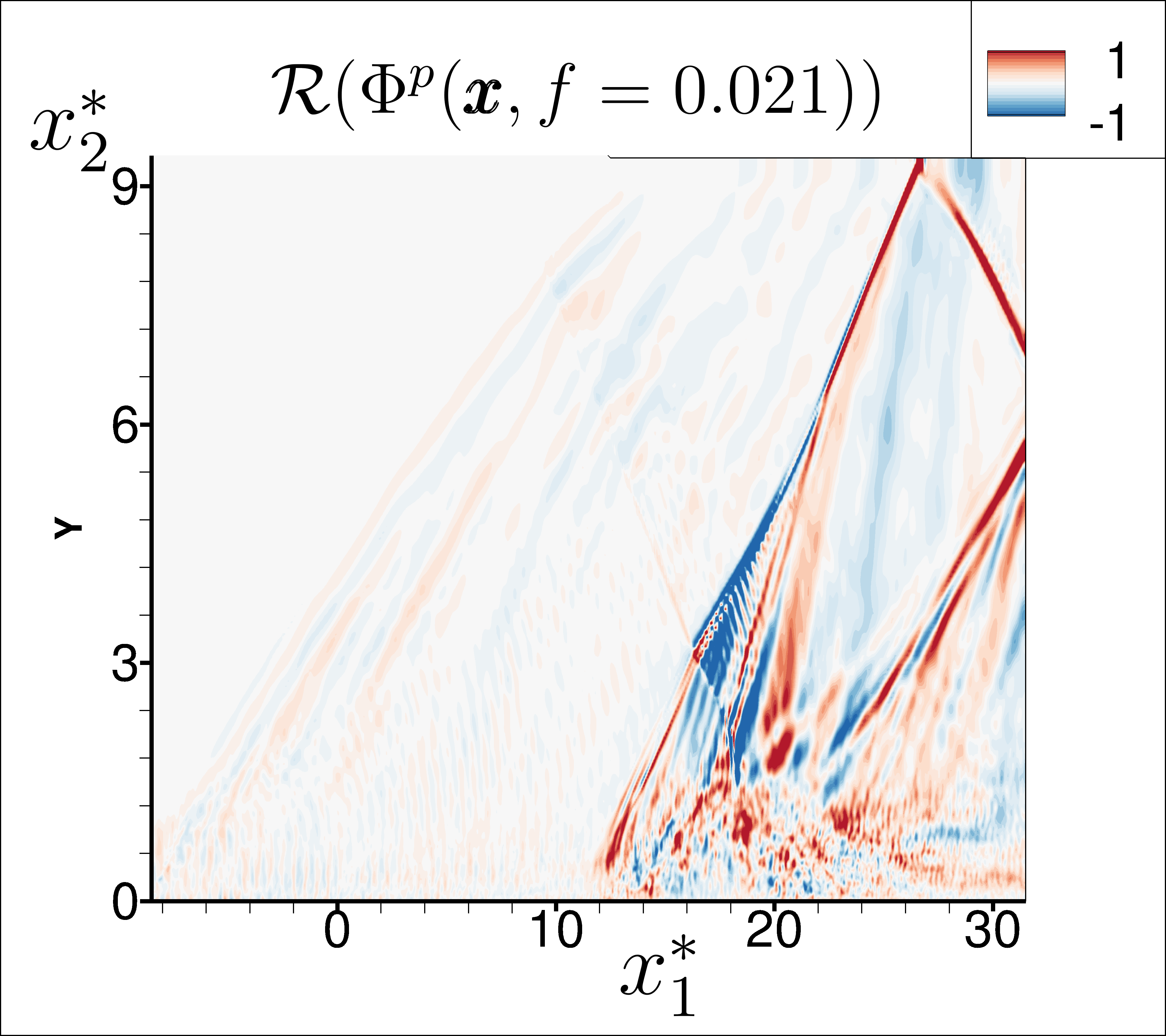}}\\(f)
\end{minipage}
\caption{Convergence of SMD modes. (a) Spectral energies of SMD modes of velocity for multiple (\#) blocks of the data (b) SMD mode of velocity at $f=0.021$ for a single block. (c) SMD mode of velocity at $f=0.021$ for eight blocks. (a) Spectral energies of SMD modes of pressure for multiple (\#) blocks of the data (b) SMD mode of pressure at $f=0.021$ for a single block. (c) SMD mode of pressure at $f=0.021$ for eight blocks.}
\label{fig:smd_conv}
\end{figure}

To improve the spectral convergence, Welch method~\citep{welch2003use} is commonly used, where the full length signal is divided into multiple time windows leading to an averaged spectrum.
In addition, the alteration of time signal to avoid spectral leakage, the ensemble averaging, and/or the zero padding of signal are some practical operations to improve the accuracy and computational efficiency of the time-frequency transformation.
The SMD procedure presented in this paper can benefit from all techniques of spectral averaging to obtain converged spectral modes.
For instance, Fig.~\ref{fig:smd_conv} presents the results of spectral averaging by means of the Welch method, selecting a rectangular window.
A total of $1024$ flowfields that are separated by a time-step of $0.4$ were utilized in a single block of data. 
A maximum of $8$ such blocks were considered to perform the convergence analysis, fixing the spectral range and resolution.
The spectral modal energies of the velocity (Fig.~\ref{fig:smd_conv}a) and pressure (Fig.~\ref{fig:smd_conv}d), to a large extent, appear converged, particularly in the higher frequency region ($f\gtrapprox 0.05$), while some discrepancies exist at the lower frequency end of the spectra.
The SMD modes at $f=0.021$ for the velocity (Fig.~\ref{fig:smd_conv}b, c) and pressure (Fig.~\ref{fig:smd_conv}e, f) are displayed here.
The single block SMD modes of velocity (Fig.~\ref{fig:smd_conv}b) and pressure (Fig.~\ref{fig:smd_conv}e) appear noisy with higher peak values compared to the eight blocks averaged spectral modes (Fig.~\ref{fig:smd_conv}c and f, respectively).

A computationally efficient strategy of 1. performing SVD of the individual block, 2. obtaining the SMD, and 3. averaging the SMD over the number of blocks is adopted here.
Alternatively one can perform the SVD of the entire series of flowfields, and them obtains the SMD modes for individual blocks by employing Fourier transform step.
In either case, an instantaneous spectrogram for the entire time duration can be obtained for the converged SMD modes by means of Eq.~\ref{eq:varphi}, where the flowfields are projected on the SMD spatial modes.

\begin{figure}
\centering
\begin{minipage}{0.48\textwidth}
\centering {\includegraphics[width=1.0\textwidth]{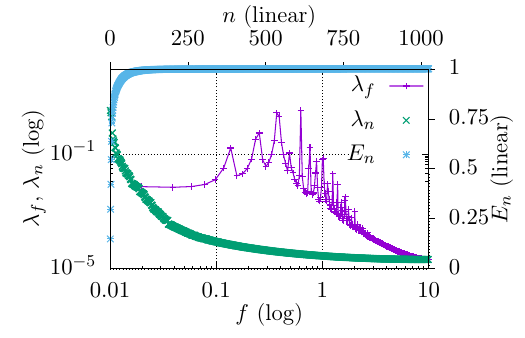}}\\(a)
\end{minipage}
\begin{minipage}{0.48\textwidth}
\centering {\includegraphics[width=1.0\textwidth]{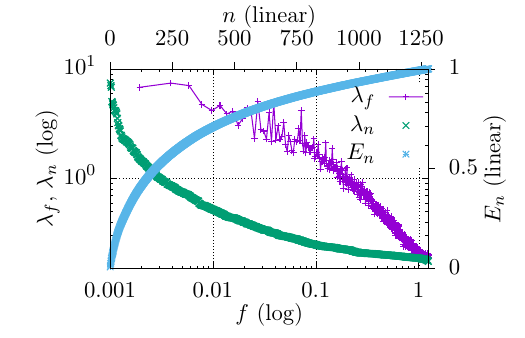}}\\(b)
\end{minipage}
\caption{Spectral modal energy as a function of frequency and mode number for (a) LDC at $Re=15000$, and (b) SBLI. $E_n={\sum_{l=1}^{n} \lambda_l}/{\sum_{l=1}^{N}\lambda_l}$ is a normalized cumulative energy of the modes.}
\label{fig:energy}
\end{figure}

A partial reconstruction of flowfields finds applications in data size reduction, denoising, reduced-order modeling (ROM), among others.
Modal decomposition techniques are popularly utilized in ROM of fluid flows, where a set of select modes leads to an efficient model that governs the system dynamics.
Linearity of the modes is an important property of a modal decomposition technique, allowing linear combinations of the essential modes while discarding the unnecessary modes of the system dynamics.
SMD inherits the linearity from both the SVD and Fourier transform, availing two possibilities of reduced-order flowfield reconstruction.
First, a reduced number of SVD modes based on their energies (squared singular values) can be utilized to compute the SMD modes, and second, a reduced number of SMD modes based on their spectral energies ($\lambda_f$).
The reconstruction of flowfields in Eq.~\ref{eq:smd_rec} represents a linear combination of the spectral modes and energies.
The mode selection in SMD for a partial reconstruction of flowfields is straightforward, where energy and/or frequency or their ranges can be used as the selection criteria.

The spectral energies of the LDC flow at $Re=15000$ and the SBLI flow are presented in Fig.~\ref{fig:energy}(a) and Fig.~\ref{fig:energy}(a), respectively, where the SMD modes have both the frequency ($f$) and the energy ($n$) ranks.
The modal energies, $\lambda_f$ and $\lambda_n$, are shown on a logarithmic scale on $y$-axis.
The corresponding frequencies and the modal ranks are marked on $x$-axis on a logarithmic scale and $x2$-axis (top horizontal axis) on a linear scale, respectively.
The sub-figures also present the total cumulative energy ($E_n$) graph for increasing modal rank as:
\begin{equation}
    E_n=\frac{\sum_{l=1}^{n}\lambda_l}{\sum_{l=1}^{N}\lambda_l},
\end{equation}
on $y2$-axis (right vertical axis) on a linear scale.

The SMD modal energy variation with frequency is evident in both sub-figures (of Fig.~\ref{fig:energy}).
For LDC flow (Fig.~\ref{fig:energy}a), most of the spectral energy is associated with the modes at $f\lessapprox 1$ with most energetic peaks at $f=0.63$ and $f=0.39$.
On the other hand, the SMD spectral energies of SBLI indicate a wide range of frequencies at higher modal energies, which is representative of a high rank system with broadband frequencies contributing to the total energy.
The low rank behavior of the LDC flow and the high rank behavior of the SBLI flow are also evident in the cumulative energies, $E_n$, that make nearly $100\%$ in a fewer number of modes ($\approx 20$ out of $1024$ SMD mdoes) for the LDC and take more number of modes (over $1000$ out of $1280$ SMD mdoes) for the SBLI flow.
Naturally, the SMD modal energy spectra provide insights into the scales of turbulence and the turbulent energy cascade that is consistent with the general theory of turbulence~\citep{lumley1970stochastic}.

\begin{figure}
\centering
\begin{minipage}{0.48\textwidth}
\centering {\includegraphics[width=1.0\textwidth]{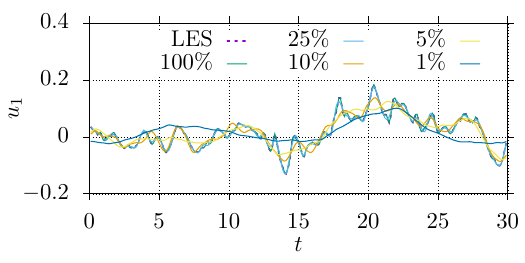}}\\(a)
\end{minipage}
\begin{minipage}{0.48\textwidth}
\centering {\includegraphics[width=1.0\textwidth]{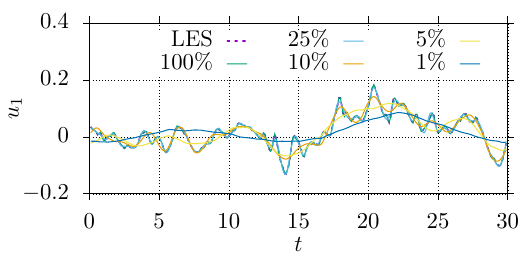}}\\(b)
\end{minipage}\\
\begin{minipage}{0.48\textwidth}
\centering {\includegraphics[width=1.0\textwidth]{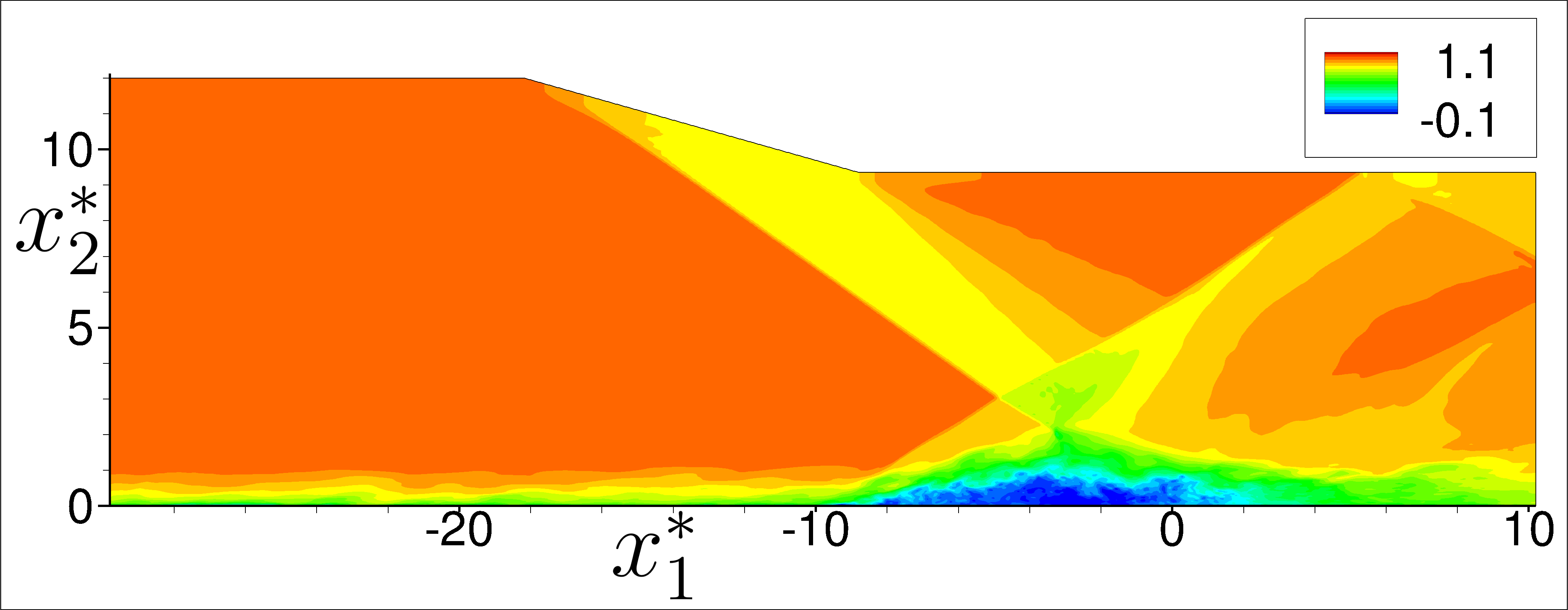}}\\(c)
\end{minipage}
\begin{minipage}{0.48\textwidth}
\centering {\includegraphics[width=1.0\textwidth]{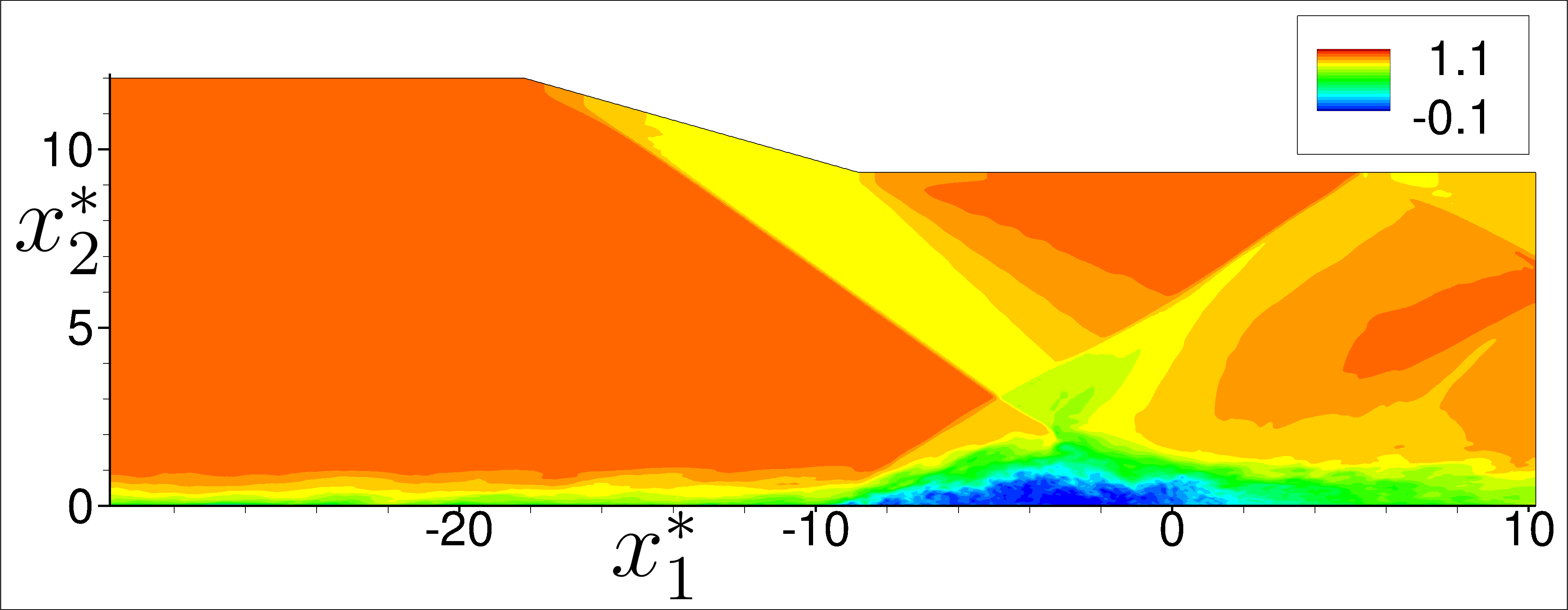}}\\(d)
\end{minipage}
\caption{Reduced order reconstruction of SBLI flowfields using SMD. Reconstruction of instantaneous probe velocity at ($x_1^\ast=0.17$, $x_2^\ast=0.1$) near the shock impingement location using (a) reduction in the number of SVD modes and retaining all SMD modes, and (b) reduction in the number of SMD modes and utilizing all SVD modes. SBLI streamwise velocity fields reconstructed using $5\%$ of the total number of modes via (c) SVD reduction as well as (d) the SMD reduction.}
\label{fig:rom}
\end{figure}

The partial reconstructions of the SBLI flow using the SVD and SMD stage mode reduction are displayed in Fig.~\ref{fig:rom}.
The figures on the left (Fig.~\ref{fig:rom}a,c) show the partial reconstruction by reducing the number of SVD modes while keeping all SMD modes, whereas the figures on the right (Fig.~\ref{fig:rom}b,d) show the partial reconstruction by keeping all SVD modes and reducing the number of SMD modes.
In either case, the flow is accurately predicted for retaining all modes ($100\%$); in addition, retaining only top $25\%$ of the total modes still reproduces the instantaneous flowfields with minor discrepancies (see the top figures of instantaneous flow velocity probe).
Further decreasing of the number of modes to $10\%$, $5\%$, and $1\%$ results in significant filtering of the turbulent features of the SBLI flow.
The top figures (Fig.~\ref{fig:rom}a, b) the velocity probe data reconstruction, for instance, using only $1\%$ of the total modes, indicating a much smoother response yet exhibiting the low frequency dynamics.
Although, the level of filtering appear similar for the SVD and SMD routes of partial reconstruction, there exist subtle differences in the instantaneous dynamics.
The partial reconstructions of the instantaneous velocity flowfield of SBLI using the top $5\%$ of the total modes are shown in Fig.~\ref{fig:energy}(c) and Fig.~\ref{fig:energy}(d) for the SVD and SMD routes, respectively.
Broadly, a large fraction of the turbulent features of SBLI are filtered out.
While the figures appear nearly identical, there exist subtle differences pertaining to the SVD or SMD route of the partial reconstruction.
Notably, the low rank characteristics of SMD are similar to SVD, where SMD also provides a possibility of frequency based mode selection, making it a suitable candidate for reduced-order modeling.

\section{Concluding remarks}\label{sec:concl}

This paper presents a spectral mode decomposition technique to perform spacetime-frequency analysis of spatiotemporal flowfields.
While the techniques for the time-frequency analysis and frequency-based modal decomposition are fairly well established in the literature, broadly these techniques have been treating the problems for either time-locality of the spectrum or a converged spatial coherence of the spectral modes.
In fluid dynamics, where high resolution spatiotemporal experimental and numerical databases are commonly encountered, a combined treatment of spatiotemporal dynamics is highly desired; the presented SMD approach seeks to addresses this point.
The existing frequency-based modal decompositions, mainly the FT, DMD, and the spectral variants of POD, consider a constant frequency time-variation of the modal time-coefficients or equivalently the modal contribution over time; this is questionable, particularly in turbulent flowfields.
Besides, the main objective of the time-frequency analysis techniques is to obtain the time-local contribution/presence of frequency.
The time-local spectral presence and its variation over time need not be at a single frequency with uniform amplitude.

The SMD procedure distills out three main features of the spatiotemporal flowfields: 1) each spectral spatial mode correspond to a single frequency, {\it i.e.}, a unique spectral mode at a unique frequency.
2) The spectral time-modes (time-coefficients) exhibit the time-local contribution/presence of the corresponding spectral spatial mode.
While a time-mode correspond to a single frequency spatial modes, its time-local contribution need not comprise a uniform amplitude over the time, instead it depends on the instantaneous flow physics and contribution of that SMD mode to the flowfield.
The non-normalized time-modes lead to an instantaneous contribution of the spectral modes - referred here as the spectrogram.
A peak on the spectrogram indicates the presence of that global mode at that specific time, providing crucial insights into the physical mechanisms of unsteadiness in the flow.
3) Lastly, the spectral modes are associated with a global spectral energy, which unambiguously defines the energy rank of the mode.
In addition to the modal frequency, the energy rank provides an alternative mode selection criterion, where the both criteria are useful in full/partial reconstruction of the flowfields, reduced-order modeling and de-noising applications.

Application of SMD to transitional (LDC at Mach $0.5$ and $Re_L=12000$)/ turbulent (LDC at Mach $0.5$ and $Re_L=15000$ and SBLI at Mach $2.7$ and $Re_{\delta_{in}}=54000$) flows manifest how SMD extracts the spectral modes that are energy dominant and their temporal presence in the flow.
In transitional flows, where the unsteadiness/instabilities are dictated by a relatively fewer number of spectral modes, SMD is particularly useful in tracing the spatiotemporal presence of those modes.
Similarly in turbulent flows, SMD can easily trace intermittent/extreme events and provide insight into their spatiotemporal presence and modal support.
For instance, the SMD spectrograms of the SBLI flow clearly exhibit the spectral energy peaks/bursts at the characteristic lower frequency range of SBLI.
Notably, the low-frequency peaks of the pressure and the streamwise velocity appear at different time instances (Figs.~\ref{fig:sg_sbli}b \&~\ref{fig:sg_sbli}d), indicating the difference in dynamics at the low-frequency range of SBLI for different variables (at least for the pressure and the streamwise velocity here).
A further detailed investigation of this temporal dynamics by considering the wall normal/spanwise velocity and temperature is expected to shed more light on the mechanisms of low-frequency unsteadiness in SBLI.

Lastly, the convergence of the spectral modes is generally enhanced by means of ensemble averaging or by employing Welch method~\citep{welch2003use}, where a long duration statistically stationary signal is divided into small-duration windows.
Other practical modifications to time signals include the shapes of windows ({\it e.g.} Hanning window) to control spectral leakage, and/or the zero padding to improve the algorithm efficiency.
The phase information is generally compromised in these operations; nonetheless, it leads to a continuous instantaneous spectrogram of the converged SMD modes.
The paper also illustrates the use of SMD in full/partial flow reconstruction based on modes at a specific frequency/energy or their ranges.
Notably, SMD comprises the desired characteristics of modal decomposition to construct reduced-order models, including the appropriate low rank behaviour, the spatial modes orthogonality, and a possibility of selecting a specific frequency dynamics.

\section*{Acknowledgment}

This work is supported by the National Aeronautics and Space Administration (NASA) under Grant No. 80NSSC24M0103.
The author acknowledges the computing resources provided by the high performance computing collaboratory (HPCC) of Mississippi State University.

\bibliographystyle{jfm}
\bibliography{jfm-instructions}

@article{sieber2016spectral,
  title={Spectral proper orthogonal decomposition},
  author={Sieber, Moritz and Paschereit, C Oliver and Oberleithner, Kilian},
  journal={Journal of Fluid Mechanics},
  volume={792},
  pages={798--828},
  year={2016},
  publisher={Cambridge University Press}
}

@article{nekkanti2021frequency,
  title={Frequency--time analysis, low-rank reconstruction and denoising of turbulent flows using SPOD},
  author={Nekkanti, Akhil and Schmidt, Oliver T},
  journal={Journal of Fluid Mechanics},
  volume={926},
  pages={A26},
  year={2021},
  publisher={Cambridge University Press}
}

@article{kosambi1943statistics,
  title={Statistics in function space},
  author={Kosambi, DD},
  journal={Journal of the Indian Mathematical Society},
  volume={7},
  pages={76--88},
  year={1943},
  publisher={Indian Mathematical Society}
}

@article{cohen1995time,
  title={Time-frequency analysis},
  author={Cohen, Leon},
  journal={Englewood Cliffs},
  year={1995}
}

@article{morales2022time,
  title={Time-frequency analysis methods and their application in developmental EEG data},
  author={Morales, Santiago and Bowers, Maureen E},
  journal={Developmental cognitive neuroscience},
  volume={54},
  pages={101067},
  year={2022},
  publisher={Elsevier}
}

@book{batchelor1953theory,
  title={The theory of homogeneous turbulence},
  author={Batchelor, George Keith},
  year={1953},
  publisher={Cambridge university press}
}

@book{tennekes1972first,
  title={A first course in turbulence},
  author={Tennekes, Hendrik and Lumley, John Leask},
  year={1972},
  publisher={MIT press}
}

@article{welch2003use,
  title={The use of fast Fourier transform for the estimation of power spectra: A method based on time averaging over short, modified periodograms},
  author={Welch, Peter},
  journal={IEEE Transactions on audio and electroacoustics},
  volume={15},
  number={2},
  pages={70--73},
  year={2003},
  publisher={IEEE}
}

@article{shinde2021lagrangian,
  title={Lagrangian approach for modal analysis of fluid flows},
  author={Shinde, Vilas J and Gaitonde, Datta V},
  journal={Journal of Fluid Mechanics},
  volume={928},
  pages={A35},
  year={2021},
  publisher={Cambridge University Press}
}

@article{ghia1982high,
  title={High-Re solutions for incompressible flow using the Navier-Stokes equations and a multigrid method},
  author={Ghia, UKNG and Ghia, Kirti N and Shin, C T},
  journal={Journal of computational physics},
  volume={48},
  number={3},
  pages={387--411},
  year={1982},
  publisher={Academic Press}
}

@article{shen1991hopf,
  title={Hopf bifurcation of the unsteady regularized driven cavity flow},
  author={Shen, Jie},
  journal={Journal of Computational Physics},
  volume={95},
  number={1},
  pages={228--245},
  year={1991},
  publisher={Elsevier}
}

@article{ramanan1994linear,
  title={Linear stability of lid-driven cavity flow},
  author={Ramanan, Natarajan and Homsy, George M},
  journal={Physics of Fluids},
  volume={6},
  number={8},
  pages={2690--2701},
  year={1994},
  publisher={American Institute of Physics}
}

@techreport{koseff1983three,
  title={Three-dimensional lid-driven cavity flow: experiment and simulation},
  author={Koseff, J R and Street, R L and Gresho, P M and Upson, C D and Humphrey, J A C and To, W M},
  year={1983},
  institution={Stanford Univ., CA (USA). Dept. of Civil Engineering; Lawrence Livermore~…}
}

@article{sheu2002flow,
  title={Flow topology in a steady three-dimensional lid-driven cavity},
  author={Sheu, Tony W H and Tsai, S F},
  journal={Computers \& fluids},
  volume={31},
  number={8},
  pages={911--934},
  year={2002},
  publisher={Elsevier}
}

@article{albensoeder2005accurate,
  title={Accurate three-dimensional lid-driven cavity flow},
  author={Albensoeder, Stefan and Kuhlmann, Hendrik C},
  journal={Journal of Computational Physics},
  volume={206},
  number={2},
  pages={536--558},
  year={2005},
  publisher={Elsevier}
}

@article{bruneau20062d,
  title={The 2D lid-driven cavity problem revisited},
  author={Bruneau, Charles-Henri and Saad, Mazen},
  journal={Computers \& fluids},
  volume={35},
  number={3},
  pages={326--348},
  year={2006},
  publisher={Elsevier}
}

@article{lopez2017transition,
  title={Transition to complex dynamics in the cubic lid-driven cavity},
  author={Lopez, Juan M and Welfert, Bruno D and Wu, Ke and Yalim, Jason},
  journal={Physical Review Fluids},
  volume={2},
  number={7},
  pages={074401},
  year={2017},
  publisher={APS}
}

@article{bergamo2015compressible,
  title={Compressible modes in a square lid-driven cavity},
  author={Bergamo, Leandro F and Gennaro, Elmer M and Theofilis, Vassilis and Medeiros, Marcello A F},
  journal={Aerospace Science and Technology},
  volume={44},
  pages={125--134},
  year={2015},
  publisher={Elsevier}
}

@article{ohmichi2017compressibility,
  title={Compressibility effects on the first global instability mode of the vortex formed in a regularized lid-driven cavity flow},
  author={Ohmichi, Yuya and Suzuki, Kojiro},
  journal={Computers \& Fluids},
  volume={145},
  pages={1--7},
  year={2017},
  publisher={Elsevier}
}

@article{ranjan2020robust,
  title={A robust approach for stability analysis of complex flows using high-order Navier-Stokes solvers},
  author={Ranjan, Rajesh and Unnikrishnan, S and Gaitonde, Datta},
  journal={Journal of Computational Physics},
  volume={403},
  pages={109076},
  year={2020},
  publisher={Elsevier}
}

@incollection{lumley67,
    address = {Moscow},
    author = {Lumley, J. L.},
    booktitle = {Atmospheric turbulence and radio propagation},
    editor = {Yaglom, A. M. and Tatarski, V. I.},
    pages = {166--178},
    publisher = {Nauka},
    title = {{The Structure of Inhomogeneous Turbulent Flows}},
    year = {1967}
}

@article{towne2017spectral,
  title={Spectral proper orthogonal decomposition and its relationship to dynamic mode decomposition and resolvent analysis},
  author={Towne, Aaron and Schmidt, Oliver T and Colonius, Tim},
  journal={arXiv preprint arXiv:1708.04393},
  year={2017}
}

@article{schmid2010dynamic,
  title={Dynamic mode decomposition of numerical and experimental data},
  author={Schmid, Peter J},
  journal={Journal of fluid mechanics},
  volume={656},
  pages={5--28},
  year={2010},
  publisher={Cambridge University Press}
}

@article{rowley2009spectral,
  title={Spectral analysis of nonlinear flows},
  author={Rowley, Clarence W and Mezi{\'c}, Igor and Bagheri, Shervin and Schlatter, Philipp and Henningson, Dans},
  journal={Journal of fluid mechanics},
  volume={641},
  number={1},
  pages={115--127},
  year={2009},
  publisher={Citeseer}
}

@article{rowley2017model,
  title={Model reduction for flow analysis and control},
  author={Rowley, Clarence W and Dawson, Scott TM},
  journal={Annual Review of Fluid Mechanics},
  volume={49},
  pages={387--417},
  year={2017},
  publisher={Annual Reviews}
}

@article{taira2017modal,
  title={Modal analysis of fluid flows: An overview},
  author={Taira, Kunihiko and Brunton, Steven L and Dawson, Scott TM and Rowley, Clarence W and Colonius, Tim and McKeon, Beverley J and Schmidt, Oliver T and Gordeyev, Stanislav and Theofilis, Vassilios and Ukeiley, Lawrence S},
  journal={Aiaa Journal},
  volume={55},
  number={12},
  pages={4013--4041},
  year={2017},
  publisher={American Institute of Aeronautics and Astronautics}
}

@inproceedings{glauser1987coherent,
  title={Coherent structures in the axisymmetric turbulent jet mixing layer},
  author={Glauser, Mark N and Leib, Stewart J and George, William K},
  booktitle={Turbulent Shear Flows 5: Selected Papers from the Fifth International Symposium on Turbulent Shear Flows, Cornell University, Ithaca, New York, USA, August 7--9, 1985},
  pages={134--145},
  year={1987},
  organization={Springer}
}

@article{braud2004low,
  title={Low-dimensional analysis, using POD, for two mixing layer--wake interactions},
  author={Braud, Caroline and Heitz, Dominique and Arroyo, Georges and Perret, Laurent and Delville, Jo{\"e}l and Bonnet, Jean-Paul},
  journal={International journal of heat and fluid flow},
  volume={25},
  number={3},
  pages={351--363},
  year={2004},
  publisher={Elsevier}
}

@article{chen2012variants,
  title={Variants of dynamic mode decomposition: boundary condition, Koopman, and Fourier analyses},
  author={Chen, Kevin K and Tu, Jonathan H and Rowley, Clarence W},
  journal={Journal of nonlinear science},
  volume={22},
  number={6},
  pages={887--915},
  year={2012},
  publisher={Springer}
}

@article{jovanovic2014sparsity,
  title={Sparsity-promoting dynamic mode decomposition},
  author={Jovanovi{\'c}, Mihailo R and Schmid, Peter J and Nichols, Joseph W},
  journal={Physics of Fluids},
  volume={26},
  number={2},
  pages={024103},
  year={2014},
  publisher={American Institute of Physics}
}

@article{citriniti2000reconstruction,
  title={Reconstruction of the global velocity field in the axisymmetric mixing layer utilizing the proper orthogonal decomposition},
  author={Citriniti, Joseph H and George, William K},
  journal={Journal of Fluid Mechanics},
  volume={418},
  pages={137--166},
  year={2000},
  publisher={Cambridge University Press}
}

@book{lumley1970stochastic,
  title={Stochastic tools in turbulence},
  author={Lumley, John L},
  year={1970},
  publisher={Academic Press}
}

@article{towne2018spectral,
  title={Spectral proper orthogonal decomposition and its relationship to dynamic mode decomposition and resolvent analysis},
  author={Towne, Aaron and Schmidt, Oliver T and Colonius, Tim},
  journal={Journal of Fluid Mechanics},
  volume={847},
  pages={821--867},
  year={2018},
  publisher={Cambridge University Press}
}

@inproceedings{shinde2022features,
  title={Features of oblique shock wave turbulent boundary layer interaction},
  author={Shinde, Vilas J and Gaitonde, Datta V},
  booktitle={AIAA SciTech 2022 Forum},
  pages={1975},
  year={2022}
}

@article{shinde2021supersonic,
  title={Supersonic turbulent boundary-layer separation control using a morphing surface},
  author={Shinde, Vilas J and Gaitonde, Datta V and McNamara, Jack J},
  journal={AIAA Journal},
  volume={59},
  number={3},
  pages={912--926},
  year={2021},
  publisher={American Institute of Aeronautics and Astronautics}
}

@article{shinde2025lagrangian,
  title={Lagrangian Stability Analysis Technique for Fluid Flows},
  author={Shinde, Vilas J},
  journal={arXiv preprint arXiv:2509.10316},
  year={2025}
}

@article{shinde2025distributed,
  title={Distributed-parallel proper orthogonal/dynamic mode decompositions of large flow data},
  author={Shinde, Vilas},
  journal={Computer Physics Communications},
  pages={109644},
  year={2025},
  publisher={Elsevier}
}

@article{gaitonde2015progress,
  title={Progress in shock wave/boundary layer interactions},
  author={Gaitonde, Datta V},
  journal={Progress in Aerospace Sciences},
  volume={72},
  pages={80--99},
  year={2015},
  publisher={Elsevier}
}

@article{sirovich1987,
  title={Turbulence and the dynamics of coherent structures. Part I: Coherent structures},
  author={Lawrence Sirovich},
  journal={Quaterly of Applied Mathematics},
  volume={XLV},
  number={3},
  pages={561--571},
  year={1987},
  publisher={Brown University}
}

@article{beresh2002relationship,
  title={Relationship between upstream turbulent boundary-layer velocity fluctuations and separation shock unsteadiness},
  author={Beresh, SJ and Clemens, NT and Dolling, DS},
  journal={AIAA journal},
  volume={40},
  number={12},
  pages={2412--2422},
  year={2002},
  publisher={AMERICAN INST OF AERONAUTICS AND ASTRONAUTICS}
}

@article{dussauge2006unsteadiness,
  title={Unsteadiness in shock wave boundary layer interactions with separation},
  author={Dussauge, Jean-Paul and Dupont, Pierre and Debi{\`e}ve, Jean-Fran{\c{c}}ois},
  journal={Aerospace Science and Technology},
  volume={10},
  number={2},
  pages={85--91},
  year={2006},
  publisher={Elsevier}
}

@article{piponniau2009simple,
  title={A simple model for low-frequency unsteadiness in shock-induced separation},
  author={Piponniau, S{\'e}bastien and Dussauge, JP and Debieve, JF and Dupont, P},
  journal={Journal of Fluid Mechanics},
  volume={629},
  pages={87--108},
  year={2009},
  publisher={Cambridge University Press}
}

@article{touber2009large,
  title={Large-eddy simulation of low-frequency unsteadiness in a turbulent shock-induced separation bubble},
  author={Touber, Emile and Sandham, Neil D},
  journal={Theoretical and Computational Fluid Dynamics},
  volume={23},
  number={2},
  pages={79--107},
  year={2009},
  publisher={Springer}
}

@article{berkooz1993proper,
  title={The proper orthogonal decomposition in the analysis of turbulent flows},
  author={Berkooz, Gal and Holmes, Philip and Lumley, John L},
  journal={Annual review of fluid mechanics},
  volume={25},
  number={1},
  pages={539--575},
  year={1993},
  publisher={Annual Reviews 4139 El Camino Way, PO Box 10139, Palo Alto, CA 94303-0139, USA}
}

@article{pasquariello2017unsteady,
  title={Unsteady effects of strong shock-wave/boundary-layer interaction at high Reynolds number},
  author={Pasquariello, Vito and Hickel, Stefan and Adams, Nikolaus A},
  journal={Journal of Fluid Mechanics},
  volume={823},
  pages={617--657},
  year={2017},
  publisher={Cambridge University Press}
}

@article{clemens2014low,
  title={Low-frequency unsteadiness of shock wave/turbulent boundary layer interactions},
  author={Clemens, Noel T and Narayanaswamy, Venkateswaran},
  journal={Annual Review of Fluid Mechanics},
  volume={46},
  pages={469--492},
  year={2014},
  publisher={Annual Reviews}
}

@article{adams2000direct,
  title={Direct simulation of the turbulent boundary layer along a compression ramp at M= 3 and Re $\theta$= 1685},
  author={Adams, Nikolaus A},
  journal={Journal of Fluid Mechanics},
  volume={420},
  pages={47--83},
  year={2000},
  publisher={Cambridge University Press}
}

@article{ganapathisubramani2009low,
  title={Low-frequency dynamics of shock-induced separation in a compression ramp interaction},
  author={Ganapathisubramani, B and Clemens, NT and Dolling, DS},
  journal={Journal of Fluid Mechanics},
  volume={636},
  pages={397--425},
  year={2009},
  publisher={Cambridge University Press}
}

@article{pirozzoli2006direct,
  title={Direct numerical simulation of impinging shock wave/turbulent boundary layer interaction at M= 2.25},
  author={Pirozzoli, Sergio and Grasso, Francesco},
  journal={Physics of Fluids},
  volume={18},
  number={6},
  pages={065113},
  year={2006},
  publisher={AIP}
}

@article{touber2009comparison,
  title={Comparison of three large-eddy simulations of shock-induced turbulent separation bubbles},
  author={Touber, Emile and Sandham, Neil D},
  journal={Shock Waves},
  volume={19},
  number={6},
  pages={469},
  year={2009},
  publisher={Springer}
}

@article{adler2018dynamic,
  title={Dynamic linear response of a shock/turbulent-boundary-layer interaction using constrained perturbations},
  author={Adler, Michael C and Gaitonde, Datta V},
  journal={Journal of Fluid Mechanics},
  volume={840},
  pages={291--341},
  year={2018},
  publisher={Cambridge University Press}
}

@article{nichols2017stability,
  title={Stability and modal analysis of shock/boundary layer interactions},
  author={Nichols, Joseph W and Larsson, Johan and Bernardini, Matteo and Pirozzoli, Sergio},
  journal={Theoretical and Computational Fluid Dynamics},
  volume={31},
  number={1},
  pages={33--50},
  year={2017},
  publisher={Springer}
}

@article{dupont2019compressible,
  title={Compressible mixing layer in shock-induced separation},
  author={Dupont, Pierre and Piponniau, S{\'e}bastien and Dussauge, JP},
  journal={Journal of Fluid Mechanics},
  volume={863},
  pages={620--643},
  year={2019},
  publisher={Cambridge University Press}
}

@article{andreopoulos2000shock,
  title={Shock wave—turbulence interactions},
  author={Andreopoulos, Yiannis and Agui, Juan H and Briassulis, George},
  journal={Annual review of fluid mechanics},
  volume={32},
  number={1},
  pages={309--345},
  year={2000},
  publisher={Annual Reviews 4139 El Camino Way, PO Box 10139, Palo Alto, CA 94303-0139, USA}
}

@inproceedings{helm2014characterization,
  title={Characterization of the shear layer in a Mach 3 shock/turbulent boundary layer interaction},
  author={Helm, Clara and Martin, M Pino and Dupont, Pierre},
  booktitle={Journal of Physics: Conference Series},
  volume={506},
  pages={012013},
  year={2014},
  organization={IOP Publishing}
}

\end{document}